\documentclass[aps,prd,twocolumn,amsmath,amssymb,amsfonts,nofootinbib,superscriptaddress]{revtex4-1}
\usepackage[utf8]{inputenc}
\usepackage{graphicx}
\usepackage{subfigure}
\usepackage{datatool}
\usepackage[scientific-notation=true]{siunitx}
\usepackage[dvipsnames]{xcolor}
\usepackage{lineno}
\usepackage{xspace}
\usepackage{rotating}

\usepackage{orcidlink}
\usepackage{url}
\usepackage{bm}

\usepackage{breakurl}
\usepackage{lineno}
\usepackage[shortlabels]{enumitem}
\usepackage[capitalise]{cleveref}

\usepackage{etoolbox}
\AtBeginEnvironment{figure}{\nolinenumbers}
\AtBeginEnvironment{figure*}{\nolinenumbers}

\def\gw{GW\xspace}
\def\gwh{GW\xspace}
\def\cwh{CW\xspace}
\def\cw{CW\xspace}
\def\uldm{ultralight dark matter\xspace}
\def\uldmh{ultralight dark-matter\xspace}
\def\tbh{TBH\xspace}
\def\tbhs{TBHs\xspace}

\def\l{\left(}
\def\r{\right)}

\def\cws{CWs\xspace}
\def\gws{GWs\xspace}

\def\ns{neutron star\xspace}
\def\nss{neutron stars\xspace}
\def\nsh{neutron-star\xspace}

\def\et{ET\xspace}

\def\ce{Cosmic Explorer\xspace}
\def\dm{DM\xspace}
\def\dmh{DM\xspace}
\def\pbh{PBH\xspace}
\def\bh{BH\xspace}
\def\bhs{BHs\xspace}
\def\pbhs{PBHs\xspace}
\def\bbhs{binary black holes\xspace}
\def\bbh{binary black hole\xspace}

\def\bns{binary neutron star\xspace}

\def\lvk{LIGO, Virgo and KAGRA\xspace}
\def\lvkg{LIGO, Virgo, KAGRA and GEO600\xspace}
\def\ifo{interferometer\xspace}
\def\ifos{interferometers\xspace}
\def\hfgw{HFGW\xspace}

\def\pn{PN\xspace}
\def\mf{matched filtering\xspace}
\def\psd{power spectral density\xspace}
\def\psds{power spectral densities\xspace}
\def\nn{\nonumber}
\def\gc{Galactic Center\xspace}
\def\msp{millisecond pulsar\xspace}
\def\msps{millisecond pulsars\xspace}
\def\bc{boson cloud\xspace}
\def\bcs{boson clouds\xspace}
\newcommand{\lpf}{LISA Pathfinder\xspace}
\def\sgwb{stochastic \gwh background\xspace}
\def\sgwbs{stochastic \gwh backgrounds\xspace}
\def\eos{equation-of-state\xspace}
\def\moi{moment of inertia\xspace}
\def\mois{moment of inertias\xspace}
\def\df{dynamical friction\xspace}
\def\imbh{IMBH\xspace}
\def\imbhs{IMBHs\xspace}
\def\imri{IMRI\xspace}
\def\imris{IMRIs\xspace}
\def\toa{ToA\xspace}
\def\toas{ToAs\xspace}

\def\snr{SNR\xspace}
\def\emri{EMRI\xspace}
\def\emris{EMRIs\xspace}
\def\smh{standard-model\xspace}
\def\sm{standard model\xspace}
\def\evc{\text{ eV}}

\newcommand{\bea}{\begin{eqnarray}}
\newcommand{\eea}{\end{eqnarray}}
\newcommand{\be}{\begin{equation}}
\newcommand{\ee}{\end{equation}}
\newcommand{\thetathr}{\theta_\text{thr}}
\newcommand{\rh}{\rho}

\newcommand{\fs}{\alpha}
\newcommand{\tinst}{\tau_{\rm inst}}
\newcommand{\tgw}{\tau_{\rm GW}}
\newcommand{\mbh}{M_{\rm BH}}
\newcommand{\mb}{m}

\newcommand{\mdm}{m_{\rm DM}}
\newcommand{\Mdm}{M_{\rm DM}}
\newcommand{\rdm}{r_{\rm DM}}

\newcommand{\dd}{\ensuremath{\mathrm{d}}}

\newcommand{\rhoDM}{\rho_{\text{DM}}}
\newcommand{\TFFT}{T_\text{FFT}}

\newcommand{\Tobs}{T_\text{obs}}
\newcommand{\Tcoh}{T_{\text{coh}}}
\newcommand{\Lcoh}{L_{\text{coh}}}

\newcommand{\Nfft}{N_\text{FFT}}

\newcommand{\ssm}{sub-solar mass\xspace}
\newcommand{\msun}{\ensuremath{M_\odot}\xspace}

\newcommand{\pta}{pulsar timing array\xspace}
\newcommand{\ptas}{pulsar timing arrays\xspace}

\newcommand{\M}{M}
\newcommand{\Mij}{\M_{ij}}

\newcommand{\hij}{h_{ij}}
\newcommand{\epij}{\vep_{ij}}

\newcommand{\gev}{\,\mathrm{GeV}}

\newcommand{\ev}{\,\mathrm{eV}}

\newcommand{\mpl}{M_{\text{Pl}}}

\newcommand{\vep}{\varepsilon}

\def\erfc{\mathrm{erfc}}

\graphicspath{{./figures/}}

\begin{document}

\title{Gravitational wave probes of particle dark matter: a review}


\author{Andrew L. Miller\,\orcidlink{0000-0002-4890-7627}}
\email{andrew.miller@nikhef.nl}
\affiliation{Nikhef -- National Institute for Subatomic Physics,
Science Park 105, 1098 XG Amsterdam, The Netherlands}
\affiliation{Institute for Gravitational and Subatomic Physics (GRASP),
Utrecht University, Princetonplein 1, 3584 CC Utrecht, The Netherlands}

\date{\today}

\begin{abstract}

Various theories of dark matter predict distinctive astrophysical signatures in gravitational-wave sources that could be observed by ground- and space-based laser interferometers. Different candidates—including axions, dark photons, macroscopic dark matter, WIMPs, and dark-matter spikes—may appear in interferometer data via their coupling to gravity or the Standard Model, altering the measured gravitational-wave strain in distinct ways. Despite their differences, these candidates share two key features: (1) they can be probed through their effects on gravitational waves from inspiraling compact objects, isolated black holes, and neutron stars, or via direct interactions with detectors, and (2) their signatures likely persist far longer than the seconds-long mergers detected today, necessitating new data analysis methods beyond matched filtering. This review outlines these dark matter candidates, their observational signatures, and approaches for their detection.

\end{abstract}

\maketitle

\begingroup
\onecolumngrid
\tableofcontents
\endgroup
\clearpage
\twocolumngrid
\section{Introduction}\label{sec:intro}

Our knowledge of physics has been incomplete for decades, if not for over a century \cite{Bertone:2016nfn}. Dark energy and dark matter (\dm) comprise 69\% and 26\% of the universe, respectively, and yet we have not been able to directly detect either. The experimental evidence for the existence of \dm is overwhelming: stars orbiting the center of the spiral galaxies would have flown out of their orbits if \dm did not hold them in place \cite{Freeman:1970mx,rogstad1972gross,whitehurst1972high,roberts1973comparison};  anisotropies in the cosmic microwave background can result from \dm potential wells in the early universe \cite{Peebles:1982ff,Burles:1997fa}; \dm can seed the formation of large-scale structures visible today, which would have been impossible with ordinary matter alone \cite{Pagels:1981ke,davis1982survey,Blumenthal:1984bp,Davis:1985rj,Navarro:1995iw}; and gravitational lensing of light by galaxies in between us and the light source can be explained by \dm inside the galaxy \cite{Paczynski:1985jf,SupernovaCosmologyProject:1993faz,Aubourg:1993wb,MACHO:2000qbb,Lasserre:2000xw,EROS-2:2006ryy}.

While the observational evidence shows that \dm exists, it does not explain what it is made of. Can it be a single new particle, such as the axion \cite{Peccei:1977hh,Peccei:1977ur,Weinberg:1977ma} or dark photon \cite{Galison:1983pa,Holdom:1985ag,Pierce:2018xmy,Filippi:2020kii}, or macroscopic, in the form of \pbhs formed within $\sim 1$ second of the Big Bang \cite{Green:2020jor}? Or, can particle \dm exist alongside \pbhs, which may comprise a portion or the totality of \dm \cite{Hawking:1971ei,Carr:2009jm,Dai:2009hx,Clesse:2015wea,Carr:2016drx,Clesse:2016vqa,Clesse:2020ghq,Miller:2024rca}?

The numerous hypotheses regarding the origins of \dm suggest that the mass of its constituents can vary across hundreds of orders of magnitude \cite{Belenchia:2021rfb}. Significant time and resources, however, have been spent on searching a somewhat narrow mass range comprising so-called Weakly Interacting Massive Particles (WIMPs) \cite{Jungman:1995df}. WIMPs have been probed through collider experiments \cite{Goodman:2010ku} and \dmh / nucleus interactions by experiments such as CDMS \cite{CDMS:2003yjm,CDMS:2005rss,CDMS:2008uih}, SuperCDMS \cite{SuperCDMS:2018gro}, XENON \cite{XENON:2007uwm,XENON10:2011prx,XENON100:2011cza,XENON:2016jmt,XENON:2024wpa}, and  LZ (LUX-Zeppelin) \cite{LZ:2011rhn,LUX:2012kmp,LUX:2016sci,Mount:2017qzi,LZ:2019sgr}. Other experiments have tried to detect annihilating \dm \emph{indirectly} through its gamma-ray emission in the \gc, galaxy clusters, or other areas, e.g. MAGIC \cite{MAGIC:2011vay,Aleksic:2014poa}, VERITAS \cite{Holder:2008ux}, HESS \cite{HESS:2006fka} and Fermi \cite{Hooper:2010mq}, or through cosmic rays, e.g. AMS \cite{AMS01:2007rrn,AMS:2013fma,Bergstrom:2013jra} or Fermi-LAT \cite{Fermi-LAT:2009yfs,Fermi-LAT:2011baq}.
Despite these efforts, however, WIMPS remain undetected\footnote{DAMA claims to see a periodic oscillation of a \dmh signal, though this result has not yet been reproduced \cite{Petriello:2008jj}}, and so it is worthwhile to ask whether other models can better explain \dm. Thus, the experimental development from the particle physics community has widened to include probing ultralight ($\ll 1$ eV) \dm through (1) torsion balance experiments, such as Eöt-Wash \cite{Su:1994gu,Schlamminger:2007ht} and the MICROSCOPE satellite \cite{Touboul:2012ui,Berge:2017ovy}, whose results can be interpreted as constraints on \dm, (2) resonant cavity experiments, such as ADMX \cite{Du:2018uak}, (3) the Event Horizon telescope \cite{Doeleman:2009te,EventHorizonTelescope:2019uob}, which can measure photons lensed by \bcs and thus affect \bh images \cite{Chen:2022kzv}, and which can measure changes in the polarization of the electric field when the light passes through the axion background \cite{Chen:2021lvo,Chen:2021lvo}, and (4) astrometry experiments, such as Gaia \cite{Gaia:2016zol} and the Roman Space Telescope \cite{Akeson:2019biv}, that can measure stochastic metric perturbations sourced by \uldm \cite{Kim:2024xcr}. In light of the diversity of experiments mentioned, it is also worth asking whether high-precision \gwh interferometers, such as \lvkg \cite{Affeldt:2014rza,Dooley:2015fpa,2015CQGra..32g4001L,2015CQGra..32b4001A,Aso:2013eba}, can be used to search for \dm, as previously reviewed in \cite{Bertone:2019irm,Bertone:2024rxe}.

While it has been demonstrated that constraints on axions can be set using current observations of \bns inspirals (GW170817) \cite{Hook:2017psm,Croon:2017zcu,Zhang:2021mks} and the Hulse-Taylor binary \cite{KumarPoddar:2019jxe}, in this review, we would like to explore how continuous gravitational-wave (\cws) can be used as a probe of \dm. 
\cws are quasi-monochromatic, quasi-infinite-duration signals that canonically arise from asymmetrically rotating, lumpy \nss. Over the last few decades, much research has focused on developing ways to probe such \nss, both those that are known electromagnetically and those that may only be emitting \gws (``gravitars'') \cite{Sieniawska:2019hmd,Tenorio:2021wmz,Riles:2022wwz,Piccinni:2022vsd,Miller:2023qyw,Wette:2023dom}. Though the signal model is simple, the unknown sky positions and the uncertain \nsh physics (e.g. the unknown equation of state, ``spin wandering'' \cite{Akmal:1998cf,Cutler:2002}, changes in the magnetosphere \cite{Lyne:2010ad}, etc.) complicate these searches, and thus methods that are robust not only to non-Gaussian noise disturbances but also to theoretical uncertainties had to be developed to look for such systems \cite{Krishnan:2004sv,Dhurandhar:2007vb,Astone:2014esa,Suvorova:2016rdc,Bayley:2019bcb}. Until recently, such searches have only yielded constraints on the maximum possible deformation of the size of known \cite{LIGOScientific:2020lkw,LIGOScientific:2021hvc,Ashok:2021fnj,Ashok:2024fts,LIGOScientific:2025kei} and unknown \nss \cite{KAGRA:2022dwb,Steltner:2023cfk}.

In this review, we show how \cwh methods have been generalized to search for the following  diverse set of \dmh models:

\begin{enumerate}
\item \textbf{Ultralight bosons (scalar, vector, or tensor)}:
Oscillating fields can couple to \smh particles, leading to periodic modulations in the \ifo arm lengths, refractive indices, \nsh \mois, or timings of atomic clocks \cite{Khmelnitsky:2013lxt,Pierce:2018xmy,Guo:2019ker,grote2019novel,Ismail:2022ukp}. \cwh methods are sensitive to these narrowband, quasi-monochromatic signals and can be applied to searches for this kind of \dm in ground-based \ifos, \ptas, or future space-based antennas. Depending on the particular \ifo, \uldm masses between $[10^{-22},10^{-11}]$ eV can be probed, showing how instruments not even designed to search for \dm are able to cover a wide range of masses and thus a variety of \dmh models. 
\item \textbf{Macroscopic dark matter objects}:
Compact dark objects of masses of $\mathcal{O}(1-10^9)$ kg passing through or near \gwh \ifos could induce transient but quasi-periodic disturbances in the test masses \cite{Hall:2016usm,Du:2023dhk}. Searches using \mf can place constraints on Yukawa-type interactions or on the rate of such transient events. Even though individually, each transient would manifest as short-lived perturbation of the detector, if multiple encounters were to occur over the observation time, the resulting signature would appear as a stochastic background of transients. Potentially, \cwh methods could be applied to look for the periodicity or stochastic background of such transients.  Moreover, networks of \ifos could help discriminate these events from terrestrial noise, providing both spatial and temporal correlations that enhance detection prospects.
\item \textbf{Boson clouds around rotating black holes}:
Ultralight bosons can grow around black holes via superradiance, accelerated when the Compton wavelength of the boson matches the size of the \bh, forming clouds that extract energy and momentum from the \bh and emit nearly monochromatic \gws through annihilation or level transitions \cite{Arvanitaki:2010sy,Arvanitaki:2014wva,Brito:2015oca,Brito:2017zvb}. \cwh searches can target these long-lived signals and test the existence of bosons in the $[10^{-13},10^{-11}]$ eV mass range in ground-based \gwh \ifos, and between $[10^{-17},10^{-15}]$ eV in future space-based \gwh antennas. 
\item \textbf{Soliton dark matter}:
Gravitational interactions can lead to the formation of solitonic cores of \uldm at the centers of galaxies \cite{Schive:2014dra,Hui:2016ltb}. Compact binaries or isolated \nss moving through solitons experience additional gravitational potentials that can modify the phase evolution of the \gws as binary systems inspiral, and as \nss spin down, respectively. For binaries, this manifests as a long-lived, cumulative dephasing effect during the inspiral \cite{Aghaie:2023lan}, while for isolated nonaxisymmetric \nss, frequency  modulations in the emitted \cws \cite{Blas:2024duy} are induced. In both cases, the effect is coherent over long timescales, making \cwh search methods particularly well suited to test these scenarios. Depending on the soliton mass and size, signatures would be detectable with \pta data (probing supermassive binary inspirals crossing galactic cores) or with next-generation terrestrial detectors such as \et and \ce (probing stellar-mass binaries and isolated pulsars embedded in soliton potentials).
\item \textbf{WIMP dark matter}:
Even though WIMPs are not inherently wave-like, their capture in stars can induce their collapse into \bhs \cite{Singh:2020wiq,Bhattacharya:2024pmp}. \cwh searches can probe this scenario indirectly, by looking for the long-lived inspirals of \bhs space-based \gwh \ifos. Additionally, null results from \cwh searches from \msps in the \gc can indirectly preference WIMP \dm as responsible for the gamma-ray GeV excess emanating from the center of the Milky Way \cite{Calore:2018sbp,Miller:2023qph}.
\item \textbf{Dark matter spikes around compact objects}:
Dense \dmh profiles near \bhs inspiraling can generate \df as the secondary object flows through the \dmh spike of the first, leading to small but measurable dephasing in these long-lived inspirals \cite{Gondolo:1999ef,Eda:2013gg,Eda:2014kra,Kavanagh:2020cfn,Coogan:2021uqv,Cole:2022ucw,Cole:2022yzw,Bertone:2024rxe}. While originally such effects were considered around extreme-mass ratio inspirals (\emris), intermediate-mass ratio inspirals (\imris) and ordinary binaries have been shown to allow the presence of \dmh spikes, and such dephasing could be measurable in \et, \ce or space-based \gwh \ifos, when inspiraling binaries will spend hours, days or years in-band. \cwh methods can thus test the presence and properties of such spikes.
\item \textbf{Atomic dark matter}: 
If \dm possesses a composite structure with a dark sector analogue of electromagnetism, it could form bound states or ``dark atoms'' \cite{Feng:2009mn,Kaplan:2009de,Buckley:2017ttd,Shandera:2018xkn}.  Such atomic \dm can exhibit dissipative dynamics, allowing it to cool and collapse differently from standard cold \dm. This can lead to the formation of compact dark objects, modify \bbh merger populations, or even produce \ssm binaries that would be invisible electromagnetically. 
While \gwh searches have constrained only the presence of such atomic \dm using \mf searches, planetary- or asteroid-mass systems (masses $<0.1\msun$) could also have been formed from atomic \dm \cite{Fan:2013tia,Fan:2013yva,Buckley:2017ttd}, and would emit longer-duration \gws. Thus, \cwh searches can be applied to look for such systems \cite{Miller:2021knj,KAGRA:2022dwb,Miller:2024fpo}.
\end{enumerate}

In each of aforementioned scenarios, considering \dm within the framework of \cws allows us to adapt methods originally designed to probe \nss to search for \dm. This is especially important given the variety and complexity of the underlying physics in different \dm models, as well as the non-stationary, non-Gaussian noise characteristic of \gwh detectors. Traditionally, matched filtering has been the primary technique for detecting \gwh signals from \bbh and \bns mergers, in which known templates are correlated with observed data to extract weak signals embedded in noise. However, this approach becomes less effective when dealing with exotic scenarios, such as those discussed in this review, where the signals may not conform to standard templates, or where the creation of non-traditional templates is computationally infeasible.

Importantly, we consider two general scenarios for the detection of \dm using \gwh interferometers:
(1) \dm interacts either gravitationally or via couplings to \sm particles, leaving a direct imprint on interferometers, similar to traditional direct detection experiments; and
(2) \dm sources or modifies \gwh emission itself, which is then observed by \ifos.

In \cref{fig:summary-dm-direct}, we present a landscape of the types of particle \dm that can be probed with current and future \ifos in the context of the first scenario. For each DM candidate, the figure maps out the \ifo component or astrophysical system it affects, the resulting observable, and the classes of interferometers capable of detecting the effect.  Note that scenario (1)  does \emph{not} lead to \gwh signals; rather, the ``signal'' arises from the coupling of \dm to the \sm.

In contrast, \cref{fig:summary-gw-dm} summarizes the second scenario, showing how \gwh signals can be sourced or modified by the presence of \dm.
Together, these plots provide a high-level overview of the key models and observables discussed throughout this review.

\begin{figure*}[htbp]
    \centering
    \includegraphics[width=\textwidth]{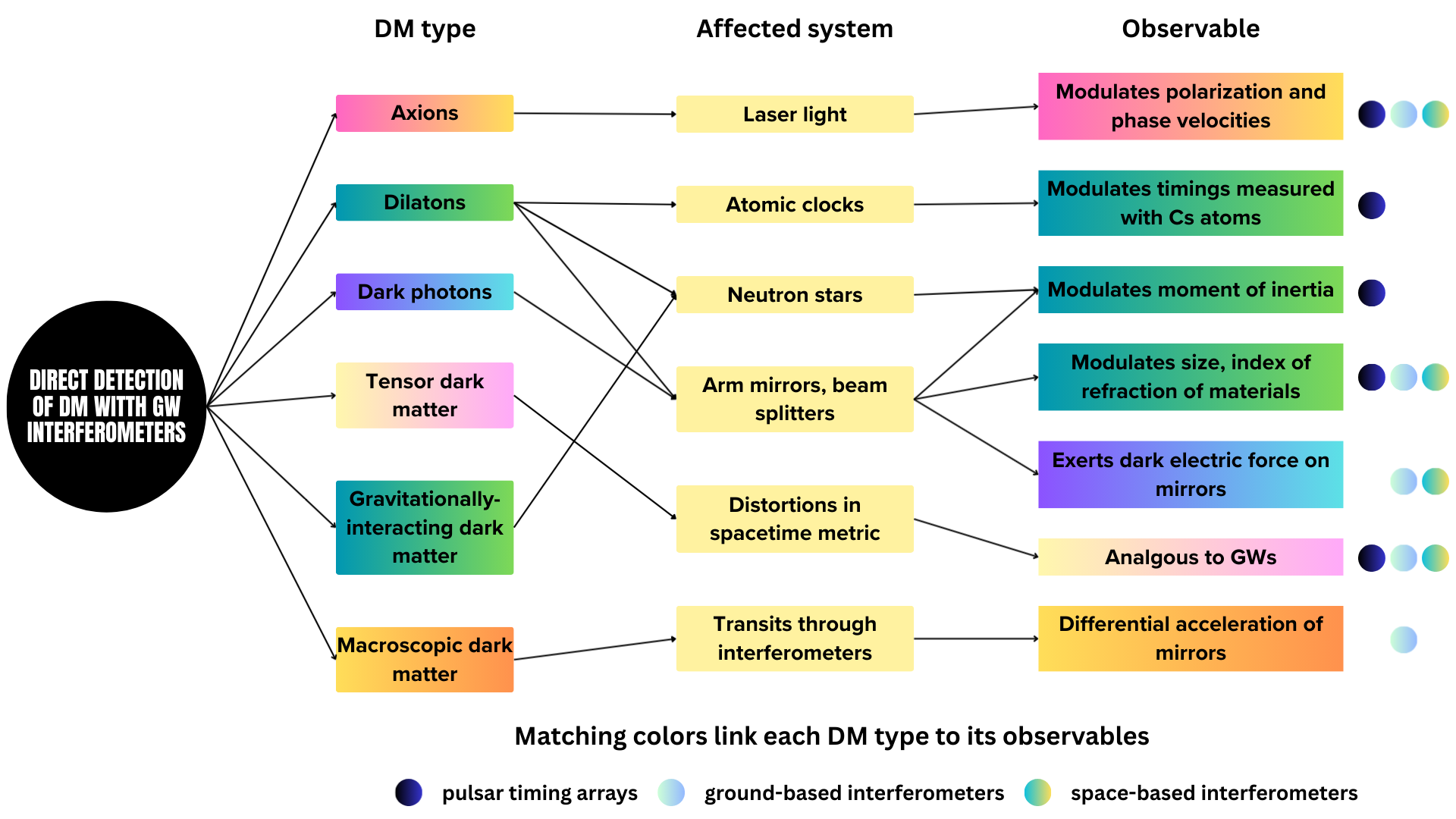}
    \caption{Landscape plot showing the various types of dark matter (\dm) that can be directly probed using gravitational-wave (\gwh) \ifos, both presently and in future experiments. The flow of logic proceeds from left to right: type of \dm $\rightarrow$ affected astrophysical system or detector component $\rightarrow$ observable effect $\rightarrow$ relevant detector(s). This framework spans a wide mass range of particle \dm, from $\mathcal{O}(10^{-23}-10^{-21})$ eV with \ptas to $\mathcal{O}(10^{-16}-10^{-13})$ eV with space-based detectors to $\mathcal{O}(10^{-13}-10^{-11})$ eV with ground-based \ifos, as well as [1,$10^9$] kg macroscopic \dm. Matching colors link each DM type to its corresponding observable.}
   \label{fig:summary-dm-direct}
\end{figure*}

\begin{figure*}[htbp]
    \centering
    \includegraphics[width=\textwidth]{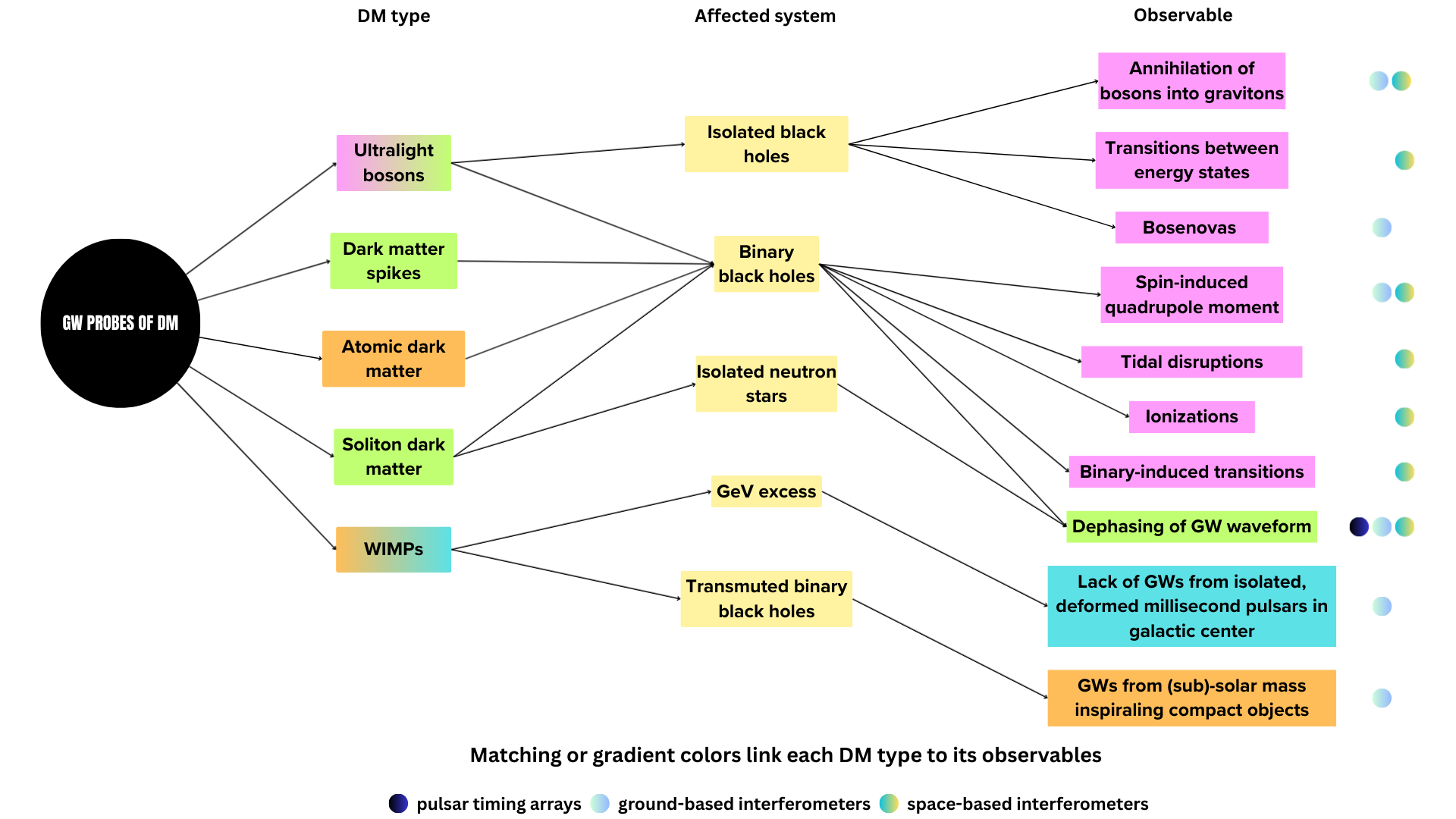}
    \caption{Similar to \cref{fig:summary-dm-direct}.  This plot shows the kinds of \dm that can be probed via their generation of \gws, both now and in the future. The logic is as follows, moving from left to right: type of \dm $\rightarrow$ astrophysical source affected $\rightarrow$ observable $\rightarrow$ which detector(s) the signal could be seen in. We note that the mass range of particle \dmh probed here ranges many orders of magnitude, from $\mathcal{O}(10^{-23}-10^{-21})$ eV with \ptas to $\mathcal{O}(10^{-16}-10^{-13})$ eV with space-based detectors to $\mathcal{O}(10^{-13}-10^{-11})$ eV with ground-based \ifos, and includes $[10^3,10^9]$ GeV WIMP \dm, probes of large-scale \dmh structure via solitons and \dmh spikes, and atomic \dmh. Matching colors imply that the source would induce those physical signatures.}
   \label{fig:summary-gw-dm}
\end{figure*}

This review article is broken into different parts depending both on the type of \dm considered and constraints on \dmh interactions with gravity or \sm particles. In \cref{sec:pdm}, we describe different ways that \uldm could couple to \sm particles in \gwh \ifos. \cref{sec:obsconstrain} details the constraints that have been derived from searches of recent \lvk data on the different ways that \dm could couple to the \ifos. \cref{sec:macdm} focuses on how macroscopic \dm could transit through the \ifos, giving rise to a measurable differential acceleration of the mirrors. In \cref{sec:gwbc}, we describe how scalar and vector bosons form around rotating \bhs, and the types of \gwh signals that could be emitted if these \bhs are isolated or are in binary systems. Search results for annihilating scalar and vector \bc systems are explained in \cref{sec:gwconsbc}. Next, we discuss how soliton \dm would impact \gwh signals from binary systems and \nss in \cref{sec:soliton}. The collection of WIMPs around \bhs and in the \gc are discussed in \cref{sec:wimp} in the context of how they can be probed with \gws. We explain how \dmh spikes around rotating \bhs could alter \gwh signals in \cref{sec:dmspike}. In \cref{sec:atomicdm}, we describe a model of atomic \dm that would give rise to ``dark atoms'' and ``dark \bhs'', and how it can be constrained with \gwh observations and non-observations of merging compact objects. We conclude in \cref{sec:concl} about the future for the burgeoning field of \gwh probes of particle \dm.

Note that we use natural units throughout this review: $c=\hbar=\epsilon_0=1$.



\section{Ultralight \dm interacting with \gwh interferometers}\label{sec:pdm}

Numerous models for \dm exist throughout the literature, and one review cannot possibly cover all of them. Instead, in this section, we focus specifically on \uldm models that have been shown to cause observable signals in \gwh \ifos. Over the last decade, the impact of more and more \dmh models on ground- and space-based instruments has been considering, allowing the field of ``direct \dmh searches with  \gwh \ifos'' to grow. The models we present here should be taken as the \textit{minimum} number of ways in which \dm could affect \gwh \ifos. In \cref{subsec:generic}, we outline some general properties of all \uldm considered in this review, and then delve into four types of \dm that couple to the \sm: axions (\cref{subsec:axions}), dilatons (\cref{subsec:dilaton}), vector, dark photon \dm (\cref{subsec:vecdm}), tensor \dm (\cref{subsec:tensor}). We end with a scalar \dmh field that couples only to gravity (\cref{subsec:grav-int}).

\subsection{Generic features of \uldm signals}\label{subsec:generic}

Cold, ultralight \dm could interact with \smh particles in many model-dependent ways. Such interactions would cause macroscopic differences in the materials, and, depending on the type of \dm, would lead to an observable signal in different components of \gwh \ifos. While the physics behind each type of \dm is different, there are some characteristics that are model-independent. First, the number of ultralight \dmh particles in a given region in space is gigantic, and can be calculated by attributing the \dmh energy density $\rho_\text{DM}$ to result from ultralight \dm:

\bea
N_{0}&=&\lambda_{\rm C}^3 \frac{\rhoDM}{\mdm}= \left(\frac{2\pi}{\mdm v_0}\right)^3\frac{\rhoDM}{{\mdm}}, \nonumber \\
&\approx& 1.69\times 10^{54}\left(\frac{10^{-12} \text{ eV}}{\mdm}\right)^4,
\eea 
where $v_0\simeq 220$ km/s is the virial velocity \cite{smith2007rave}, $\mdm$ is the mass of the \uldm particle, $\lambda_{\rm C}$ is the Compton wavelength of \dm, and $\rhoDM=0.3\gev/$cm$^3$ is the \dm energy density \cite{deSalas:2020hbh}.
The large occupation number $N_0$ implies that ultralight \dm can be approximated as a plane wave that oscillates at a fixed frequency $f_0$, which is given by the \dmh mass:

\be
f_0=\frac{\mdm}{2\pi} \simeq 241\text{ Hz}\left(\frac{\mdm}{10^{-12} \evc}\right), \label{eqn:f0dp} 
\ee
In other words, the observable effect is that the \ifo components will forever oscillate at a fixed frequency since they always sit in the \dmh field \cite{Pierce:2018xmy}. 

Despite the fixed frequency implied by \cref{eqn:f0dp}, we note that in reality, the \dmh field is not a perfectly monochromatic wave because the constituent particles have a finite Maxwell-Boltzmann velocity dispersion set by the virial velocity $v_0$ of the galactic halo. This spread in velocities causes a small but finite spread in the oscillation frequency, which in turn limits the coherence of the field. As a result, the \dmh wave can be treated as coherent only over a characteristic time scale $\Tcoh$, after which the accumulated phase dispersion between different velocity components causes the signal to lose coherence \cite{Carney:2019cio}. The wave coherent time $\Tcoh$ is given by \cite{Carney:2019cio}:

\begin{equation}
T_\text{coh}=\frac{4\pi}{\mdm v_0^2}=1.4\times 10^4 \text{ s} \left(\frac{10^{-12}  \text{ eV}}{\mdm}\right),\label{eqn:tcoh}   
\end{equation}
Practically speaking, we can observe the interaction of \dm with ground-based \gwh \ifos for as long as the detectors are on. Since observing runs of \lvk span around one year, we will always be in a regime in which the signal will not be monochromatic and thus have its power stochastically distributed at frequencies slightly higher than the characteristic frequency of the \dmh particle. The following expression sets the range of frequencies for which this happens:

\be
\Delta f_{v}=\frac{1}{2}v_0^2 f_0\approx 2.94\times 10^{-7}f_0. \label{eqn:deltafv}
\ee
Additionally, a finite coherence time also implies a finite coherence length $\Lcoh$:

\begin{equation}
\Lcoh=\frac{2\pi}{\mdm v_0}=1.6\times10^9 \text{ m}  \left(\frac{10^{-12}  \text{ eV}}{\mdm}\right). \label{eqn:lcoh}
\end{equation}
This length scale characterizes the maximum separation over which two detectors can still observe the \dmh-induced oscillations with the same phase. Since the arm lengths and baselines of current ground-based \gwh \ifos are much smaller than $\Lcoh$, we expect them to record correlated signals at the same frequencies. However, $\Lcoh$ increases with decreasing $\mdm$, which will not be a problem for spaced-based \gwh \ifos ($\Lcoh$ will still exceed arm separation and $\Tcoh$ will greatly exceed the observation time), but does require careful treatment when using \ptas, in which $\Lcoh$ is of the same order as the separation between pulsars. This will be discussed further in \cref{subsec:ptas}.

The finite $\Lcoh$ and $\Tcoh$ naturally imply that cross-correlation and other \cwh methods are natural choices to look for this kind of \dm, which will be detailed in \cref{subsec:meth-pdm}. Additionally, we show a concrete example of a potential \uldmh signal in \cref{fig:dp_hoft_asd}. Here, simulated vector \dmh particles are interacting with one of the LIGO detectors, which produces a stochastic, narrow-band signal in the frequency domain.

\begin{figure*}[ht!]
     \begin{center}
        \subfigure[ ]{%
            \label{hoft}
            \includegraphics[width=0.5\textwidth]{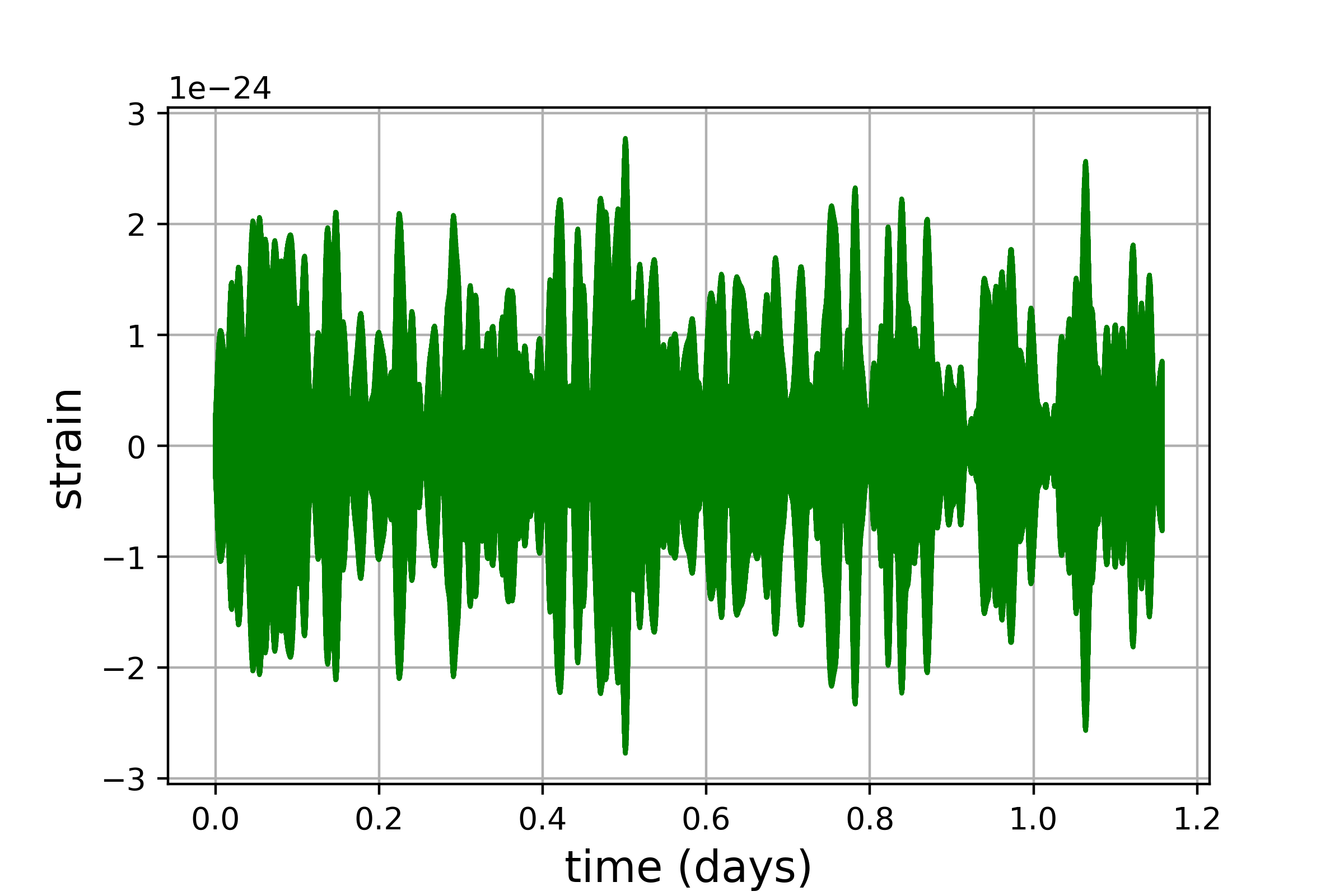}
        }%
        \subfigure[]{%
           \label{asd}
           \includegraphics[width=0.5\textwidth]{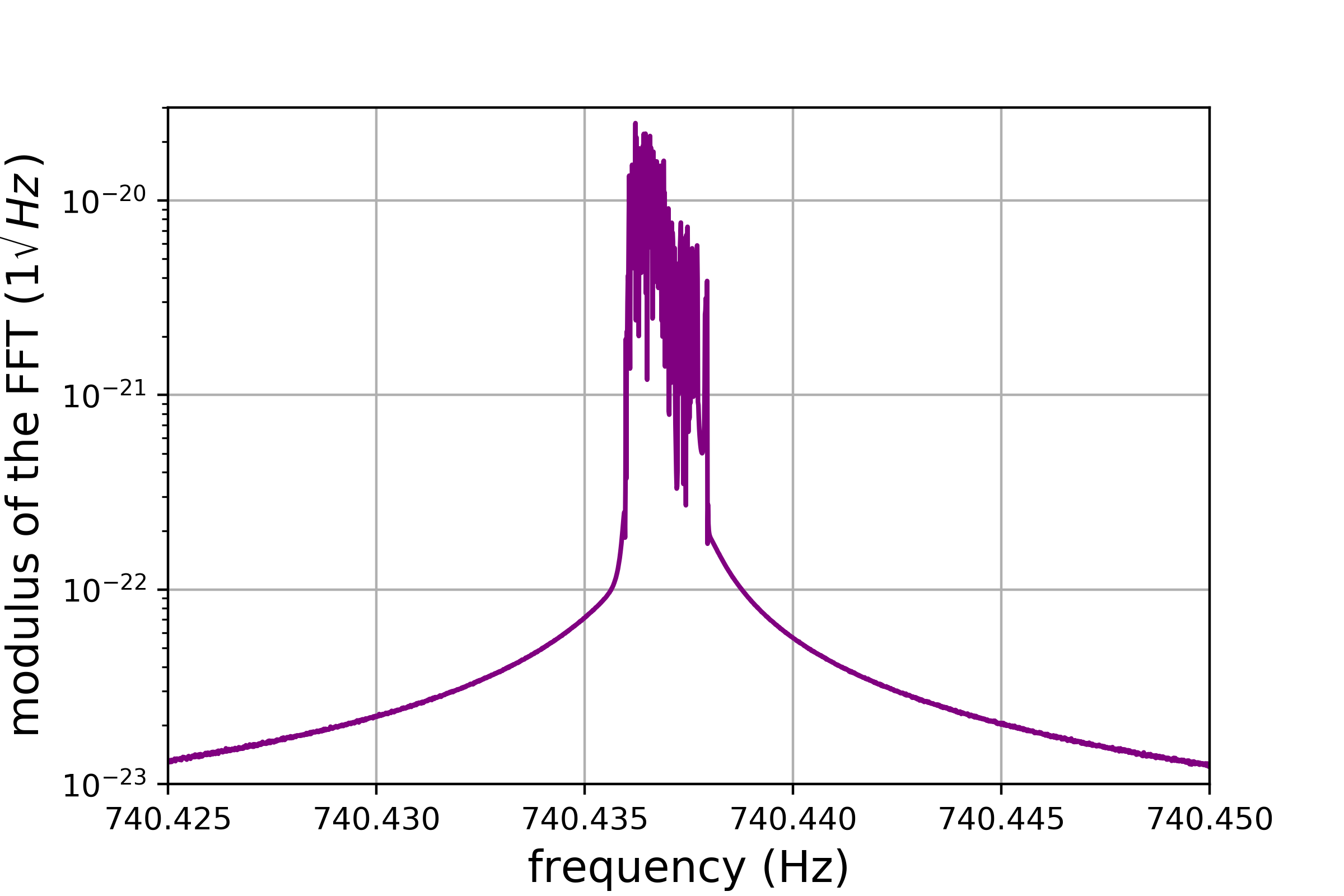}
        }\\ 
    \end{center}
    \caption[]{%
   Taken from \cite{Miller:2020vsl}. We show the strain time series $h(t)$ (left) and the Fourier transform of it (right) of a simulated dark photon \dm signal. The structure in the amplitude spectrum density in the right-hand plot arises because we have taken the length of the Fourier transform to exceed that of the coherence time of the signal. This dark photon \dmh signal is a superposition of $1000$ dark photons traveling with distinct Maxwell-Boltzmann-distributed velocities, which cause small frequency deviations away from the minimal frequency $f_0=740.436$ Hz ($m_A=3.06\times 10^{-12}$ eV). This plot shows why it is important to choose $\TFFT\sim \Tcoh$ for the analysis of \dmh signals. Though we have simulated a dark photon signal, the power spectrum will appear similarly for the other kinds of \dmh interaction signals. Here, $\Tobs\sim 10^5$ s, meaning the frequency resolution is $\delta f=10^{-5}$ Hz. The coherence time and length of this signal are: $T_\text{coh}=4595$ s and $L_\text{coh}=5.28\times 10^8$ m; the coupling strength is $\epsilon=3\times10^{-21}$. The signal is actually simulated for $\sim 233$ days, though we only show the first day of its time evolution.
     }%
   \label{fig:dp_hoft_asd}
\end{figure*}

We will now describe different ways in which \dm could interact with particles in the \ifos. We will assume that we are working within $\Tcoh$, in order to avoid technical difficulties from working with particles of slightly different frequencies. However, we note that in practice, this consideration is necessary when assessing the sensitivity of \gwh \ifos to \dmh/\smh interactions \cite{Pierce:2018xmy,Miller:2020vsl,Vermeulen:2021epa,Gottel:2024cfj}.

\subsection{Axions}\label{subsec:axions}
 
Axions were originally proposed as a solution to the strong CP problem in QCD \cite{Peccei:1977hh,Peccei:1977ur}, and more generally, axion-like particles arise in many extensions of the \sm. If sufficiently light and weakly interacting, they can also constitute a viable component of \dm. A key feature of axions is their ability to couple to \smh fields, in particular photons, through a Chern–Simons interaction. This coupling is described by the Lagrangian \cite{Chern:1974ft}:

\begin{align}
    \mathcal L_I &= \frac{ g_{\text{a}\gamma} }{4} a(t) F_{\mu\nu} \tilde F^{\mu\nu} \nn  \\ &= g_{\text{a}\gamma} \dot{a}(t) \epsilon_{ijk} A_i \partial_j A_k,
\end{align}
where $a(t)$ is the axion field, $g_{\text{a}\gamma}$ is the axion-photon coupling constant, $F_{\mu\nu}$ is field strength of electromagnetic field, $\tilde F^{\mu\nu}\equiv \epsilon^{\mu\nu\rho\sigma}F_{\rho\sigma}/2$ is its Hodge dual with the Levi-Civita anti-symmetric tensor $\epsilon^{\mu\nu\rho\sigma}$, the dot denotes the time derivative, $\epsilon_{ijk}$ is Levi-Civita tensor, and $A_\mu$ is 
the electromagnetic vector potential. $F_{\mu \nu} \equiv \partial_\mu A_\nu - \partial_\nu A_\mu$. Because the coherence length of the axion is much larger than typical experimental setups, the axion field can be treated as homogeneous; thus, all spatial derivatives of the axion field here are neglected.

In Fourier space, we can write $A_i$ in terms of two circular polarization modes:
\begin{equation}
A_i(t, \bm{x}) = \sum_{\lambda = \text{L}, \text{R}}\int\dfrac{\dd^3k}{(2\pi)^3}A_\lambda(t,\bm k)\,e^\lambda_i(\hat{\bm{k}})\,e^{i\bm{k}\cdot\bm{x}} \ ,
\end{equation}
where $\bm{k}$ is the wave number vector, $k \equiv |\bm{k}|$, and ``L'' and ``R'' denote the left- and right-hand polarizations, and $\bm{x}$ is a spatial point at which the \dm field is measured.
By completely fixing gauge degrees of freedom in which the divergence of the spatial part of the field is zero and its temporal component vanishes ($\partial_i A_i = 0$ and $A_0 = 0$), and noting that the following conditions are satisfied:
\begin{enumerate}
    \item The two circularly polarized states given by $\pm k $ are related through complex conjugation to ensure $A_i$ is real in position space: $e^\lambda_i(\hat{\bm{k}}) = e^{\lambda*}_i(-\hat{\bm{k}})$
    \item The two circularly polarized states are orthonormal: $e^\lambda_i(\hat{\bm{k}})e^{\lambda'*}_i(\hat{\bm{k}}) = \delta^{\lambda\lambda'}$
    \item The two circularly polarized states are eigenvectors of the curl operator in Fourier space with eigenvalues of $\pm k$: $i\epsilon_{ijm}k_je^{\text{L/R}}_m(\hat{\bm{k}}) = \pm k e^{\text{L/R}}_i(\hat{\bm{k}})$,
\end{enumerate}
the equations of motion for $A_{\text{L/R}}(t,\bm k)$ can be derived \cite{Nagano:2019rbw}:

\begin{align}
\ddot{A}_{\text{L/R}}+\omega^2_{\text{L/R}}A_{\text{L/R}}&=0 \\ 
\omega_{\text{L/R}}^2&=k^2\left(1\mp g_{\text{a}\gamma} \dot{a}/k\right),
\end{align}
where $k$ is the momentum and $\omega_{\rm L/R}$ is the angular frequency of the light. Here, we can see a modified dispersion relationship based on the strength of the axion-photon coupling.

Noting that the axion field can be written as (assuming to work within one $\Tcoh$:
\begin{equation}
    a(t) = a_0\cos(\mdm t) = \left(\frac{\sqrt{2 \rhoDM}}{\mdm }\right)\cos(\mdm t)
\end{equation} 
where $a_0$ is the axion field amplitude normalized by the \dmh energy density, and $t$ is time. 
Using the ordinary photon dispersion relationship is $c_0=\frac{\omega}{k}$, we can derive how the velocities of different polarizations of light will differ:

\begin{align}
\frac{c_\text{\text{L/R}}(t)}{c} &\simeq 1 \pm \delta c(t) 
\equiv 1 \pm \delta c_0 \sin(\mdm t), \label{eq:SoLofL} 
\end{align}
where $\delta c_0=g_{\text{a}\gamma}a_0m/(2k)$ is the maximum change in the phase velocity of light induced by the axion, and $\delta c_0\ll 1$ is assumed. In physical quantities, $\delta c_0$ is the amplitude of the signal we expect, and is equal to:
\begin{align}
\delta c_0 \simeq 1.3 \times 10^{-24} \left( \frac{\lambda}{1550 \text{ nm}}\right) 
\left( \frac{g_{\text{a}\gamma}}{10^{-12} \text{ GeV}^{-1}}\right),
\end{align}
in which the laser light is assumed to have a wavelength that relates to $k$ as: $\lambda = 2\pi/k$.

\cref{eq:SoLofL} implies that in the presence of an axion field, the phase velocities of circularly polarized photons in \ifos are modulated. In order to detect this effect in \lvk instruments, additional but simplistic optical components are required in order to measure the optical path difference between two orthogonal polarizations of light ($p$- and $s$-polarizations) \cite{Nagano:2021kwx}. In practice, \gwh \ifos use linearly polarized light ($p$-polarized), which becomes partially $s$-polarized due to axion-induced modulation. Similar deviations in polarization can also be probed with \ptas \cite{Xue:2024zjq,EPTA:2024gxu}, and experimental constraints on axions from them will be discussed in \cref{subsubsec:axion-constr}. Notably, ground-based \ifos and \ptas are sensitive to very different axion mass ranges, roughly $[10^{-13},10^{-11}]$ eV and $[10^{-23},10^{-21}]$ eV, respectively, but can achieve comparable absolute constraints on the axion-photon coupling $g_{\text{a}\gamma}$, which vary strongly with the axion mass. However, no constraints exist yet for this effect from ground-based \ifos.

\subsection{Dilaton scalar dark matter}\label{subsec:dilaton}

Another class of \uldm candidates are dilatons, scalar fields that can couple to \smh particles and induce time-dependent variations in fundamental constants \cite{Preskill:1982cy,Abbott:1982af,Dine:1982ah,Cho:1998aa,Cho:2007cy,Arvanitaki:2014faa,Stadnik2015a,Stadnik2015b,Stadnik2016}. In particular, a dilaton field can cause the electron mass and other physical constants to oscillate at the dilaton frequency. 
Similarly to the axion, the scalar \uldmh field $\phi$ can be written as \cite{Arvanitaki:2014faa,Derevianko:2016vpm,Vermeulen:2021epa}:
\begin{equation}
        \phi(t,\vec{x}) = \left(\frac{\sqrt{2 \rho_{\rm{DM}}}}{\mdm }\right) \cos\left(\mdm t - \vec{k} \cdot \vec{x} + \Psi\right),\label{eqn:dilatonphi}
\end{equation}
where $\vec{x}$ is a position vector, $\vec{k} =
\mdm\vec{v}_0$ is the wave vector, and $\Psi$ is a phase factor, and has a corresponding Lagrangian $\mathcal{L}_\mathrm{int}$ \cite{Vermeulen:2021epa}:

\begin{equation}\label{eqn:L_int}
    \mathcal{L}_\mathrm{int} \supset \frac{\phi}{\Lambda_\gamma} \frac{F_{\mu\nu}F^{\mu\nu}}{4}  - \frac{\phi}{\Lambda_e} m_e \bar{\psi}_e \psi_e,
\end{equation}
where $\psi_e$ and $\bar{\psi}_e$ are the standard-model electron field and its Dirac conjugate, $m_e$ is the mass of the electron, and $\Lambda_\gamma$ and $\Lambda_e$ denote the scalar \dmh coupling parameters to the photon and electron, respectively. 

We will now show how the fundamental constants will change in response to the dilaton field. To begin, we write a part of the \sm Lagrangian that will be compared to \cref{eqn:L_int}:

\begin{equation}
\label{eqn:SM_Lagrangian}
\mathcal{L}_\textrm{SM} \supset - \frac{F_{\mu\nu}F^{\mu\nu}}{4} - m_e \bar{\psi}_e \psi_e - e J_\mu A^\mu,
\end{equation}
where $J^\mu$ is a current term. From comparing \cref{eqn:L_int} and \cref{eqn:SM_Lagrangian} term-wise, we can see that the rest mass of the electron effectively changes in the presence of \dm:

\begin{equation}
    m_e' = m_e\l1+\frac{\phi}{\Lambda_e}\r\label{eqn:meprime}
\end{equation}
Additionally, the four-vector potential must be canonically normalized to reproduce Maxwell's equations, so:

\begin{equation}
    A'_\mu = A_\mu \sqrt{1- \frac{\phi}{\Lambda_\gamma}},
\end{equation}
which further implies that the charge of the electron $e$ changes as:

\begin{equation}
    e' = \frac{e}{\sqrt{1- \frac{\phi}{\Lambda_\gamma}}}
\end{equation}
and thus the fine structure constant changes as:

\begin{equation}
    \alpha' = \frac{e'^2}{4\pi} = \alpha \l1- \frac{\phi}{\Lambda_\gamma}\r^{-1} \simeq \alpha \l 1+\frac{\phi}{\Lambda_\gamma}\r \label{eqn:alphaprime}
\end{equation}
Thus, the fundamental constants oscillate in response to the dilaton field. In the following subsections, we will discuss how these variations induce measurable strain in ground-based interferometers and pulsar timing arrays through their effects on solids, neutron stars, and atomic clocks.

\subsubsection{Changes to size and indices of refraction of solids}
\label{subsubsec:size-dilaton}

The oscillations in fundamental constants induced by a dilaton field lead to modulations of the Bohr radii of atoms in \ifo components \cite{grote2019novel}, leading to measurable changes in the size and refractive index of the beam splitter \cite{Vermeulen:2021epa}, and the arm mirrors \cite{Gottel:2024cfj}.
Oscillations in the beam splitter size cause photons traveling down each arm to traverse slightly different paths along its surface, producing a differential phase shift. Initially, it was assumed that the beam-splitter effect was the dominant contribution \cite{Vermeulen:2021epa}, but more recent work has shown that small differences in the thicknesses of the mirrors in each arm also lead to non-negligible path-length variations in the interferometer \cite{Gottel:2024cfj}.

We will now calculate the impact of dilatons on \gwh \ifos. A typical solid length can be estimated as roughly $l\sim N_A a_B$, where $N_A$ is the number of atoms in the body and $a_B = (m_e\alpha)^{-1}$ is the Bohr radius. Small changes in the electron mass or fine-structure constant then produce a fractional change in length, or strain, given by

\begin{align}
    \delta l &= -N_A(m_e\alpha)^{-2}(m_e\delta\alpha +\alpha \delta m_e) \nn \\
    \frac{\delta l}{l} &=  -\left(\frac{\delta\alpha(t)}{\alpha} + \frac{\delta m_{e}(t)}{m_e} \right) \nn \\
    \delta l &= l\phi\l \frac{1}{\Lambda_\gamma} + \frac{1}{\Lambda_e}\r
\end{align}
where $\delta$ refers to the change in each parameter:

\begin{align}
    \delta m_e &= m_e' - m_e = m_e \frac{\phi}{\Lambda_e}  \label{eqn:delta_m_e} \\ 
    \delta \alpha &= \alpha'-\alpha = \alpha\frac{\phi}{\Lambda_\gamma} \label{eqn:delta_alpha}
\end{align}
In the context of \gwh \ifos, this effect translates into a differential displacement of the beam splitter along the two arms. For the 
$x$-arm, assuming light incidence at 45 degrees and 50\% mirror reflectivity, the induced change in arm length, $\delta L_x$, can be computed as in \cite{grote2019novel}:

\begin{equation}
    \delta L_x \approx \delta [\sqrt{2} n l - l/ (2\sqrt{2}) - w/2],
\end{equation}
where $w$ is the thickness of the mirrors, $n$ is the index of the beamsplitter, and $l$ is the length of the beamsplitter. Along the $y$-arm, the change in arm length is \cite{Vermeulen:2021epa}:

\begin{equation}
    \delta L_y = -\delta l / (2\sqrt{2}) - \delta w/2,
\end{equation}
and finally, the differential length change is:

\begin{align}\label{eq:deltaL}
    \delta(L_x - L_y)  &\approx \sqrt{2}\left[\left(n-\frac{1}{2}\right)\delta l\right] \\ 
     &\approx \left(\frac{1}{\Lambda_\gamma} + \frac{1}{\Lambda_e} \right)\left(\frac{n\,l\,\sqrt{2\,\rho_{\mathrm{DM}}}}{\mdm\, }\right)\cos\left(\mdm t\right),
     \label{eqn:scalar}
\end{align}
We have only considered the contribution arising from the change in length of the beamsplitter, not from the change in index of refraction, because the former is much larger than the latter \cite{grote2019novel}.

In addition to dilatons interacting with atoms in the beamsplitter, the reference cavity of the detector can provide a region in which dilatons interact appreciably with light, allowing for potential detection. Laser light locked to an optical cavity made out of a solid material will have its frequency modulated due to the presence of a dilaton \dmh field, compared to light locked to a free-space
suspended cavity, whose frequency would remain stable \cite{Derevianko:2016vpm,Geraci:2018fax,Hall:2022zvi}. The strain amplitude of such a signal, which would be distinct from that in \cref{eqn:scalar}, can be written as:

\begin{align}\label{eqn:hdm-ref-cavity}
   {h}_\text{DM}(t)= \left[\frac{1}{\mdm}\sqrt{8\pi \rhoDM G} (d_e + d_{m_{e}})\right] \cos\mdm t,
\end{align}
where $d_e$ and $d_{m_e}$ relate to the other couplings as:

\begin{equation}
d_{e,m_e}=\frac{M_\mathrm{Pl}}{(\sqrt{4\pi}\Lambda_{\gamma,e})}, 
\end{equation}
where $M_\mathrm{Pl}$ is the Planck mass, and $m_e$ is the electron mass. 

The effects described in \cref{eqn:scalar} and \cref{eqn:hdm-ref-cavity} can both produce measurable signals in \ifos. Currently, however, only the spatial strain affecting mirrors and the beamsplitter has motivated dedicated search methods for dilaton dark matter (see \cref{subsec:meth-pdm}) and has been constrained using data from GEO600 and LIGO (see \cref{subsec:ground-space}). A search strategy targeting the temporal modulation of the laser frequency is still under development \cite{Hall:2022zvi}.

\subsubsection{Changes to neutron-star \moi}\label{subsubsec:change-moi}

Beyond laboratory-based detectors, dilaton-induced fluctuations in fundamental constants could also have observable consequences in astrophysical systems. For example, they could alter \moi of neutron stars \cite{Kaplan:2022lmz}, which would produce deviations in the arrival time of pulses, to be discussed in \cref{subsec:ptas}. In order to quantify the extent to which the \ns \moi will change, we must know how protons and neutrons will change in response to dilatons. This motivates us to consider the following interactions in QCD of quarks in the presence of dilatons \cite{Damour:2010rp,Kaplan:2022lmz}:
\begin{align}\label{eq:qcd_couplings}
	\mathcal{L}_{\phi, \rm QCD}\supset\frac{\phi}{\Lambda}\left(\frac{d_g\beta_3}{2g_3}G_{\mu\nu}^AG_A^{\mu\nu}-\sum_{q=u,d}(d_q+\gamma_q d_g)m_q\bar qq\right)
\end{align}
where $\beta_3$ is the QCD beta function, $g_3$ is the QCD gauge coupling, $G_{\mu\nu}^A$ is the gluon field strength tensor, $\gamma_q$ are the light quark anomalous dimensions, $\Lambda=\mpl/\sqrt{4\pi}$, $m_q$ is the quark mass, and $d_g$ and $d_q$ are the dilaton coupling constants to the gluon and the quarks, respectively.

Similarly to \cref{subsec:dilaton}, we can write the \smh Lagrangian for quarks:
\begin{equation}
\mathcal{L}_\text{quark}^\text{SM} = - \sum_q m_q \bar q q,
\end{equation}
which, upon comparing with \cref{eq:qcd_couplings}, allows us to read off the change in quark mass as
\begin{align}
    \frac{\delta m_q}{m_q} & = \frac{d_q}{\Lambda} \phi.
    \label{eq:light_quark_mass_shift}
\end{align}
The shift in nucleon masses can then be obtained by noting that nucleons are composite objects whose masses arise mostly from QCD binding energy, with a smaller contribution from the quark masses. The dilaton couples to both the quark masses and the gluon field strength, so the fractional change in a nucleon mass is a weighted sum of these effects \cite{Damour:2010rp,Kaplan:2022lmz}:
\begin{align}
    \frac{\delta m_{p,n}}{m_{p,n}} \simeq \frac{1}{\Lambda} \Big( d_g + C_n\, d_{\hat m} \Big) \phi \,,
    \label{eq:light_nucleon_mass_shift}
\end{align}
where $C_n$ accounts for the fraction of the nucleon mass due to the light quarks, and 
\begin{equation}
d_{\hat m} = \frac{m_u d_u + m_d d_d}{m_u + m_d}
\end{equation} 
is the effective quark coupling. In other words, the nucleon mass responds both to the direct change in quark masses induced by the dilaton and to the change in the gluon's binding energy, allowing us to propagate the fundamental dilaton couplings to macroscopic observables such as the neutron-star moment of inertia.


\cref{eq:light_quark_mass_shift,eq:light_nucleon_mass_shift} show how dilaton \dm will alter the masses of quarks and nucleons, which are the primary constitutions of \nss. Thus, these changed masses will impact the rotational frequency and moment of inertia of \nss, which will affect the times of arrivals (\toas) of pulses that come from them. To calculate these effects, let us recall that conservation of angular momentum requires that the fluctuations in the spin frequency $\delta\omega$ relate to changes in the pulsar's \moi $\delta I$ as:

\begin{align}\label{eqn:deltaI}
   \frac{ \delta \omega}{\omega_0} =- \frac{ \delta I}{I_0}\,.
\end{align}
where $I_0$ is the unperturbed \moi and $\omega_0$ is the unperturbed angular frequency of the \ns.
Assuming a simplistic model of a non-rotating, spherically symmetric \ns composed only of neutrons with a polytropic equation-of-state, the radius and \moi can be written as \cite{Shapiro:1983du}:

\begin{align}
R &\propto M_0^{-1/3} m_n^{-8/3} \label{eqn:R} \\
    I_0&\propto M_0R^2 = M^{1/3}m_n^{-16/3}, \label{eqn:I0}
\end{align}
where $R$ and $M_0$ are the radius and unperturbed mass of the \ns, respectively.  Plugging in \cref{eqn:R,eqn:I0} into \cref{eqn:deltaI}, we can derive how the \moi changes as a function of the \nsh mass and neutron mass:
\begin{align}
    \frac{ \delta I}{I_0} = \eta \frac{\delta M}{M_0} + \delta \eta \frac{\delta m_n}{m_n}\,.
\end{align}
where $\delta M$ is the change in the pulsar's mass, and $\eta = 1/3$ and $\delta \eta = -16/3$ in this simplistic model\footnote{Solving explicitly the Tolman–Oppenheimer–Volkoff (TOV) equations for a realistic \eos would cause $\mathcal{O}(1)$ deviations from the simple scalings. These ``unknown'' scalings can be parameterized in terms of $\eta$ and $\delta\eta$}.

Because of the changes in the pulsar's \moi, the pulsar's spin frequency will change. For \pta experiments, the measured quantities are the \toas of pulses from each pulsar, which will change in response to the oscillating \moi. In practice, the observable here is the residual $h$ of the \toas of pulses from the neutron star, which is given by:
\begin{align}\label{eq:df2h}
	h= -\int\frac{\delta\omega}{\omega_0}dt=\int\frac{\delta I}{I_0}dt\,.
\end{align}
Now, the changes in the particle mass must be related to the macroscopic change in the pulsar mass

\begin{align}\label{eq:p_mass_shift}
    \frac{\delta M}{M_0} & = \sum_{f \in \{ e, \mu, p, n \} } Y_f \frac{m_f}{m_n} \frac{\delta m_f}{m_f}\,.
\end{align}
The sum is over the different types of particles present in the pulsar: electrons, muons, protons and neutrons, $Y_f \equiv N_f/(N_n + N_p)$ is a weighting function that gives the number of $f$ particles relative to the sum of protons $N_p$ and neutrons $N_n$ (which dominate the pulsar's mass). For \nss observed by \ptas, the following is assumed for $Y_f$: $Y_n=0.9$, $Y_p=Y_e=0.1$ and $Y_\mu=0.05$ \cite{Bell:2019pyc,Kaplan:2022lmz}.

By plugging in the relations in \cref{eqn:alphaprime} and \cref{eqn:meprime} to \cref{eq:p_mass_shift}, and subsequently integrating over time in \cref{eq:df2h}, the changes in the timing residuals due to the \uldmh particles can be obtained:
\begin{align}
    h(t) = & \frac{\sqrt{2 \rhoDM}}{\mdm^2 \Lambda} \left( \vec{y} \cdot \vec{d} \,\right) \hat{\phi}_P \sin{\big( \mdm t + \gamma(\vec{x}_P) \big)}\,,
        \label{eq:pulsar_spin_fluctuation_signal}
\end{align}
where $\vec{x}_P$ is the location of the pulsar, $\hat{\phi}_P$ is the pulsar normalized signal amplitude, $\gamma$ is a phase, and $\vec{y} \cdot \vec{d} \equiv \sum_{i} y_i d_i$ where $i \in \{ m_e, m_\mu, g, \hat{m} \}$ and:
\begin{equation}
    \label{eq:y_PTA_PSF}
    \begin{split}
        \{y_g,\,y_{\hat m}, \,y_\mu,\,y_e\}=&\,\eta\left\{1,\,C_n,\, 6\times10^{-3},\, 5\times 10^{-5}\right\} \\ 
        +&\delta \eta\left\{1,\,C_n,\,0,\,0\right\} \, .
    \end{split}
\end{equation}
Using the above formalism, dilatons have been constrained with recent \pta experiments \cite{Porayko:2018sfa,EuropeanPulsarTimingArray:2023egv,NANOGrav:2023hvm}, which will be extensively discussed in \cref{subsec:ptas}.

\subsubsection{Changes to atomic clock times}\label{subsubsec:change-atomic}

Changes in fundamental constants would also affect the way in which atomic clocks measure time. Since \pta experiments reference their measurements of the \toas of pulses to terrestrial time (TT) using atomic clocks composed primarily of cesium, any oscillation of the fundamental constants will affect the atomic clocks, and thus the times measured \cite{Flambaum:2004tm}. Likewise, a shift in the frequency of the atomic clock will give rise to an apparent shift in the measured pulsar spin frequency, given by \cref{eq:df2h}.

From \cite{Flambaum:2004tm,Kaplan:2022lmz}, the frequency of an atomic clock is:

\begin{align}
    f \propto \Big(m_e \alpha^2 \Big)\Big[\alpha^2 F_{\rm rel}(Z\alpha)\Big]\left(\mu \frac{m_e}{m_p}\right)^\zeta
    \label{eq:clock_scaling}
\end{align}
where $Z$ is the nuclear charge, $F_{\rm rel}(Z\alpha)$ is a relativistic correction to an atom's energy levels, $\mu$ is the nuclear magnetic moment, and $\zeta=1$ if the clocks that have hyperfine transitions (the case of cesium), and $\zeta=0$ for optical transitions\footnote{This equation arises from solving the non-relativistic Schrodinger equation for the valence electron wavefunction and evaluating it at the nucleus of the atom.}.

The fluctuations in fundamental constants will cause modulations in the frequencies of atomic clocks composed of atoms $A$. Using previously derived expressions for variations in the electron mass, fine-structure constant, and quark and nucleon masses, the fractional frequency shift can be obtained \cite{Flambaum:2004tm,Kaplan:2022lmz}:
\begin{equation}\label{eq:clock_f_shift}
\begin{split}
	\frac{\delta f_A}{f_A}\simeq&\Bigg[\frac{\delta m_e}{m_e}+(4+K_A)\frac{\delta\alpha}{\alpha}\\
	 &+\zeta\Bigg(\frac{\delta m_e}{m_e}+C_A\sum_{q=u,d}\frac{\delta m_q}{m_q}-\frac{\delta m_p}{m_p}\Bigg)\Bigg]\, ,
\end{split}
\end{equation}
where $\delta F_{\rm rel}/F_{\rm rel}=K_A$, and the derivative of the nuclear magnetic moment is $\delta \mu/\mu=C_A\,\delta m_q/m_q$. For Cesium, $K_A=0.83$ depends on $Z$ and $\alpha$,  and $C_A=0.110$ \cite{Flambaum:2004tm}. The index $q$ runs over up and down quarks. 

Finally, by substituting the aforementioned equations for fluctuations of the fundamental constants (\cref{eqn:delta_m_e,eqn:delta_alpha,eq:light_quark_mass_shift,eq:light_nucleon_mass_shift}) into \cref{eq:clock_f_shift}, and then further plugging this result into \cref{eq:df2h}, the timing residuals induced by the \uldmh can be obtained:
\begin{align}
    h(t) = \frac{\sqrt{2 \rhoDM}}{\mdm^2 \Lambda} \left( \vec{y} \cdot \vec{d} \right) \hat{\phi}_E \sin{\big( \mdm t + \gamma(\vec{x}_E) \big)} \, .
    \label{eq:reference_clock_shift_signal}
\end{align}
where $\vec{x}_E$ is the position of the Earth, and the sensitivity parameters of this search are given by:
\begin{align}
    \label{eq:y_PTA_clock}
	\left\{y_g,\,y_\gamma,\,y_{\hat m},\,y_{m_e}\right\}\simeq\left\{\zeta,\,\xi_A,\,\zeta\left( C_n + \hat C_A \right),1+\zeta\right\}\,,
\end{align}
where $\xi_A\equiv4+K_A$, and $\hat C_A=C_A (m_u+m_d)^2/ 2m_u m_d$. This type of \dmh interaction would leave imprints on pulsar timing array experiments \cite{Porayko:2018sfa,EuropeanPulsarTimingArray:2023egv,NANOGrav:2023hvm}, which will be discussed in \cref{subsec:ptas}.

We would like to draw a parallel between \cref{eq:pulsar_spin_fluctuation_signal,eq:y_PTA_clock}. Both equations have a similar form, though the underlying physics is different, as indicated by the phase factors: the former depends on the phase at the \ns, while the latter depends on the phase on earth (at the atomic clocks). Of course, the sensitivity parameters of the search are different, which encode which couplings that each effect depends on. These two effects highlight an important point that will be made throughout this review: while the physical signatures of different kinds of \dm are different, often the impact on the signal (the residuals in \ptas, the phase shifts in ground- and spaced-based \ifos) are similar, and only the \emph{interpretation} of the lack of a detection is different.


\subsection{Vector dark matter}\label{subsec:vecdm}

So far, we have considered explain how spin-0 particles would affect \gwh \ifos; however, \dm can also have nonzero spins. Here, we consider a model for \dm that has spin-1, i.e. vectors, which we refer to as ``dark photons''. Like dilatons, dark photons could explain the entirety of the relic abundance of \dm, which could arise from the misalignment mechanism \cite{nelson2011dark,arias2012wispy,graham2016vector}, parametric resonance or the tachyonic instability of a scalar field \cite{agrawal2020relic,Co:2018lka,Bastero-Gil:2018uel,Dror:2018pdh}, or from cosmic string network decays \cite{long2019dark}. They could couple to \smh particles, either to baryon or baryon-lepton number. In the \ifos, these interactions would occur everywhere, but would be most pronounced in the four \ifo test masses in the main resonant cavities of the instrument. The dark photons\footnote{These are different dark photons than those that kinetically mix with the ordinary photon, which will be discussed later in the context of luminous superradiance \cite{Siemonsen:2022ivj}.} would cause a ``dark'' force on the mirrors, causing quasi-sinusoidal oscillations \cite{Pierce:2018xmy,Miller:2020vsl}, as will be shown below.

To derive the impact of dark photons on \gwh \ifos, we will begin by describing dark photons analogously to ordinary photons, i.e. with a four-vector potential:

\be
A_\mu(t, \vec{x})=(A_0)_\mu\sin(\mdm t -\vec{k} \cdot\vec{x}+ \Upsilon), \label{eqn:Amu}
\ee
where $(A_0)_\mu$ is the four-amplitude of $A_\mu$ $\Upsilon$ is a random phase, $\vec{k}$ is the wavevector, and $\vec{x}$ is the position at which $A_\mu$ is measured. The index $\mu={0,1,2,3}$ refers to both time and spatial components.

As always, we have the freedom to choose a gauge, and usually, the Lorenz gauge is easiest to work in ($\partial^\mu A_\mu=0$). After making this choice, we can compare the time component to magnitude of the spatial components of the four-potential:

\be
\frac{(A_0)_0}{|\vec{A}_0|} = v_0\simeq 7.667\times 10^{-4}, \label{eqn:A00}
\ee
where $|\vec{A}_0|=\left(\frac{\sqrt{2\rhoDM}}{\mdm}\right)$ is the magnitude of the spatial components of $A_\mu$, normalized by the present \dmh energy density of the universe. In \cref{eqn:A00}, we see that the dark scalar potential is suppressed by about three orders of magnitude compared to the the dark three-vector potential. Therefore, the time-component of the four-vector potential can be safely neglected, leaving only the three-vector potential:

\be
\vec{A}= \vec{A}_0\sin\left(\mdm t -\vec{k} \cdot\vec{x}+ \Upsilon\right),
\label{eqn:vecpot}
\ee

With the four-potential, we can write the Lagrangian $\mathcal{L}$ that characterizes the dark photon coupling to a number current density $J^\mu$ of baryons or baryons minus leptons:

\begin{equation}
    \mathcal{L} = -\frac{1}{4} F^{\mu \nu} F_{\mu \nu} + \frac{1}{2}\mdm^2 A^\mu A_\mu - \epsilon e J^\mu A_\mu, \label{eqn:dplagrangian} \\
\end{equation}
where $\epsilon$ is the strength of the particle/dark photon coupling normalized by the electromagnetic coupling constant.

Using \cref{eqn:Amu,eqn:vecpot}, we can compute the  ``dark'' electric and magnetic fields, analogously to electromagnetism, which will subsequently allow us to compute the ``dark'' force:

\begin{align}
\vec{E}&=\partial_0\vec{A}-\vec{\nabla} A_{0} \simeq \mdm  \vec{A}_0\cos(\mdm t -\vec{k}\cdot\vec{x}+\Upsilon), \\ 
\vec{B}&=\vec{\nabla} \times \vec{A}= -\vec{k} \times \vec{A}_0\cos(\mdm t -\vec{k}\cdot\vec{x}+\Upsilon),
\end{align}
It is important to understand the relative magnitudes of the electric and magnetic fields. Comparing them, we see that the electric field is much stronger than the magnetic one:

\be
\frac{|\vec{E}|}{|\vec{B}|}\sim \frac{\mdm}{|\vec{k}|}=\frac{1}{|\vec{v}|}\sim10^3,
\ee
which implies that we need to only consider the electric field. This dark electric field causes the test masses to oscillate at a quasi-fixed frequency given by the dark photon mass. The acceleration of a given mirror can thus be derived using \cref{eqn:dplagrangian,eqn:vecpot} \cite{Guo:2019ker,Pierce:2018xmy}:

\begin{align}
 \vec{a}_j(t,\vec{x}_j)&=
 \simeq\epsilon e \frac{Q_{D,j}}{M_j}\mdm |\vec{A}_0|\hat{A}\cos(\mdm t -\vec{k} \cdot\vec{x}_j+ \Upsilon), \nonumber \\
\label{eqn:accel}
\end{align}
where $Q_{D,j}$ is the total charge in the $j$th mirror of mass $M_j$, and $\hat{A}$ is a unit vector of the vector potential. If dark photons couple to the baryon number, $q_j$ is the number of protons and neutrons in each mirror; if they couple to the difference between the baryon and lepton numbers, $q_j$ is the number of neutrons in each mirror. Each mirror is positioned differently with respect to the incoming dark photon \dmh field, and thus is accelerated at a slightly different amount over time. Thus, the \ifo experiences a measurable differential acceleration, and thus a differential strain.

Assuming that all mirrors have the same charge-to-mass ratio, integrating \cref{eqn:accel} twice over time, and averaging over random polarization and propagation directions, the strain on the \ifos caused by a dark photon \dmh signal can be computed \cite{Pierce:2018xmy}:

\begin{align}
\sqrt{\langle h^2_D\rangle}&=C\frac{Q}{M}e \sqrt{2\rho_\text{DM}} v_0 \frac{\epsilon}{f_0}, \nonumber \\
&\simeq 6.56\times 10^{-26}\left(\frac{\epsilon}{10^{-22}}\right)\left(\frac{100 \text{ Hz}}{f_0}\right),
\label{eqn:dph0}
\end{align}
where $C=\sqrt{2}/3$ (for an interferometer with two perpendicular arms) is a geometrical factor obtained by averaging over all possible dark photon propagation and polarization directions, and different detector geometries (the calculation for $C$ is shown 
in the appendix of \cite{Pierce:2018xmy} and in \cite{Morisaki:2020gui}). 

A second strain results due to the so-called ``finite light travel time effect, in which the mirrors in the \ifos have moved in the time that it takes the light to reach them from the beamsplitter. This strain can actually be larger than that given in \cref{eqn:dph0} and can be derived \cite{Morisaki:2020gui}:

\begin{align}
\sqrt{\langle h_C^2\rangle}&=\frac{\sqrt 3}{ 2} \sqrt{\langle h_D^2\rangle}\frac{2\pi f_0 L}{v_0}, \nonumber \\
&\simeq 6.58\times 10^{-26}\left(\frac{\epsilon}{10^{-23}}\right).
\label{eqn:dpfiniteh0}
\end{align}
The total power is: $\langle h_{{\rm total}}^2\rangle = \langle h_D^2\rangle+\langle h_C^2\rangle$.

Both of these effects can be constrained by analyzing data from \lvk and \ptas, which will be discussed in \cref{subsec:ground-space,subsec:ptas}, though in different mass regimes.

\subsection{Tensor dark matter}\label{subsec:tensor}

Modifications to gravity due to an additional spin-2 particle could act as \dm. Specifically, bimetric gravity \cite{Hassan:2011zd,Schmidt-May:2015vnx}, a theory in which a massless and massive spin-2 field interact, provides a plausible candidate for \uldm \cite{Marzola:2017lbt}.

Here, we consider a massive spin-2 field \( \M_{\mu\nu} \) described by the Fierz-Pauli Lagrangian density \cite{Fierz:1939ix,Armaleo:2020yml}:

\begin{align}\label{PF}
  {\mathcal{L}} &\, := \frac{1}{2} \M_{\mu\nu}\mathcal{E}^{\mu\nu\rh\sigma}\M_{\rh\sigma} - \frac{1}{4} \mdm^2 \left( \M_{\mu\nu}\M^{\mu\nu} - \M^2 \right) \,, 
\end{align}
where \( \M := g^{\mu\nu} \M_{\mu\nu} \) and \( \mathcal{E}^{\mu\nu\rh\sigma} \) is the Lichnerowicz operator, defined in equation 2.2 of \cite{Armaleo:2020yml}.

For the Friedman-Lemaitre-Robertson-Walker (FLRW) background metric, the equations of motion for this field, assuming it is ultralight, can be derived at late times as in ~\cite{Marzola:2017lbt,Aoki:2017cnz,Kun:2018ino}. The field spatial component can be written in a similar way as for dark photons and scalar bosons at given position:

\begin{align}\label{eq:mij}
	\Mij & = \frac{\sqrt{2\rhoDM(\vec{x})}}{\mdm}\cos{\left(\mdm t+\vec{k}\cdot\vec{x}+\Upsilon(\vec{x})\right)}\vep_{ij}(\vec{x}) \,, 
\end{align}
where and $\vep_{ij}$ encodes the polarizations of the massive spin-2 field.

The strain on \gwh detectors arises analogously to that from \gws: a stretching of space-time in the presence of the field, since the massive spin-2 metric and its coupling constant can be absorbed into the definition of the massless spin-2 metric in the linear regime of the coupling constant $\alpha_{\rm TB}$\footnote{In the literature, $\alpha_{\rm TB}=\alpha$ simply, but to disentangle it from the fine structure constants, we rename it. TB is tensor boson.} \cite{Armaleo:2020yml}; thus, the perturbation $\hij$ is: 

\begin{align}
    \hij(t) &= \frac{\alpha_{\rm TB}}{\mpl}\Mij(t) = \frac{\alpha_{\rm TB}\sqrt{2\rhoDM}}{\mdm\mpl}\cos{\left(\mdm t+\Upsilon\right)}\epij(\vec{x}) \,,
\end{align}
where $\alpha_{\rm TB}$ is a dimensionless constant that quantifies the difference between the strengths of each of the spin-2 fields, and $\mpl$ is the Planck mass. 

To obtain the strain on \gwh \ifos, we note that the detector response function can be parameterized in terms of a tensor, often written as $D^{ij}$ \cite{Romano:2016dpx}. By contracting $\hij$ with $D^{ij}$, the resulting strain can be derived \cite{Armaleo:2020efr}.


\begin{equation}
    h(t) = D^{ij}h_{ij} = \frac{\alpha_{\rm TB}\sqrt{\rhoDM}}{\sqrt2 m\mpl}\cos{\left(\mdm t+\Upsilon\right)}\Delta\vep, \label{eq:signal}
\end{equation}
where $\Delta\vep := \,\epij (n^i n^j - m^i m^j)$ and $n,m$ are unit vectors pointing along directions $i,j$.

We note that an interesting discrepancy arises when comparing the results from Refs. \cite{Aoki:2016zgp,Manita:2023mnc} with those from Refs. \cite{Marzola:2017lbt,Armaleo:2020efr} regarding the strain induced by tensor dark matter. The two sets of works differ cause a factor of 2 difference in the strain, which can be traced back to a factor of eight difference in their respective Lagrangians. This difference stems from variations in the normalization factors and scaling of the tensor field amplitude, as well as differences in how the tensor field’s energy density is treated. Such discrepancies are not uncommon in the study of dark matter and may have implications for the interpretation of experimental constraints and the modeling of astrophysical signals. However, we note that no constraints on tensor \dm have been placed as of yet using \gwh \ifos. 

\subsection{Gravitationally interacting dark matter}\label{subsec:grav-int}

The previous sections assumed a coupling of \dm to \smh particles. However, in principle \dm need not interact at all with the \sm; thus, models of \dm have been devised in which only a minimal coupling to gravity is assumed. This is, in fact, the simplest assumption that can be made, since we only have evidence for the existence of \dm through gravitational interactions. Below, we consider two kinds of scalar \dm that couple only to gravity.

\subsubsection{Minimally coupled dark matter}

In minimal coupling models, \dm interacts with the \sm exclusively through gravity by contributing to the stress-energy tensor. The gravitational field responds to this energy density and pressure according to Einstein’s field equations, where the presence of dark matter modifies the curvature of spacetime in the same manner as any other form of energy or matter. Here, \uldm can source time-varying gravitational potentials, leading to observable effects such as oscillations in pulsar timing residuals, which will be discussed in \cref{subsec:ptas}.

A commonly studied example of a minimally coupled scalar \dmh field $\phi$ is described by the Lagrangian \cite{Khmelnitsky:2013lxt}:

\begin{equation}
    \mathcal{L}=\sqrt{-g}\left[\frac{1}{2}g^{\mu\nu}\partial_\mu\phi\partial_\nu\phi-\frac{1}{2}\mdm^2\phi^2\right]\,.
    \label{eq:action}
\end{equation}
where $g^{\mu\nu}$ is the metric tensor, and $g$ is the determinant of the metric tensor. In this model, the scalar field evolves:
\begin{equation}
\phi(\vec{x}, t)=\frac{\sqrt{2 \rhoDM}}{\mdm} \hat{\phi}(\vec{x}) \cos \left(\mdm t+\gamma(\vec{x})\right),
\label{eq:phi}
\end{equation}
where  
$\gamma(\vec{x})$ is a phase that depends on position $\vec{x}$, and $\hat{\phi}(\vec{x})$ describes the  variation in space of the \uldmh field. As before, the scalar field density is normalized by $\rhoDM$.

Scalar \dm can cause the arrival times of pulses from \msps on earth to change, in an analogous way that \gws do \cite{Khmelnitsky:2013lxt}, i.e. introduce stochastic fluctuations of the metric \cite{Kim:2023kyy,Kim:2023pkx}. Such a difference in the arrival times can be written as~\cite{Khmelnitsky:2013lxt, Porayko:2018sfa}:
\begin{equation}
    \delta t_\text{DM} = \frac{\Psi_\text{c}(\vec{x})}{2\mdm} [\hat{\phi}^2_\text{E}\sin{(2\mdm + \gamma_\text{E} )} - \hat{\phi}^2_\text{P}\sin{(2\mdm + \gamma_\text{P} )} ], 
    \label{eq:st}
\end{equation} 
where
\begin{equation}
    \Psi_\text{c}(\vec{x}) \approx 6.52 \times 10^{-18}\left(\frac{10^{-22~}\text{eV}}{\mdm}\right)^2 \left(\frac{\rho_\phi}{0.4~\text{GeV}/\text{cm}^3}\right),
\label{eq:psi_c}
\end{equation}
where $\gamma_\text{P} \equiv 2\gamma(\vec{x}_\text{p}) - 2 \mdm d_p$ ($\gamma_\text{E} \equiv 2\gamma(\vec{x}_\text{e})$) parameterize random phases evaluated at the pulsar (P) or Earth (E), and $d_p$ is the distance between the pulsar and the Earth. The \dmh energy density is assumed to be constant when calculating
$\Psi_\text{c}(\vec{x})$, though the possibility of deviations, due to the coherently oscillating \uldmh field, are parametrized in terms of two phase factors: one for the pulsar  $\hat{\phi}^2(\vec{x}_\text{p})\equiv \hat{\phi}^2_\text{P} $ and one for the Earth,$\hat{\phi}^2(\vec{x}_\text{e}) \equiv \hat{\phi}^2_\text{E}$ \cite{Porayko:2018sfa}. 

\subsubsection{Non-minimally coupled (conformal) dark matter}

Another form of \dm that couples only to gravity would be conformal \dm. This kind of \dm typically involve scalar fields that couple to the trace of the energy-momentum tensor. These models are closely related to the dilaton-type scalar DM discussed in \cref{subsec:dilaton}, but with specific couplings motivated by scalar–tensor theories of gravity, such as Fierz–Jordan–Brans–Dicke (FJBD) and Damour–Esposito–Farese (DEF) theories \cite{Fierz:1939ix,Fierz:1956zz,Dicke:1961gz,Jordan:1959eg,Damour:1992we,Damour:1993hw}.

In such models, \dm couples universally to the gravitational sector through a conformal rescaling of the metric. This coupling can induce a gravity-mediated force between \nss and the \uldm field, altering the \moi of rotating pulsars -- a manifestation of the Nordtvedt effect \cite{Nordtvedt:1968qs}. These variations lead to deterministic changes in pulsar \toa and constitute a violation of the equivalence principle. This kind of conformal scalar DM has been extensively studied and constrained using binary pulsar timing measurements \cite{Damour:1991rd,Kramer:2006nb}, which will be discussed in\cref{subsec:ptas}.

In the case of a linear FJBD coupling, the pulsar timing residuals take the form:

\begin{align}
    \Delta t(t) 
    &= 2\alpha s_I \left.\frac{\sqrt{\rhoDM}}{\mpl \mdm^2}\hat{\phi}(\vec{x}) \sin(\mdm t + \theta(\vec{x}))\right|_{t_{\text {start }}-d} ^{t_{\text {end }} -d},
    \label{eq:alphadep}
\end{align}
where $s_I$ is the angular moment sensitivity parameter computed in \cite{Kuntz:2024jxo} and is of $\mathcal{O}(1)$ in FJBD theory, $\alpha$ is the strength of the coupling, $\theta(\vec{x})$ is a random phase, and $\hat{\phi}$ is a stochastic parameter whose value depends on whether the correlated, uncorrelated or pulsar correlated scenarios are considered (see \cref{subsec:ptas} for a description of these terms).

In summary, both dilaton-type and conformal scalar \dmh predict couplings that can modify either the properties of matter or the effective gravitational interaction, leading to observable signatures. In ground-based \ifos, these effects manifest primarily as strain signals induced by oscillations in material properties or cavity frequencies, while in \ptas they appear as deterministic modulations in pulse arrival times due to variations in \nsh \mois or equivalence-principle–violating forces. These complementary channels allow terrestrial and astrophysical detectors to probe overlapping but distinct regions of parameter space. In the next section, we review the current observational constraints on such models, focusing on limits derived from \gwh \ifo data.




\section{Observational constraints on \dm directly interacting with \gwh interferometers} \label{sec:obsconstrain}

\cref{sec:pdm} outlined several theoretically motivated models of \dm that could leave detectable imprints in \gwh interferometer data, and we showed how these models map on to induced strains on the detectors. The natural next question is whether current and future \ifos are actually sensitive to such weak signals. In this section, we turn to the search strategies themselves: the methods used to look for \dm interactions with gravitational-wave detectors, the observational constraints that have already been obtained, and the prospects for upcoming instruments.

Each method introduced in \cref{subsec:meth-pdm} can be characterized by a sensitivity, i.e. the minimum detectable strain amplitude at a given confidence level. This quantity is what ultimately gets translated into limits on the physical couplings between \dm and the \sm, and does not typically depend on the \dmh model considered. In other words, these search techniques are model-agnostic: they look for correlated, excess power at each frequency the dataset, independent of the underlying nature of \dm. The interpretation comes later, when the observed sensitivities are framed as constraints on particular coupling constants and interaction scenarios.

After describing the methods, we present current observational constraints set by ground- and space-based instruments in \cref{subsec:ground-space}, and those set by \ptas in \cref{subsec:ptas}. We conclude in \cref{subsec:prospect-pdm} by discussing how future \gwh \ifos could set tighter constraints on, or potentially detect, a wider variety of \dmh models, as outlined in \cref{sec:pdm}.

\subsection{Methods}\label{subsec:meth-pdm}

As explained in \cref{subsec:generic}, the expected signal frequency is fixed by the \dmh mass, but has some stochastic variations of $\mathcal{O}(10^{-6})$ \cite{Centers:2019dyn}. Because the \ifos always exist in the \dmh field, the signal is always present. Thus, we can describe the expected signal as: quasi-monochromatic and quasi-infinite duration. Essentially, we are looking for a resonance at a particular frequency fixed by the \dmh mass.

If the signal were purely monochromatic, we could simply take a single fast Fourier transform of the data and look for peaks in the power spectrum. However, the stochastic frequency variations prevent us from doing that. If we observe for a duration $\Tobs$ longer than $\Tcoh$, the signal will not be sinusoidal and thus its power will be spread among different frequency bins, as shown in \cref{fig:dp_hoft_asd}, which would inhibit a possible detection. Thus, the following methods have been designed to combine signal power across chunks of data of length $\TFFT\ll \Tobs$ in such a way that avoid power loss and thus optimizes sensitivity towards particle \dmh interactions with the \sm and gravity. Essentially, these methods vary the length of the chunk of data $\TFFT$ to match the coherence time of \dm $\Tcoh$. 

Each of the methods described below formulates its sensitivity, detection statistic, and search strategy differently. As such, applying multiple methods to search for the same type of \dm is worthwhile, since each independently defines how to assess the significance of candidates, compute upper limits, set detection thresholds, and veto outliers that historically have arisen from instrumental or environmental noise. Using diverse approaches provides a valuable cross-check and increases the robustness of any potential detection or constraint.

\subsubsection{Cross-correlation}\label{subsubsec:crosscorr}

Conceptually, cross-correlation \cite{Pierce:2018xmy,Guo:2019ker} requires at least two separate time-domain datasets that are Fourier transformed and multiplied together to compute the cross-power in each frequency bin. The cross power is then divided by the auto-power (the \psd) of each detector, and then summed over all the Fourier transforms to arrive at a measure of power at each frequency. It is then divided by the standard deviation of the noise to compute the signal-to-noise ratio (\snr).

Mathematically, the cross-correlated signal strength for detector pair $IJ$ is \cite{Pierce:2018xmy,Guo:2019ker}:
\be
S_{IJ,j}=\frac{1}{N_{\rm FFT}}\sum_{i=1}^{N_{\rm FFT}}\frac{z_{I,ij}z_{J,ij}^*}{P_{I,ij}P_{J,ij}},
\ee
in the $j^{\rm th}$ frequency bin at the $i^{\rm th}$ time. ``$*$'' denotes the complex conjugate, and $z_{I,ij}$ and $z_{J,ij}$ is the Fourier transform of the time-domain data from detectors $I$ and $J$, respectively, and $\Nfft$ is the total number of FFTs taken over the observing run. If the observing run lasts $\Tobs$ and the FFT length is $\TFFT$, then $\Nfft=\Tobs/\TFFT$. $P_{I,ij}$ and $P_{J,ij}$ are the individual detector \psds, which are typically estimated using a running median over each FFT. Typically, $\TFFT=1800$ s, independently of the \dmh mass, which ensures a relatively straightforward analysis but implies the sensitivity towards \dmh interactions is not optimal at each mass analyzed.

In addition to the signal strength, we have to obtain an estimate of the variance of the noise data. This can be computed from the individual \psds in each FFT as \cite{Guo:2019ker}:
\be
\sigma_{IJ,j}^2=\frac{1}{N_{\rm FFT}}\left\langle\frac{1}{2P_{I,ij}P_{J,ij}}\right\rangle_{N_{\rm FFT}},
\ee
where $\langle...\rangle_{N_{\rm FFT}}$ is the average over $N_{\rm FFT}$ time segments. We note that the average is over inverse noise-weighted \psds, (analogous to adding resistors in parallel), which helps to suppress spurious power due to large noise disturbances.

The \snr, the detection statistic in each frequency bin, is then simply:
\be
{\rm SNR}_{IJ,j}=\frac{S_{IJ,j}}{\sigma_{IJ,j}}.\label{eqn:snr}
\ee
In the presence of pure Gaussian noise, the \snr will follow a normal distribution, with a mean of 0 and a standard deviation of 1. If the \snr exceeds a certain threshold, which is set both theoretically assuming Gaussian noise and by the trials factor (accounting for the size of the parameter space), then a particular frequency is classified as being ``significant''.

Depending on the type of \uldm being searched for, the detectability of a signal in cross-correlation searches between interferometers depends on the spin of the DM: spin-0 (scalar), spin-1 (vector), or spin-2 (tensor) \cite{Manita:2023mnc}. The \snr for a given detector pair is determined by the \dmh energy density fraction in that spin state, the strength of its couplings to \smh particles, the observation time, and the relative orientation, location and geometry of the detectors. The latter is characterized by an ``overlap reduction function'' (ORF), which determines the sensitivity loss when applying cross-correlation to non-collocated, non-aligned detector pairs. For these models of \dm, two kinds of ORFs arise: one from the spatial displacement of mirrors, and another from the finite-light time travel effects of the signal on the mirrors, a previously explained in \cref{subsec:vecdm} \cite{Morisaki:2020gui}. In particular, cross-correlation is not nearly as sensitive to the finite-light travel time effect as to the spatial effect when it was applied to search for spin-1 \dm, which will be shown in \cref{subsec:ground-space}.

Larger couplings, longer effective observation times, and favorable overlap between detectors increase the \snr and thus the probability of detecting a signal. Conversely, weaker couplings, shorter observations, or misaligned detectors reduce the \snr, making a signal more difficult to distinguish from instrumental noise. After an \snr is computed via \cref{eqn:snr-general}, it can be translated directly from \snr to coupling strengths using Eq. 120-122 in \cite{Manita:2023mnc}). 

It is also useful to calculate a minimum detectable strain amplitude, which can also be used to place constraints on the DM coupling constants for scalar, vector, or tensor \uldm. Given the \snr, the corresponding strain amplitude can be estimated as:

\begin{equation}
h_{0,j} = \left(\frac{ 2\, \mathrm{SNR}}{ \gamma} \right)^{1/2} \left(\frac{\, P_{I,j}\, P_{J,j} }{\Tobs\, \TFFT}\right)^{1/4},
\label{eqn:snr-general}
\end{equation}
where $\gamma$ is the ORF that depends on the way that different kinds of \dm couple to the \ifos \cite{Manita:2023mnc}. By inserting the \snr returned from a search (\cref{eqn:snr}), we can obtain an estimate of the minimum detectable strain amplitude for a potential \dmh signal, and then map those amplitude values to constraints on \dmh couplings in spin-0, spin-1 or spin-2 using the strain amplitude equations derived in \cref{sec:pdm}, analogously to translating the \snr directly to coupling strengths in \cite{Manita:2023mnc}.

\subsubsection{BSD Excess power method}\label{subsubsec:excesspower}

Cross-correlation is able to capture the phase information in each FFT; however, at the moment the software has not yet been implemented to change $\TFFT$ to match $\Tcoh$, which would result in optimal sensitivity to each \dmh mass. Thus, another method \cite{Miller:2020vsl} called ``BSD excess power'' was developed that can vary $\TFFT$ as a function of frequency that employs Band-Sampled Data (BSD) structures that allow easy changes of $\TFFT$ \cite{piccinibsd}, and that is more robust against noise disturbances. It relies on creating time-frequency spectrograms in which frequencies in each FFT are only kept if their equalized powers are (1) above a given threshold $\theta_{\rm thr}=2.5$ and (2) local maxima. The equalized power is computed at each time and frequency bin by calculating the ratio $R_{ij}$ of the square modulus of the FFT with a running median estimation of the \psd, as in \cref{subsubsec:crosscorr}: 

\begin{equation}
    R_{ij} = \frac{|FFT|_{I,ij}^2}{P_{I,ij}}
\end{equation}
which, on average, takes on $\mathcal{O}(1)$ values.
After applying these cuts, a time-frequency ``peakmap'' is created, which is a collection of ones that indicates particular time/frequency points at which the aforementioned two conditions are met (the value $R_{ij}$ does not enter into any subsequent part of the analysis). A peakmap can be created every Hz, and $\TFFT$ can be varied to match $\Tcoh$ in each 1 Hz window. Because $\TFFT\sim \Tcoh$, the signal is expected to be sinusoidal. Thus, we can sum, in each frequency bin, the ones that are present, in essence creating a histogram. On this histogram, at each frequency, we compute a detection statistic called the ``critical ratio'' $CR$:

\begin{equation}
    CR = \frac{n - \mu}{\sigma}
\end{equation}
where $n$ is the number of peaks at a given frequency, and $\mu$ and $\sigma$ are the mean and standard deviation of the number of peaks in the histogram. The $CR$, like the \snr, follows a normal distribution. Using the $CR$ and assuming Gaussian noise, the minimum strain amplitude of a sinusoidal signal can be derived that would, in a frequentest interpretation, produce a detectable signal in a fraction $\ge \Gamma$ of a repeated number of experiments \cite{Astone:2014esa,Miller:2020vsl}:

\begin{align}
h_\text{0,min}&\approx 
\frac{\mathcal{G}}{\Tobs^{1/4}\TFFT^{1/4}}\sqrt{\frac{P_{I}(f)}{ \thetathr}}\nn \\ &\times \left(\frac{p_0(1-p_0)}{p_1^2}\right)^{1/4}\sqrt{CR_\text{thr}-\sqrt{2}\erfc^{-1}(2\Gamma)}, \nonumber \\
p_0&=e^{-\thetathr}-e^{-2\thetathr}+\frac{1}{3}e^{-3\thetathr}, \nonumber \\
p_1 &=  e^{-\thetathr} - 2e^{-2\thetathr} + e^{-3\thetathr}
\label{h0min}
\end{align}
where $\mathcal{G}$ depends on whether \dm is scalar, vector or tensor: 

\begin{align}
\mathcal{G}&=\sqrt{\frac{2\pi}{2.4308}}\frac{\sqrt{2}}{3}\times
\begin{cases}
     1, \text{\quad \, dilatons} \\
    \sqrt{\frac{9}{2}}, \text{ dark photons} \\
    \sqrt{\frac{5}{2}}, \text{ tensor bosons} \\
    \frac{3}{\sqrt{2}}\frac{5}{2}, \text{ \cws}
\end{cases} \\
& = 
\begin{cases}
    1.31,\text{ \, dilatons} \\
    2.78,\text{ dark photons} \\
    2.08,\text{ tensor bosons} \\
    4.02,\text{ \cws}
\end{cases}
\end{align}
The $\mathcal{O}(1)$ differences arise from averages over the spin-1 or spin-2 polarizations, and are valid for ground-based L-shaped \gwh \ifos (for triangular shapes, the averages will be slightly different). The $\frac{\sqrt{2}}{3}$ factor comes from an average over the \dmh Maxwell-Boltzmann distributed velocities as discussed in \cref{subsec:vecdm}, while the $\frac{2\pi}{2.4308}$ factor comes from convolving a sinusoid with a rectangular window function when computing the sensitivity of \gwh \ifos to a sinusoidal signal \cite{Astone:2014esa}.

\subsubsection{Logarithmic power spectral density (LPSD) method}
\label{subsubsec:lpsd}

To maximize sensitivity to \uldmh signals, it is important to let $\TFFT$ match $\Tcoh$ as the \dmh mass varies. The logarithmic power spectral density (LPSD) method \cite{trobs2006improved,Vermeulen:2021epa,Gottel:2024cfj} implements this idea by constructing spectra that are logarithmically spaced in frequency, in contrast to the BSD excess power method in \cref{subsubsec:excesspower} that use linearly spaced spectra every 1 Hz. While this approach is computationally intensive due to the need for a separate implementation from the standard discrete FFT, it captures subtle features in the data that could otherwise be missed. In particular, since \uldm induces fractional frequency variations on the order of $\mathcal{O}(10^{-6})$, a logarithmic spectral estimate is well-suited to resolving such tiny modulations.

Once the LPSD is computed, it can be combined with the \ifo response to convert measured strains into constraints on DM couplings. Accounting for the strain induced on the test masses and beam splitter (see \cref{subsubsec:size-dilaton}), the expected signal can be written as

\begin{align} \label{eq:full_signal1}
h(\omega) \approx \left(\frac{1}{\Lambda_\gamma} + \frac{1}{\Lambda_e}\right) \cdot\left(\frac{\sqrt{2\,\rho_{\mathrm{DM}}}}{\mdm}\right) A_{\rm cal}^{-1}(\omega)\,,
\end{align}
where $A_{\rm cal}(\omega)$ encodes the full interferometer transfer function. This function accounts for the response of the beam splitter and test masses to the \dmh signal, which is amplified by the difference in mirror thicknesses as discussed in \cref{subsubsec:size-dilaton}, mirror transmissivities, and the overall canonical \gwh detector transfer function.

By applying \cref{eq:full_signal1} to the logarithmically estimated spectra, one can extract an estimate of the strain $h(\omega)$ from the data. This estimate can then be mapped to constraints on the coupling constants $\Lambda_\gamma$ and $\Lambda_e$, providing a direct connection between the interferometer measurement and the underlying ultralight dark matter model \cite{Gottel:2024cfj,Cowan:2010js}. In this way, the LPSD method offers a way to detect logarithmically-spaced frequency modulations that might be missed by linear spectral analyses.


\subsubsection{Stochastic summing method}\label{subsubsec:stochsumm}

The stochastic summing method \cite{Nakatsuka:2022gaf} exploits the fact that when $\TFFT > \Tcoh$, the signal power from ultralight dark matter is spread stochastically over a range of frequencies rather than concentrated at a single frequency. Instead of searching for a narrow peak, this method sums the power across all frequency bins where the signal is expected, producing a detection statistic that captures the total available signal power.

\begin{equation}
\rho(f_c) \equiv \sum_i^{\Nfft} \sum_{f_0 \leq f_n \leq f_0(1 +\kappa^2v^2_{0})}\frac{4|\tilde{d}(f_n ; t_i)|^2}{\TFFT S(f_n; t_i)},\label{eqn:stochsum}
\end{equation}
where $\tilde{d}(f_n; t_i)$ is the FFT of $i^{\rm th}$ data chunk, $S(f_n; t_i)$ is the one-sided noise power spectral density at $t = t_i$,  and $\kappa$ = 3.17 is chosen to ensure that no more than 1\% of signal power is lost by limiting the frequencies summed over. The outer sum is over all chunks of length $\TFFT$; the inner sum is over the frequency spread of the signal $\delta f_v$.

The sensitivity of the stochastic summing method has been studied for both axions and dark photons \cite{Nakatsuka:2022gaf}, using a combination of analytical estimates and numerical simulations. In the absence of a signal and assuming Gaussian noise, a detection threshold can be established based on a chosen false-alarm probability. This threshold can then be used to determine, at a given confidence level, the minimum detectable signal amplitude, similar to \cref{h0min} in \cref{subsubsec:excesspower}.

The exact sensitivity depends on the spectral shape of the signal, which in turn is influenced by the relative motion of the \dm with respect to the \ifo. Conservative and optimistic assumptions about the direction of the \dmh velocity lead to $\mathcal{O}(1)$ variations in the sensitivity, which are accounted for in this method \cite{Nakatsuka:2022gaf}.  

Once the minimum detectable strain amplitude is determined, it can be translated into upper limits on the relevant coupling constants for each type of dark matter interaction, similarly to what has been discussed in \cref{subsubsec:crosscorr,subsubsec:excesspower}. These include axion-photon couplings as well as scalar or vector couplings that induce displacements or charge effects on the interferometer mirrors. In practice, because data are typically divided into segments with $\TFFT \sim \Tcoh$, the analytic likelihood approach can become unstable. Consequently, numerical simulations are used to estimate the sensitivity and derive robust upper limits, as done in a search on O3 KAGRA data \cite{KAGRA:2024ipf}, which will be discussed in \cref{subsec:ground-space}.

\subsubsection{Distinguishing amongst dark-matter models}

Only a few works \cite{Miller:2022wxu,Manita:2023mnc} have explored ways to determine whether a particular type of \dm interacted with \gwh detectors. In \cite{Miller:2022wxu}, the authors employ the Wiener filter \cite{wiener1964extrapolation} to compute the cross power across different detector pairs during follow-up stages, using different $\TFFT$ to exploit the fact that different types of \dm couple differently to different detector baselines. 

In Wiener filtering, the goal is to find a waveform that, when subtracted from the data, leaves only noise. The \emph{residual} quantifies the mismatch between the data, $h(t)+n(t)$, and a candidate signal, $h(t)$. Low residuals indicate that the candidate waveform closely matches the data, whereas high residuals indicate poor agreement.

\cite{Miller:2022wxu} showed that the Wiener filter produces small residuals when the data contain a signal consistent with the model and large residuals when the wrong signal is applied. This property allowed the method to confidently veto spurious candidates in real O3 data. Conversely, this means that the method has the potential to \emph{confirm} the detection of a certain type of \dm interaction signal. This is in contrast to other methods and previous searches \cite{Guo:2019ker,Miller:2020vsl,LIGOScientific:2021odm}, which focused mostly on rejecting outliers that were shown to be compatabile with noise.

After the work of \cite{Miller:2022wxu}, Ref. \cite{Manita:2023mnc} uses standard cross correlation to argue that the overlap reduction function can be used as way to distinguish amongst different spins of \dmh particles, since the cross power will be different. In particular, Ref. \cite{Manita:2023mnc} how the cross-correlation \snr would change as a function of the type of \dm and which baseline was considered. In both cases, cross power is important, since the individual power spectra from one detector are indistinguishable for all types of \uldm, which arises because the phase evolutions for the considered types of \uldm here are the same.

\subsection{Ground- and space-based interferometers}\label{subsec:ground-space}

Searches for \dmh particles interacting with the \lvkg \ifos have been performed using data from the most recent observing run, O3, resulting in constraints on dilatons and dark photons \cite{LIGOScientific:2021odm,Vermeulen:2021epa,Gottel:2024cfj,KAGRA:2024ipf}. Furthermore, \lpf data were also used to set constraints on dark photon \dm \cite{Miller:2023kkd,Frerick:2023xnf}. The theoretical framework for each particle is described in \cref{subsec:dilaton} and \cref{subsec:vecdm}, respectively. In this subsection, we summarize key constraints from existing ground-based interferometers on these two particles.

\subsubsection{LIGO, Virgo and GEO600}

In Ref. \cite{LIGOScientific:2021odm}, two complementary methods were employed to search for spin-1 \uldm that can couple to baryons in the test masses of LIGO in O3 data -- cross-correlation and BSD excess power, both of which are described in \cref{subsubsec:crosscorr} and \cref{subsubsec:excesspower}, respectively. Cross-correlation benefits from utilizing the phase information of the signal, something that is lost when using the time-frequency BSD excess power method. On the other hand, the BSD excess power method matches its coherence time $\TFFT$ to the signal coherence time $\Tcoh$ in every one-Hz band analyzed, allowing an improved sensitivity across the frequency range compared to cross-corelation, which fixed $\TFFT=1800$ s.

Both searches returned no significant \dmh candidate signals; thus, 95\% confidence-level upper limits were set on the coupling strength $\epsilon^2$ of dark photons to baryons for both methods, as shown in \cref{fig:uls}.. We can see, across a broad frequency range, that results from \gwh \ifos surpass by orders of magnitude constraints from Eöt-Wash and the MICROSCOPE experiments. These upper limits were derived for cross-correlation and BSD excess power following the procedure outlined in \cref{subsubsec:crosscorr} and \cref{subsubsec:excesspower}, respectively. Both methods employed the Feldman-Cousins approach \cite{Feldman:1997qc} to set upper limits, which is robust against non-Gaussian noise and has been shown to produce conservative upper limits with respect to those that would have been obtained through simulations \cite{Miller:2020vsl}. 

\begin{figure}[htbp]
    \centering
    \includegraphics[width=\columnwidth]{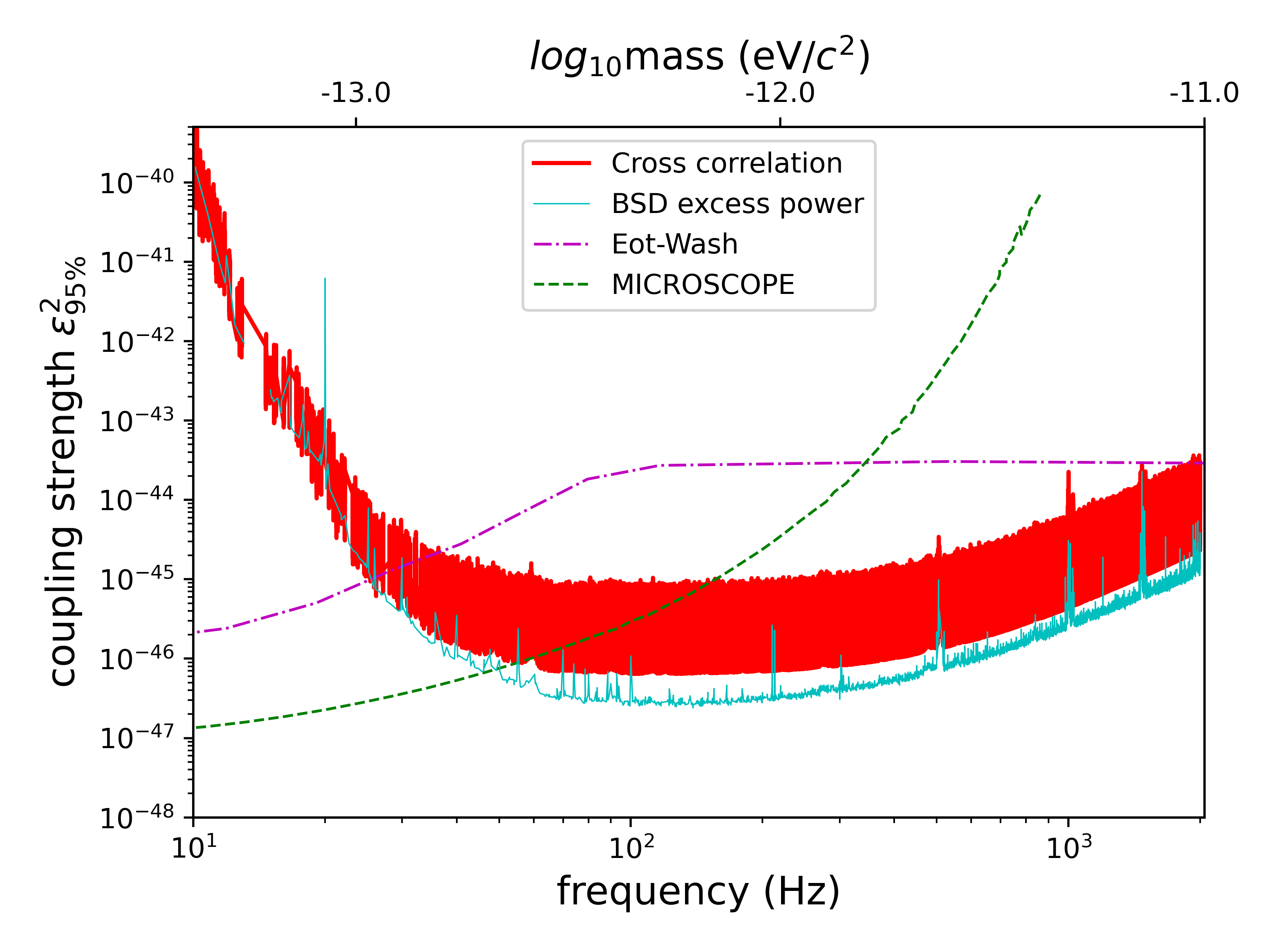}
    \caption{Adapted from \cite{LIGOScientific:2021odm}. Upper limits obtained from analyzing LIGO O3 data on the square of the coupling of dark photons to baryons $U(1)_{\rm B}$ in the LIGO mirrors. The physics behind this form of \dm has been discussed in \cref{subsec:vecdm}. The limits derived from each of the methods discussed in \cref{subsubsec:crosscorr,subsubsec:excesspower} are show in red (cross-correlation) and cyan (BSD excess power), respectively. MICROSCOPE  \cite{Berge:2017ovy} and Eöt-Wash torsion balance upper limits are plotted as a comparison to the results here \cite{Schlamminger:2007ht}. To produce limits on the square of the dark photon/baryon-lepton coupling, $U(1)_{\rm B-L}$ in the LIGO mirrors, these limits should be multiplied by four.}
    \label{fig:uls}
\end{figure}

While cross-correlation and BSD excess power analyses were used for dark photon searches, the LPSD technique (described in \cref{subsubsec:lpsd}) was employed in Ref.~\cite{Gottel:2024cfj} to search for scalar dilaton dark matter in the LIGO O3 dataset. This analysis placed strong limits on the coupling of scalar dark matter to electrons and photons.  In \cref{fig:results}, we can see, in blue, the constraints that have been derived, which supersede, at least at low masses, those that were derived in a GEO600 search for the same kind of \dm \cite{Vermeulen:2021epa}. The enhancement comes from the improved low-frequency sensitivity of LIGO relative to GEO600, as well as incorporating the interaction of scalar \dm with the mirrors, as well as the beam splitter, into the search. However, at high frequencies, we can see that GEO600 still outperforms LIGO, primarily because of sophisticated quantum technologies that were employed in its design. These upper limits were derived by following the procedure outlined in \cref{subsubsec:lpsd} and \cref{subsubsec:excesspower}.

\begin{figure}[htbp]
    \centering
    \includegraphics[width=.5\textwidth]{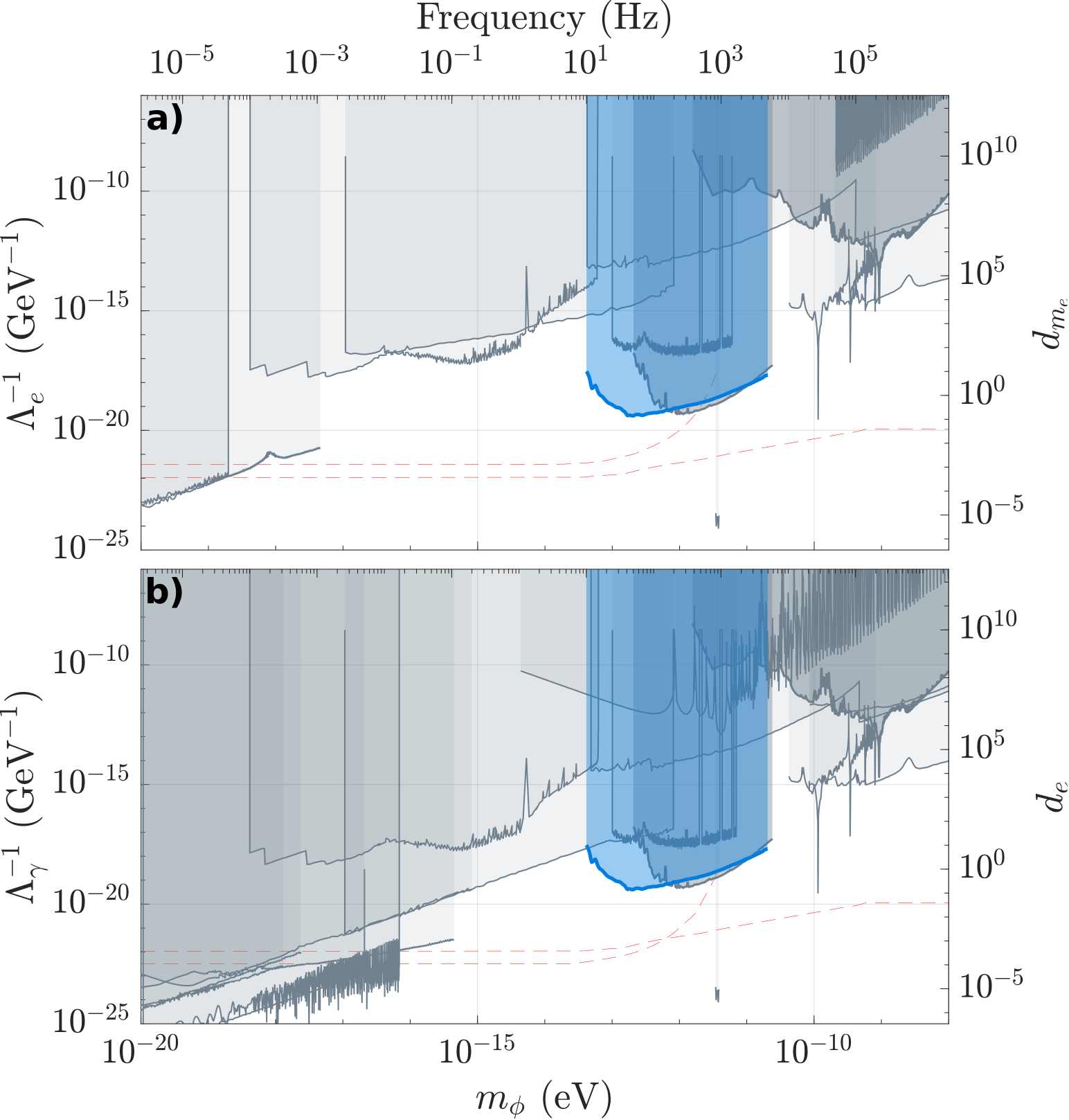}
    \caption{Taken from \cite{Gottel:2024cfj}. 95\% confidence-level upper limits on $\Lambda_i^{-1}$, the coupling of dilaton \dm to electrons or photons, as a function of mass and frequency from LIGO O3 data, which would have cause time-dependent oscillations of the sizes and indices of refraction of the beamsplitter and the LIGO mirrors. The physics behind this form of \dm has been discussed in \cref{subsubsec:size-dilaton}.\textbf{a}) and \textbf{b}) show these constraints compared to existing ones on $\Lambda_e$ and $\Lambda_{\gamma}$, respectively. The results from the LIGO O3 search are shown by the thick blue line, constraints from direct experimental searches for DM~\cite{Vermeulen:2021epa,Aiello:2021wlp,Aharony:2019iad,Savalle:2020vgz,Kennedy:2020bac,Antypas:2019qji,Antypas:2020rtg,Tretiak:2022ndx,Oswald:2021vtc,Campbell:2020fvq,BACON:2020ubh,Zhang:2022ewz,Fukusumi:2023kqd,Sherrill:2023zah} are shown in thin grey, and constraints from searches for ``fifth 
    forces''~\cite{Berge:2021yye,Hees:2018fpg} are depicted by the dashed red lines.}
    \label{fig:results}
\end{figure}

\subsubsection{KAGRA}\label{subsubsec:kagra}

KAGRA offers unique opportunities to search for \dm. In particular, the different materials used in KAGRA’s mirrors result in significantly different charge-to-mass ratios for baryon-lepton number, enhancing the sensitivity to vector \dm coupling to this quantum number \cite{Michimura:2020vxn}. While LIGO and Virgo currently set the strongest limits on dark photon coupling to baryon-lepton number -- obtainable from the results in \cref{fig:uls} by multiplying the limits on baryon number coupling by four \cite{LIGOScientific:2021odm} -- this difference in mirror materials means that once KAGRA reaches comparable sensitivity levels, it will outperform LIGO in constraining such couplings \cite{Michimura:2020vxn}.

This enhancement is particularly apparent in KAGRA’s auxiliary length control channels, PRCL and MICH, which record displacements of mirrors made from both fused silica (beam splitter and power recycling mirrors) and cryogenic sapphire (test masses). The baryon-lepton charge-to-mass ratios differ by approximately \(9 \times 10^{-3}\) between these materials, compared to only about \(4 \times 10^{-5}\) for baryon number alone, significantly amplifying the \dmh interaction signal in these channels -- see \cref{eqn:dph0}.

The raw outputs of PRCL and MICH correspond to time-varying displacements of the form \cite{KAGRA:2024ipf}:
\begin{align}
 \delta L_{\rm MICH} &= \delta (l_{\rm x} - l_{\rm y}) \label{eq:mich}, \\
 \delta L_{\rm PRCL} &= \delta \left[ \frac{l_{\rm x} + l_{\rm y}}{2} + l_{\rm p} \right],  \label{eq:prcl}
\end{align}
where \(l_x = 26.7\,\text{m}\) and \(l_y = 23.3\,\text{m}\) are the distances between the beam splitter and the input mirrors, and \(l_p = 41.6\,\text{m}\) is the length of the power recycling cavity. All these lengths are much shorter than the \dmh coherence length \(\Lcoh\). See Fig.~1 of \cite{KAGRA:2024ipf} for a schematic of the \ifo.

No significant signals were found in this analysis of KAGRA O3 data \cite{KAGRA:2024ipf}, and upper limits on the coupling strength were derived following the procedure outlined in \cref{subsubsec:stochsumm}. These limits, shown in \cref{fig:UL} for each channel, exhibit many narrow-band noise artifacts that limit sensitivity at low frequencies. Although not yet competitive with LIGO constraints, future KAGRA searches are expected to provide powerful limits on this type of \dm due to the pronounced difference in baryon-lepton charge-to-mass ratios among its mirrors \cite{Michimura:2020vxn}.

\begin{figure}[htbp]
  \centering
  \includegraphics[width=\columnwidth]{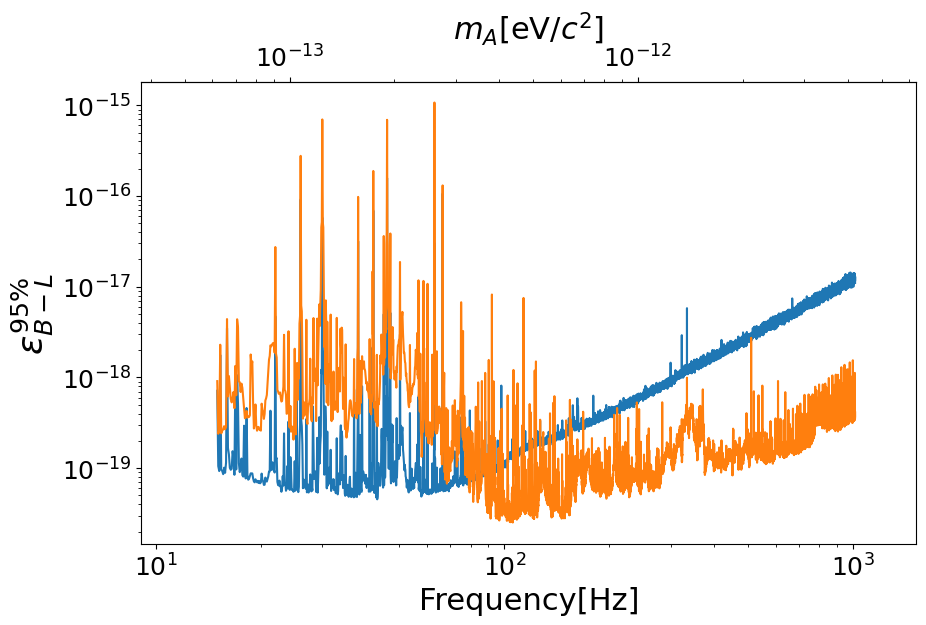}
  \caption{Taken from \cite{KAGRA:2024ipf}. 95\% confidence-level upper limit on the strength of the coupling between dark photons and baryon-lepton number using KAGRA O3 data and the method described in \cref{subsubsec:stochsumm}, derived from two particular \ifo channels: MICH (blue line) and PRCL  (orange line). The physics behind this form of \dm has been discussed in \cref{subsec:vecdm}. Because the KAGRA input mirrors, end mirrors, and power recycling mirrors are made of different material, the coupling strength of dark photons to baryon-lepton number is enhanced (\cref{eqn:dph0}), and appears more strongly in these particular channels than in the ordinary strain one. In the low-mass (low-frequency) range, many narrow-band noise disturbances of unknown origin are seen.}
  \label{fig:UL}
\end{figure}




\subsubsection{LISA Pathfinder}

Spaced-based \gwh \ifos, such as LISA \cite{danzmann2003lisa,Babak:2021mhe}, Taiji \cite{Hu:2017mde}, Tianqin \cite{TianQin:2015yph} and DECIGO \cite{Hu:2017mde}, will hopefully fly within the next 10-15 years. The exquisite sensitivity of \gws in the $\mu$Hz to mHz band will permit sensitivity to \uldm with masses of [$10^{-19},10^{-15}$] eV, a few orders of magnitude lower than those currently searched for in ground-based \ifos \cite{Kim:2023pkx}. Until now, however, only a protoype for LISA, called \lpf, has been flown as a ``proof-of-concept'' mission \cite{Armano:2016bkm,Vetrugno:2017oft,Armano:2018kix}. Nevertheless, pilot searches can be performed on \lpf data as a way to prepare for when space-based detectors fly, which allows for the development of robust data analysis pipelines and for handling of the peculiarities of future data, e.g. glitches, downtimes, etc.

Such a search was preformed using \lpf data \cite{Miller:2023kkd} using the BSD excess power method described in \cref{subsubsec:excesspower} for ultralight dark photons coupling to baryons, however, no physical constraints on the coupling strength were obtained, since the noise level was too high, the arm length was too short (only $40$ cm compared to the expected LISA arm lengths of $\mathcal{O}(10^6)$ km), and only one arm existed. However, the search procedure designed will serve as a roadmap for future analyses of spaced-based detector data, since many problems, including limited sampling at low frequency, gaps, noise non-stationarities and glitches had to be handled in this analysis of data from a space-based detector. 

After this search was performed, however, it was realized that data from a different channel would produce tighter constraints, particularly in the case of baryon-lepton coupling. Here, the relative acceleration of the spacecraft and one of the test masses would produce a much stronger signal than that which was searched for previously because the charge-to-mass ratios of each are different \cite{Frerick:2023xnf}. This echoes the logic presented in \cref{subsubsec:kagra} to use KAGRA to search for dark photons coupling to baryon-lepton number. 

A conservative upper limit on the baryon-lepton coupling was set without performing a search, assuming no signal would have been detected in \lpf data, as shown in \cref{fig:ohare}, by using a reference acceleration power spectral density of the spacecraft with respect to one of the test masses. We can see that the upper limits from the relative acceleration of space-craft/ test mass are much better than that from \cite{Miller:2023kkd}, labeled (decoherence). The constraint on B-L coupling is stronger with respect to that of baryon coupling because the test mass and spacecraft are primarily made of gold, and gold/carbon, respectively.

\begin{figure*}[htbp]
    \centering
    \includegraphics[scale=0.4]{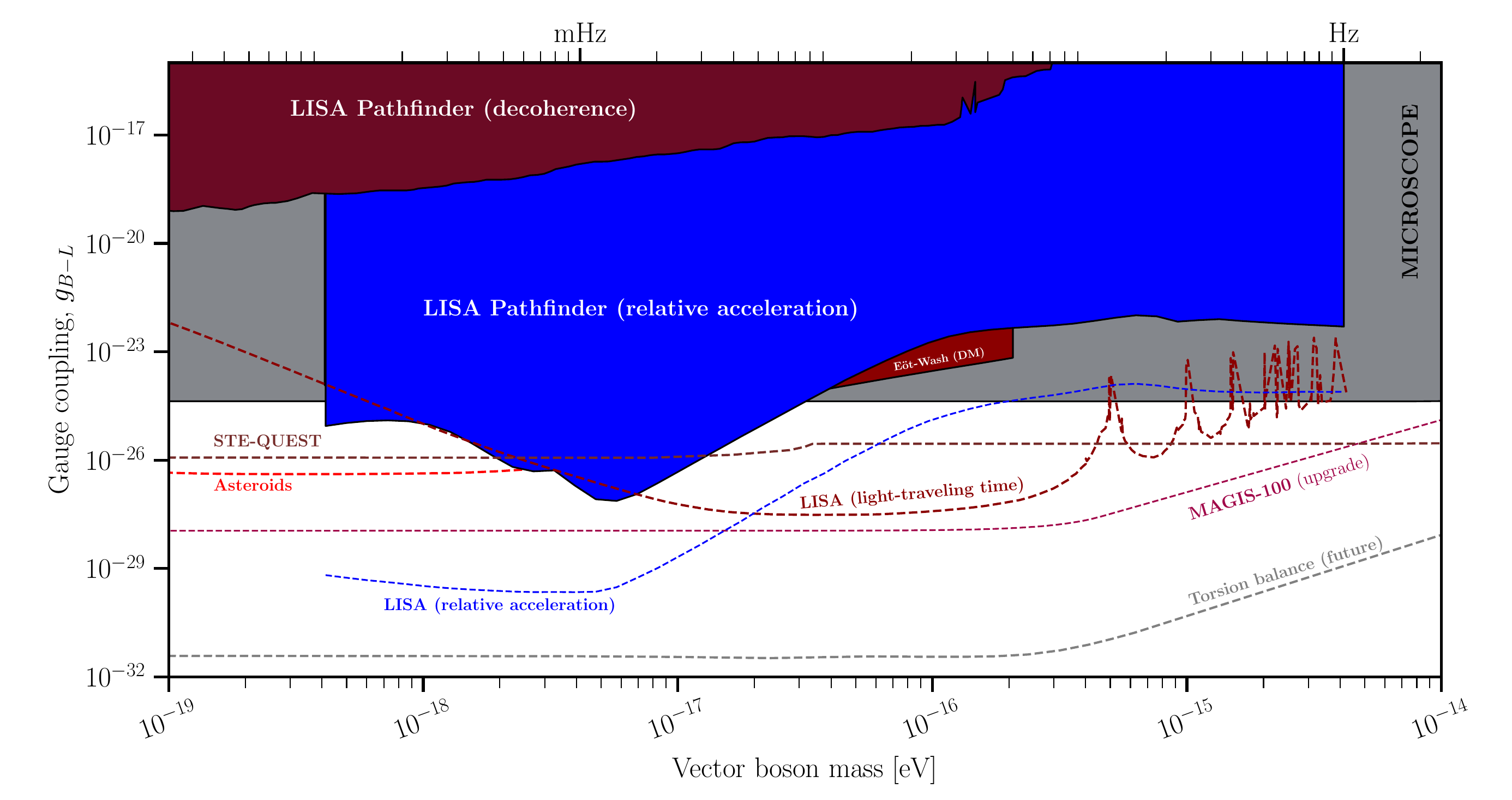}
    \caption{Taken from \cite{Frerick:2023xnf}. Upper limits on the coupling of dark photons to baryon-lepton number considering the relative acceleration of the spacecraft and one of the test masses in \lpf (dark blue region). The physics behind this form of \dm has been discussed in \cref{subsec:vecdm}, and constrains the same model as in \cref{fig:UL,fig:uls}, though at much lower boson masses. The constraints derived from a search for the induced acceleration of the two test masses by dark photons in \lpf data is shown as ``decoherence'' (dark red region). Using the relative acceleration caused by dark photon interaction with baryon-lepton number allows for stringent constraints to be placed on the coupling, since the spacecraft and test masses are made of different materials. Constraints from other experiments (Eöt-Wash and MICROSCOPE) and future projections for LISA, MAGIS-100 and torision balance are shown as well. Details on the curves coming from STE-QUEST and Asteroids can be found in \cite{ciaran_o_hare_2020_3932430}. }
    \label{fig:ohare}
\end{figure*}

Only B-L constraints are shown from \cite{Frerick:2023xnf} because those arising from coupling to baryons are not yet strong enough to probe a tighter constraint than that provided by MICROSCOPE. However, many conservative assumptions were made in \cite{Frerick:2023xnf}, including the composition of the spacecraft and the use of a reference acceleration power spectral density, both of which could, if studied in more detail, lead to more stringent constraints for both the baryon and B-L coupling scenarios.

\subsection{Pulsar timing arrays}\label{subsec:ptas}

Because pulsars are extremely stable astrophysical clocks, they can be used to search for \gws, which perturb the spacetime between Earth and pulsars. In practical terms, \gws induce fluctuations in the \toas of radio pulses, relative to what would be expected in their absence. However, interpreting these \toa shifts is complicated by other astrophysical and instrumental effects -- such as dispersion and scattering in the interstellar medium, and intrinsic rotational instabilities of the pulsars themselves -- which makes precision measurements challenging. Nevertheless, these effects can be mitigated by cross-correlating data from many pulsars in the sky to form a \pta.

As described in \cref{subsec:dilaton}, ultralight scalar \dm can also induce variations in pulse \toas \cite{Khmelnitsky:2013lxt}, though the relevant mass range lies many orders of magnitude below that probed by ground-based interferometers. Such scalar fields can cause coherent oscillations in the gravitational potential or effective constants of nature, and may also generate stochastic perturbations in the metric \cite{Kim:2023kyy}, which could lead to further timing variations over long baselines \cite{Kim:2023pkx}.




Throughout this section, we will see that, independently of the physical mechanism of coupling of \dm to gravity or the \sm, the shapes of the constraints on the coupling constant appear similar. This is because in these cases, we are looking for an almost sinusoidal signal embedded in \pta data with particular noise characteristics that are independent of the \dmh model. Though the underlying physics behind each \dmh model is different, the observable -- shifts in \toas -- is identical, though with different amplitudes that relate to the coupling constant of \dm. This connects to our previous comments in the introductory paragraphs of \cref{sec:obsconstrain}: the signal arising from \dm coupling to gravity or the \sm is the same -- a quasi-sinusoidal oscillation of spacetime, of materials, of test masses, etc. -- but the \emph{cause} and therefore the \emph{interpretation} of time-varying \toa in \ptas or time-varying displacements of test masses in ground- or space-based \gwh \ifos is fundamentally different, which leads to different constraint plots whose curves look similar.

In \cref{subsec:dilaton}, we discussed the theoretical mechanisms by which scalar \dm could leave signatures on \pta data, and in \cref{subsec:ground-space} we described how related effects could be searched for with ground-based interferometers. In contrast to that section, here we organize existing constraints from \ptas according to the specific physical mechanism by which scalar \dm couples to the \sm or to gravity. Each mechanism leads to a distinct signature in the pulsar timing residuals. We treat these mechanisms independently: if we put constraints on \dm coupling to dilatons in the context of one mechanism, we assume that all other mechanisms do not contribute

\subsubsection{Gravitationally-interacting \uldm}\label{subsubsec:grav-int-uldm-pta}

As discussed in \cref{subsec:grav-int}, \dm could interact only via gravity, and alter the distances between the pulsars and earth, analogously to \gws. Using European \pta data, a search was performed in the residuals for this kind of \dm interaction. In \cref{fig:25yrULDM}, we show recent constraints on ultralight scalar \dm in two ways: (1) the minimum detectable strain amplitude as a function of \dmh mass, and (2) the fraction of \dm that \uldm could compose. We can see that the masses below $\sim 10^{-23.2}$ eV can be well constrained by pulsar timing arrays, and that \uldm cannot make up all of \dm at these masses.
\cref{fig:25yrULDM} presents upper limits in three scenarios: (1) uncorrelated, (2) correlated and (3) pulsar-correlated. 
\begin{enumerate}
\item ``Uncorrelated'' refers to the case when the \uldmh $\Lcoh$ is less than the average pulsar separation and the earth-pulsar distance, and means that the pulsars experience different phases of the \uldmh signal. Thus, $\hat{\phi}_P^2$ and $\hat{\phi}_E^2$ are independent. 
\item ``Correlated'' means that the earth-pulsar distance and the average distance between pulsars is smaller than $\Lcoh$, and that the \dmh $\Lcoh$ comprises the inner 20 kpc region of the galaxy for which galaxy rotation curves have been used to test the \dmh hypothesis. This means that each pulsar experiences the same phase of the \uldmh signal. In this case, the same coherence patch and local \dmh energy density are measured: the amplitude of the \uldmh interaction signal can be attributed completely to measurement of $\rhoDM$ coming from galaxy rotation curves\footnote{This case is directly analogous to measurements of \uldmh couplings in ground- and space-based \gwh \ifos, discussed in \cref{subsec:ground-space}, where the arm lengths are negligible compared to $\Lcoh$, so all test masses experience the same \uldmh phase.}.
\item ``Pulsar-correlated'' means that $\Lcoh$ is larger than the earth-pulsar and inter-pulsar separations, so $\hat{\phi}_P^2 = \hat{\phi}_E^2$, but smaller than the galactocentric radius sampled by galaxy rotation curves. In this case, pulsars and the Earth probe the same coherence patch, but the \uldmh amplitude measured with pulsars cannot be identified one-to-one with the \dmh energy density inferred from galaxy rotation curves, since the latter averages over many independent coherence patches.
\end{enumerate}
Each type of analysis optimally probes a different mass regime depending on the \dmh coherence length. Therefore, the ``correlated'' curve is valid for masses less than $\sim 2 \times 10^{-24}$ eV; the ``pulsar-correlated'' curve can be applied for $ 2 \times 10^{-24}\, \text{eV} \lesssim \mdm \lesssim 5 \times 10^{-23}\, \text{eV} $ and the ``uncorrelated'' curve holds for $\mdm \gtrsim 5\times 10^{-23}~\text{eV}$.  In both correlated cases, $\hat{\phi}_P^2=\hat{\phi}_E^2$.

\begin{figure*}[htbp]
     \begin{center}
        \subfigure[ ]{%
            \label{fig:pe_xg:m}

        \includegraphics[width=0.5\textwidth]{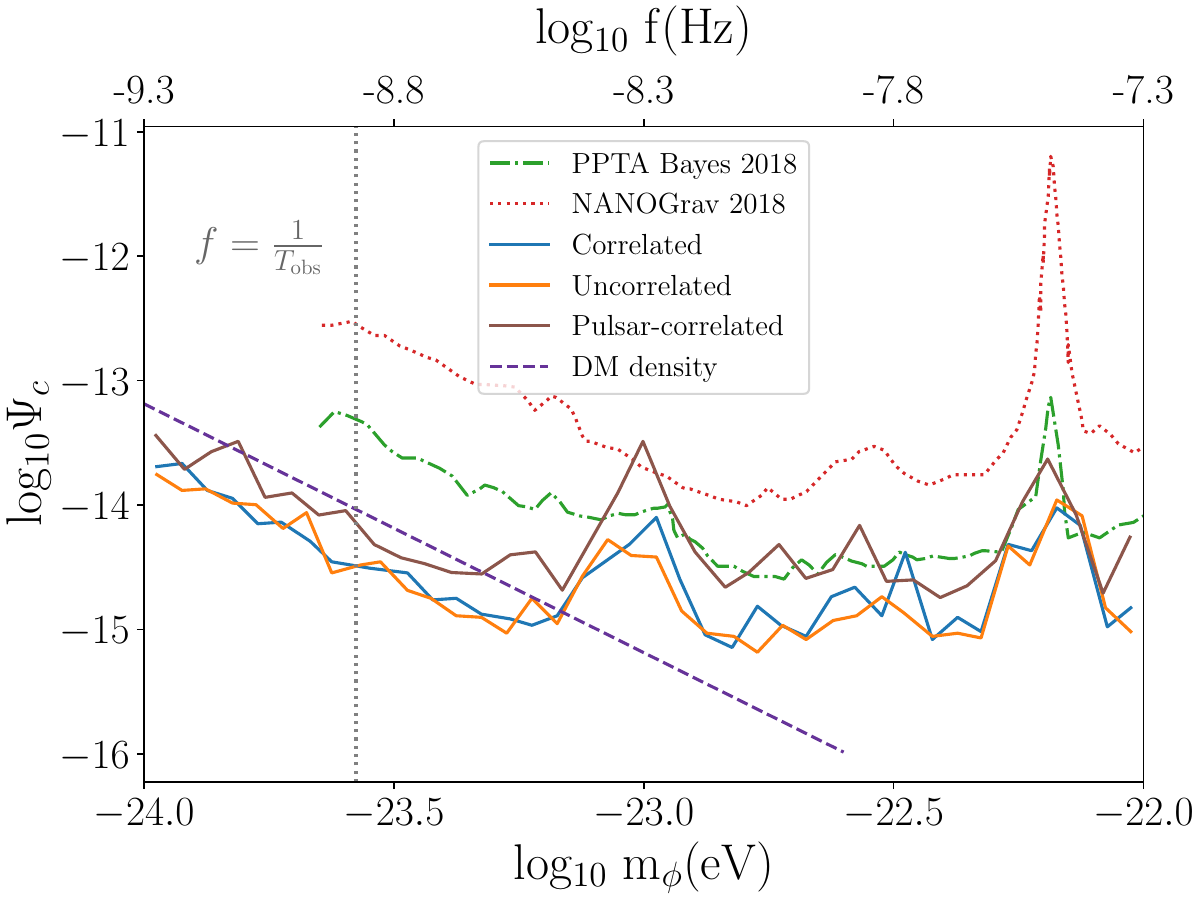}
        }%
        \subfigure[]{%
           \label{fig:pe_xg:nom}
           \includegraphics[width=0.5\textwidth]{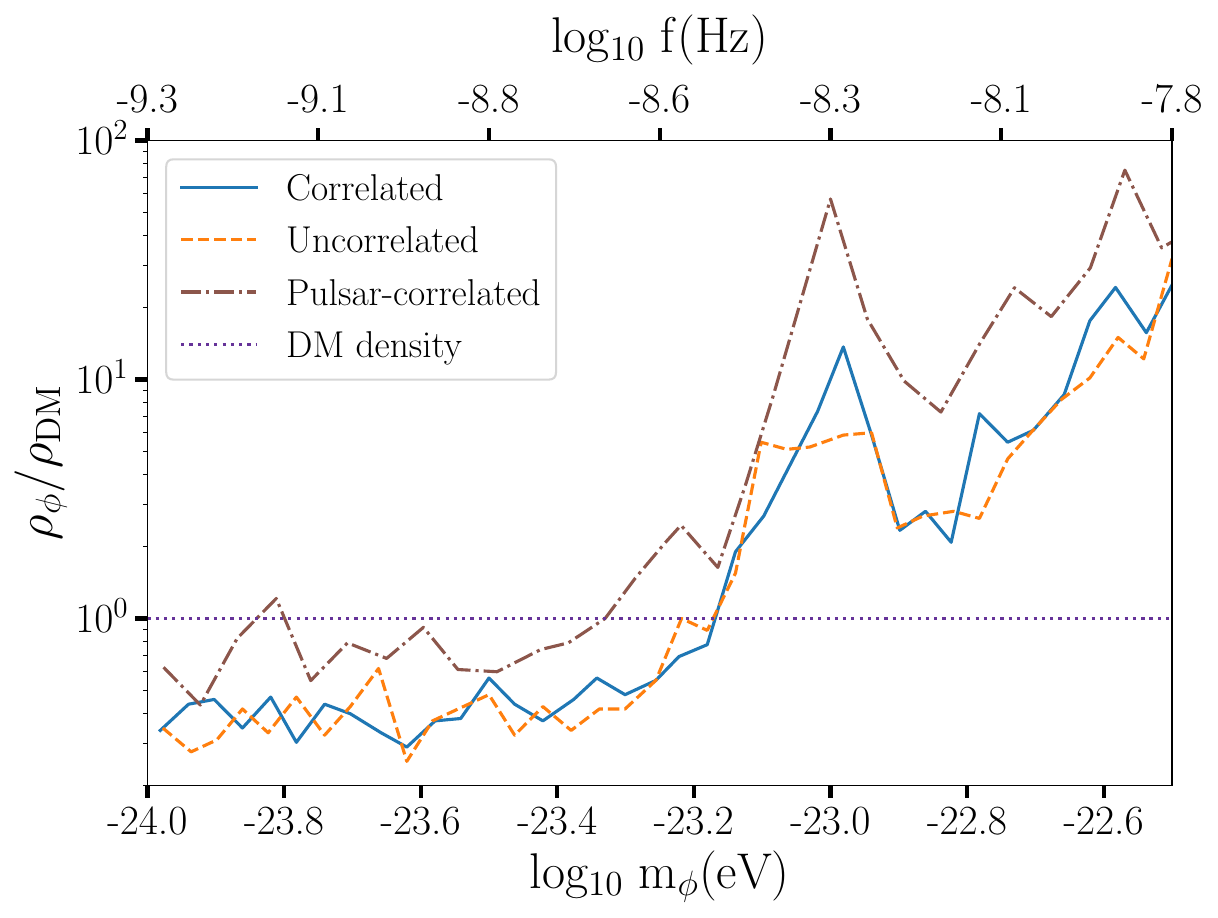}
        }\\ 
    \end{center}
    \caption{Taken from \cite{EuropeanPulsarTimingArray:2023egv}. 95\% confidence-level upper limits from analyses of European \pta data on \uldm interacting only gravitationally and changing the \toas of pulses in an analogous way that \gws do. The physics behind this form of \dm has been discussed in \cref{subsec:grav-int}. The panels show (a) the dimensionless strain amplitude $\Psi_\text{c}$ and (b) the fraction of the \dmh energy density that scalar \dm could compose  as a function of \dmh mass and frequency. $\rhoDM = 0.4 \text{GeV}/\text{cm}^3$.
    For comparison, upper limits on these quantities are shown from previous searches (PPTA Bayes 2018, NANOGrav 2018) \cite{Porayko:2014rfa,Porayko:2018sfa}. Furthermore, the dashed line indicates the signal amplitude assuming that this form of \dm comprises all of \dm; thus, curves that lie above this dashed line do not represent physical constraints. Upper limits on the amplitude are shown for the three cases outlined in \cref{subsubsec:grav-int-uldm-pta}.
    In the right panel, we zoom in on the excluded \uldm masses. The horizontal dotted line represents the scenario in which \uldm is responsible for all of the local \dmh energy density. The right panel shows that
    scalar \uldmh particles with mass $-24.0 < \text{log}_{10}\text{~}(\mdm/\text{eV}) < -23.7$ can comprise at most $30-40\%$ of $\rhoDM$, while particles with masses $-23.7< \text{log}_{10}\text{~}(\mdm/\text{eV}) < -23.3$ could comprise up to  $ \sim 70$ \% of $\rhoDM$.
    } 
     \label{fig:25yrULDM}
\end{figure*}
Additionally, NANOGrav data has been be used to constrain this kind of \uldm \cite{NANOGrav:2023hvm}, though the constraints are weaker than from the EPTA. In particular, NANOGrav cannot yet probe a physical constraint for the fraction of \dmh energy density that this model of \dm could compose, likely because their dataset only spans 15 years, while the European \pta collaboration analyzed almost 25 years of data.

\subsubsection{Constraints on scalar conformal dark matter}

While the previous subsection considered generic strain-like gravitational interactions of scalar \uldm, scalar conformal couplings represent a distinct class of models in which ULDM interacts through the trace of the energy–momentum tensor, As discussed in \cref{subsec:grav-int}. This model of \dm can also be constrained with \pta data, and in particular Ref.~\cite{Smarra:2024kvv} uses European \pta data to constrain both linear (FJBD-type) and quadratic (DEF-type) conformal couplings.

In \cref{fig:comp}, we show the upper limits on the linear coupling parameter $\alpha$ using European \pta data. However, despite showing upper limits (that assume no signal), Ref.~\cite{Smarra:2024kvv} notes that there is some evidence for additional signal power at two masses: $m \sim 10^{-22.7}$ eV and $m \sim 10^{-21.4}$ eV, though other physical processes could be responsible for these excesses. 
It is worth noting that the effective change in the \nsh \moi in this scenario arises from the universal rescaling of the metric due to the conformal coupling, and is therefore distinct from the composition-dependent \moi variations discussed in \cref{subsec:dilaton}, where scalar fields couple directly to quarks and nucleons. The constraints arising from the interaction of dilatons with \smh particles in \nss will be discussed in the next subsection.

\begin{figure}[htbp]
\centering
\includegraphics[width=0.5\textwidth]{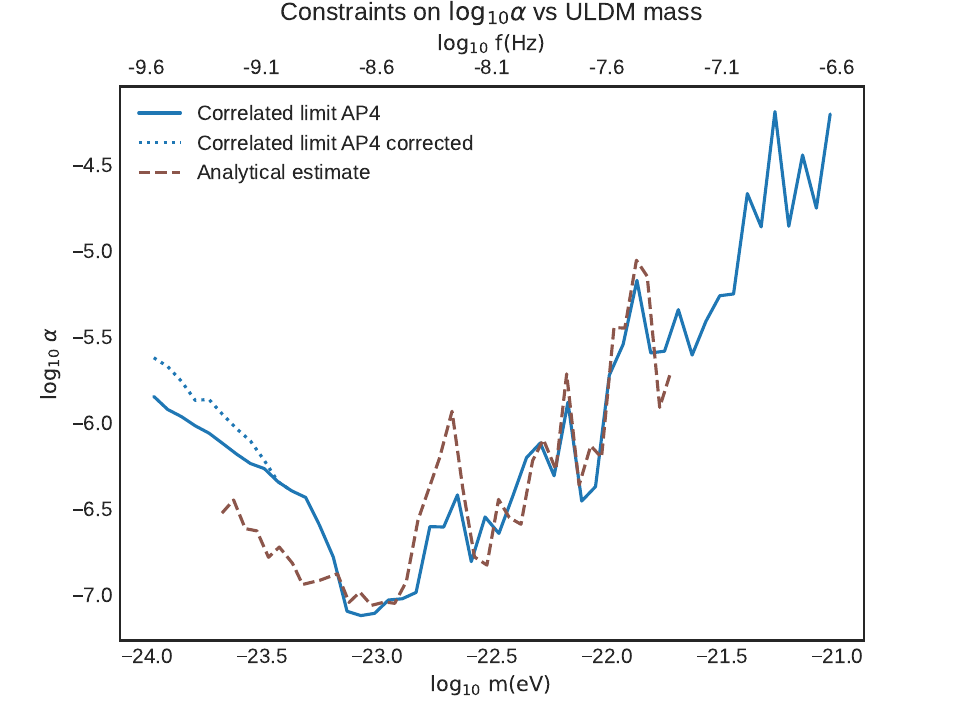}
\caption{Taken from \cite{Smarra:2024kvv}. Constraints on the coupling $\log_{10}\alpha$ of conformal  \dm to gravity at the 95\% confidence-level (solid line) for the correlated case compared to the analytical estimate obtained by using the upper limits from the European \pta search in \cite{Smarra:2024kvv} to estimate $\alpha$ (brown dashed lines). The physics behind this form of \dm has been discussed in \cref{subsec:grav-int}. $\rho_\text{DM} = 0.4 \, \text{GeV/cm}^3$. Here, this model of \dm would change the \moi of \nss  used in \ptas, resulting in changes of the \toa of the pulses.  If conformal \dm does not comprise all of \dm, the limits will weaken, which is shown by the dotted line, for values of $\rho$ given in \cite{EuropeanPulsarTimingArray:2023egv}. AP4 indicates that these limits are derived assuming the AP4 \eos of \nss \cite{Akmal:1998cf}.}
\label{fig:comp}
\end{figure}

\subsubsection{Variation of fundamental constants constraints}

As discussed in \cref{subsec:dilaton}, scalar, dilaton \dm can induce oscillations in fundamental constants, which would have macroscopic effects on the \moi of a \ns. 
Changes in the \moi of pulsars that are observed would induces changes in the \toa of pulses in their spin frequencies that NANOGrav measures. Thus, constraints on the coupling of scalar \dm to the \sm can be set for masses below $10^{-22}$ eV \cite{NANOGrav:2023hvm}. \cref{fig:uldm_direct} depicts constraints on the couplings of the scalar \dmh particle to up- and down- quarks weighted by their masses ($\hat{m})$, electrons $(d_e)$, muons $(d_\mu)$, photons $(d_\gamma)$ and gluons $(d_g)$. To derive these constraints, the simplest model of the equation-of-state of the \ns is assumed, as outlined in \cref{subsubsec:change-moi}, and, when constraining one particular coupling constant, the rest are assumed to be zero. 

The strongest constraints from \ptas are on the couplings to the electron and muons. This is in contrast to constraints that come from atomic clock experiments, which also aim to find \uldm but are insensitive to the electron coupling, since changing the electron mass does not affect the spacing between energy levels \cite{Kaplan:2022lmz}. In terms of (the lack of) other constraints on muons, laboratory experiments do not have enough muons to harness on earth to study the coupling of \dm to them, while, in contrast, \nss have an abundance of muons.

In \cref{fig:uldm_direct}, there is a gray-shaded region that indicates the situation in which $A<A_{\rm grav}$, where $A$ is given by the amplitude of \cref{eq:pulsar_spin_fluctuation_signal} or \cref{eq:reference_clock_shift_signal}. In other words, these constraints on \uldm coupling to \smh particles are valid when the coupling of \dm to gravity can be neglected, i.e. when \cite{Kaplan:2022lmz}:

\begin{align}
    d_i \gtrsim \frac{4.5 \times 10^{-9}}{y_i} \left( \frac{10^{-23}\,\text{eV}}{\mdm} \right) \, ,
\end{align}
In the gray-shaded region, the interaction between gravity and \dm is stronger, and thus constraints on \uldm coupling to \smh particles alone would not be valid in that region. Furthermore, these constraints assume that scalar particles constitute all of \dm, but if they only comprise a fraction $f$ of \dm, then the constraints will be weakened by $\sqrt{f}$, because of the $\sqrt{\rhoDM}$ term in the amplitude of these interaction signals.

\begin{figure*}[htbp]
	\centering
    \includegraphics[width=\textwidth]{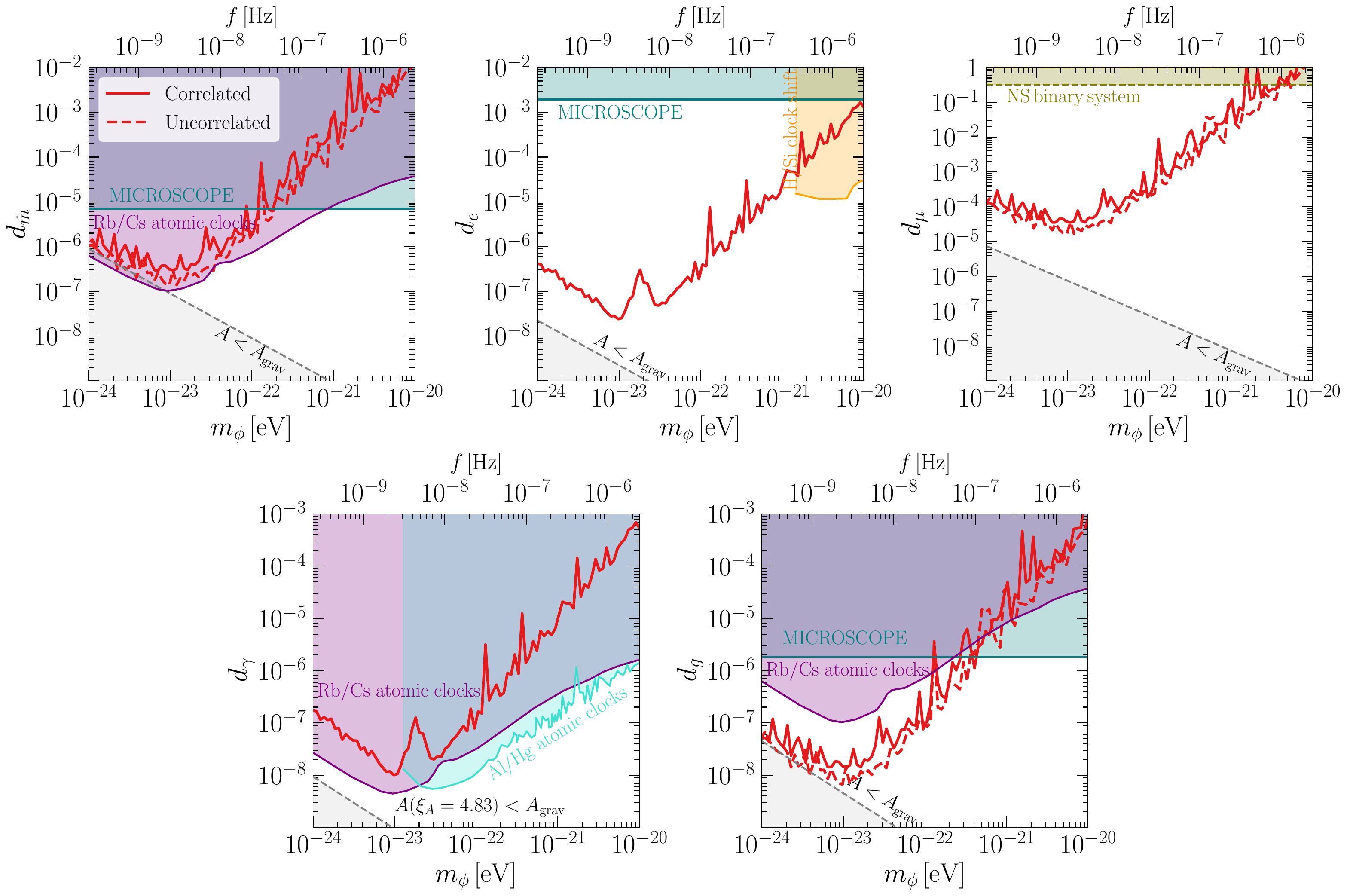}
	\caption{Taken from \cite{NANOGrav:2023hvm}. 
 Upper limits from an analysis of NANOGrav data on the coupling of \uldm to various \smh particles: quarks, electrons, muons, photons and gluons, shown in red for the correlated case, and red dashed for the uncorrelated case.  In this model, \dm would alter the moment of inertia of \nss and thus affect the \toas of pulses in \ptas. The physics behind this form of \dm has been discussed in \cref{subsubsec:change-moi}. The constraints on one coupling constant assume that all other coupling constants are set to 0. The black dashed line, and the gray shaded region, indicate where the amplitude of the signal would be less than the amplitude of the signal arising from \dm coupling only to gravity.  Current constraints ``Rb/Cs atomic clocks" (purple) are from~\cite{Hees:2016gop}, ``Al/Hg atomic clocks" (turquoise) are from~\cite{BACON:2020ubh}, ``MICROSCOPE" (teal) are from~\cite{Berge:2017ovy}, ``H/Si clock shift" (orange) are from~\cite{BACON:2020ubh}, and ``NS binary system" are from~\cite{KumarPoddar:2019ceq} and~\cite{Dror:2019uea}. The shape of these constraints is similar to those in \cref{fig:vec_pta_constraints}, since both searches look for nearly sinusoidal signals embedded in \pta data with similar noise characteristics, even though the underlying \dm coupling mechanisms differ. \label{fig:uldm_direct}}
\end{figure*}

The scalings of these constraints with the \dmh mass should be discussed. At high masses, the signal is essentially deterministic, and the \snr scales as the square of the signal amplitude, which means that $d_i\propto \mdm^2$. At low masses ($\mdm<1/\Tobs)$, the signal is observed in different coherence patches, since the coherence length is shorter than the pulsar separation. The signal is no longer a sinusoid but can be expressed as a polynomial in $mt$, where $t$ is time. The first two terms of this polynomial ($\mdm t$ and $(\mdm t)^2$) are degenerated with pulsar timing terms, and thus the first observable term is $(\mdm t)^3$, meaning that the \snr is proportional to $\mdm ^3$. Thus, $d_i\propto 1/\mdm $, which explains the scaling at low masses. Still, though, red noise causes the constraints to flatten as the mass decreases, which, at first, hides this scaling.

\subsubsection{Constraints on vector dark matter}

As discussed in \cref{subsec:vecdm}, \dm could also be composed of spin-1 bosons that interact with \sm particles. In the case of \ptas, the \uldm would cause oscillatory forces on the earth and on the pulsars themselves with strengths proportional to the charge-to-mass ratio of these objects.

\cref{fig:vec_pta_constraints} shows constraints on the coupling constants $g_B=\epsilon_B$ and $g_{B-L}=\epsilon_{B-L}$ as a function of the \dmh mass and oscillation frequency. Again, the gray-shaded region indicates the regime in which the coupling of \dm to gravity becomes stronger than the coupling of \dm to \smh particles. We can see that the NANOGrav constraints greatly surpass those that come from MICROSCOPE at very low masses. Furthermore, these constraints assume that vector bosons constitute all of \dm, but if they only comprise a fraction $f$ of \dm, then the constraints will be weakened by $\sqrt{f}$.

\begin{figure*}[htbp]
	\centering
        \includegraphics[width=\textwidth]{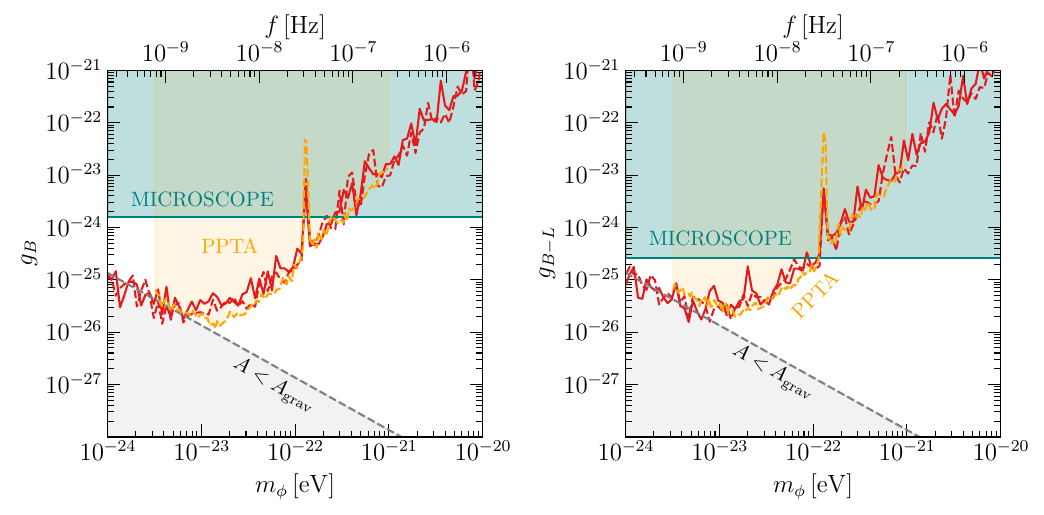}
	\caption{Taken from \cite{NANOGrav:2023hvm}. Upper limits from analyses of NANOGrav \pta data on the coupling of \dm to baryon and baryon-lepton number, shown for the correlated (solid lines) and uncorrelated (dashed lines) cases. In this scenario, dark photon \dm couples to \sm particles in both the Earth and the pulsars. The physics behind this form of \dm has been discussed in \cref{subsec:vecdm}. The black dashed line and gray shaded region indicate where the amplitude of the signal would be less than that from gravitational coupling alone. Constraints from equivalence principle tests (``MICROSCOPE") are shown in teal~\citep{Berge:2017ovy}, and previous limits from the PPTA Collaboration are shown in yellow~\citep{PPTA:2021uzb}. The similarity in shape between the curves in this figure and those in \cref{fig:uldm_direct} arises because both searches target nearly sinusoidal signals embedded in \pta data with similar noise characteristics. Although the underlying \dm models differ—scalar versus vector, moment of inertia changes versus direct forces—the observable effect in each case is the same: modulations in pulse \toas. The coupling strength to \smh particles determines the amplitude of these signals.}\label{fig:vec_pta_constraints}
\end{figure*}

Furthermore, recent evidence for a \sgwb reported by NANOGrav has also been interpreted in terms of vector \uldm. In particular, Ref.~\cite{Chowdhury:2023xvy} showed that a vector \uldm model coupled to muons ($L_\mu-L_\tau$) can provide an even better fit to NANOGrav’s observations than a conventional \sgwb. In this scenario, pulsars undergo oscillations due to their muon content, leading to a spin-dependent frequency shift in the pulses. This produces angular correlations that differ from the standard Hellings–Downs curve expected for a \sgwb, thereby offering a way to test the vector \uldmh hypothesis against future \pta data.

\subsubsection{Constraints on axions through polarization measurements}\label{subsubsec:axion-constr}

As described in \cref{subsec:axions}, axion-like dark matter induces an oscillatory modulation of the polarization plane of linearly polarized light—cosmic birefringence—via its coupling to photons. 

Recent analyses of European \pta data, particularly from the Parkes Pulsar Polarization Array (PPPA), have performed the first dedicated searches for these axion-induced polarization oscillations \cite{Xue:2024zjq,EPTA:2024gxu}. By tracking polarization angles over long time baselines, these searches place upper limits on the axion-photon coupling constant, $g_{a\gamma}$, for axion masses in the ultra-light regime.

\cref{fig:results-epta-pol} shows the resulting constraints from this analysis. The red and blue curves correspond to different statistical treatments of the data, including models that account for spatial correlations between pulsars. Notably, these PTA constraints improve upon previous laboratory and astrophysical bounds at very low axion masses (below approximately $10^{-22}$ eV), where the long coherence time of the axion field aligns well with the decade-scale observation windows of PTAs.

These results highlight the complementary sensitivity of PTAs compared to ground-based interferometers, which probe higher axion masses but with comparable constraints on $g_{a\gamma}$. Together, they cover a broad mass range, making multi-messenger searches a powerful strategy for testing axion dark matter scenarios.

\begin{figure}[htbp]
	\centering
        \includegraphics[width=0.5\textwidth]{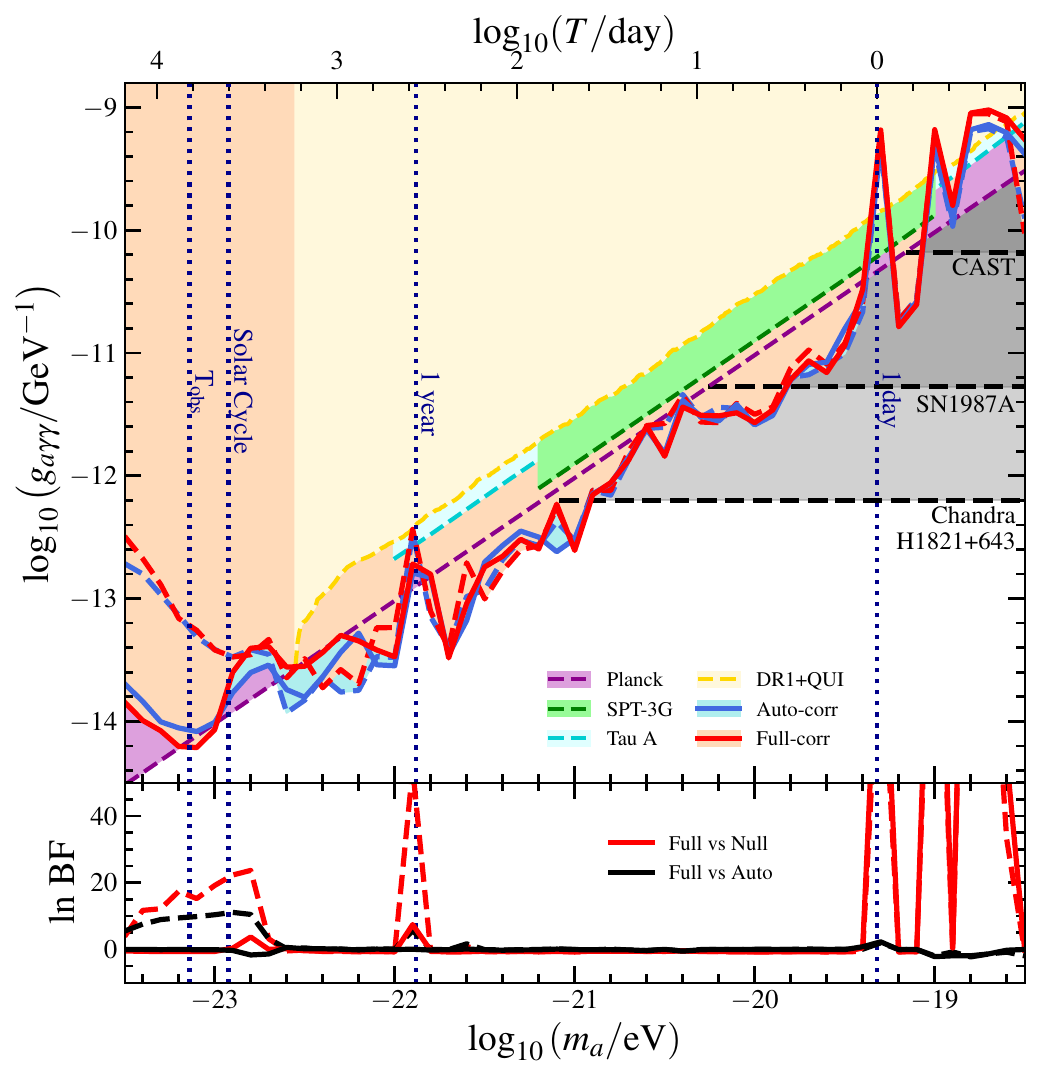}
	\caption{Taken from \cite{Xue:2024zjq}. Upper limits from analyses of European \pta data on the coupling of axion-like \dm to photons. $\rhoDM=0.4\,\textrm{GeV}/\textrm{cm}^3$. The physics behind this form of \dm has been discussed in \cref{subsec:axions}. The blue and red curves correspond to the limits derived using the auto-correlation-only and full-correlation signal models, respectively. Excluded portions of the axion mass /coupling parameter space lie above these curves, and constraints from different experiments are also shown. Key timescales are shown with vertical dotted lines, included $\Tobs=18$ years. Smoothed constraints are shown as dashed lines in cyan, yellow, green, and purple, representing limits from polarization angle (PA) variations observed in the Crab Nebula by POLARBEAR~\cite{POLARBEAR:2024vel}, analyses of PPTA DR1 and Crab Nebula data from QUIJOTE~\cite{Castillo:2022zfl}, measurements of CMB polarization by SPT-3G~\cite{SPT-3G:2022ods}, and studies of axion-like \dm-induced washout effects during recombination from the Planck mission~\cite{Fedderke:2019ajk}, respectively. Additional constraints from the CAST experiment~\cite{CAST:2017uph}, as well as observations of SN1987A~\cite{Payez:2014xsa}, updated in \cite{Fiorillo:2025gnd}, and X-ray spectral distortions in quasar H1821+643~\cite{Reynes:2021bpe}, are shown as black dashed lines. The lower panel shows the Bayes factors (BFs, essentially, the preference for \dm to be in the data versus noise) for each of the scenarios. The BF stays around zero, although some significant spikes are apparent, which has been found to be due to imprecise noise modelling, not astrophysical signals. }
    \label{fig:results-epta-pol}
\end{figure}

\subsection{Prospects for detecting the coupling of ultralight DM to \gwh \ifos}\label{subsec:prospect-pdm}

Though no searches so far have yielded conclusive evidence of any \dm /\sm coupling with the \ifos, impressive constraints, relative to existing \dmh experiments, have been set on these interaction models. The current searches and their constraints have been framed only in terms of upper limits on dilatons, dark photons and \dm that couples to gravity; however, they are actually sensitive to \emph{any} kind of \dm interaction that would cause a differential strain in the \ifos. In fact, to derive these constraints, we first compute the minimum detectable signal amplitude for a generic quasi-monochromatic signal as a function of frequency, as outlined in \cref{subsec:meth-pdm}, and then map these values to constraints on the \dm/\sm coupling constants. It is thus simply a choice that these results are interpreted in terms of two relatively well-motivated \dmh models.

Of course, \uldm could take on any mass a priori, and ground-based \gwh detectors only allow us to probe a small region of this parameter space, from $\sim 10^{-14}-10^{-12}$ eV. Space-based \gwh detectors will permit being sensitive to masses a few orders of magnitude smaller, while high-frequency gravitational-wave (\hfgw) detectors could see \dmh particles a few orders of magnitude larger \cite{Aggarwal:2020olq}. However, it should be noted that when placing constraints on \uldm, the stochasticity of the signal could affect the limits: if we happen to ``get unlucky'' and observe the signal when the phase offset reduces the signal amplitude to zero, as can be seen at the zero points in \cref{fig:dp_hoft_asd}, we would have no constraint. Refs. \cite{Nakatsuka:2022gaf,Boddy:2025oxn} have shown that for ground- and space-based \ifos, that this effect has an $\mathcal{O}(1)$ impact on upper limits, but needs to be mitigated for \ptas by averaging over the signals coming from many pulsars.

The prospects for future ground-based \gwh \ifos, such as \et and \ce, to observe \uldm interactions are bright. Within a decade or so of observation, such detectors could actually be competitive at masses between $[10^{-15},10^{-14}]$ eV, thus permitting an extremely strong probe of new physics \cite{Pierce:2018xmy,Morisaki:2020gui}.

Though \gwh \ifos can probe extremely small \dm/\sm couplings, we do not have an estimate of a minimum strength of these interactions. We thus have no guarantee, in any search that we do, that we are getting closer to probing the true coupling value between ordinary matter and the dark sector. Such a caveat is balanced by the relatively model agnostic nature of our searches, which make minimal assumptions about the signal model, and essentially just look for correlated noise. While we probably cannot believe any one model for \dm/\sm interactions, we can be sure that our techniques are sensitive to a wide range of theories, even ones that have yet to be thought of.




%

\section{Macroscopic \dm transiting through \gwh interferometers}\label{sec:macdm}

As mentioned in \cref{sec:intro}, the mass of \dm  could span many orders of magnitude. In particular, \emph{macroscopic} dark matter, with masses in the range $\Mdm \sim [1, 10^9]$ kg, has been proposed as a viable candidate~\cite{Hall:2016usm}. These candidates could include composite objects such as nuclear-density nuggets \cite{Itoh:1970uw,Witten:1984rs} or Q-balls \cite{Coleman:1985ki},

If such objects were to transit through or near a \gwh \ifo, they could exert measurable gravitational forces on the test masses. In most models, the dominant interaction would be gravitational: the passage of a massive, compact object would cause a time-varying, localized gravitational acceleration of the test masses, leading to a transient displacement signal in the detector.

The resulting signature would be a short-duration, impulsive signal, potentially localized in time and space depending on the trajectory of the \dm object. Unlike continuous-wave or stochastic \dm signals discussed in \cref{sec:pdm}, this effect would appear as a \emph{transient}, with a characteristic waveform determined by the mass, size, velocity, and trajectory of the object.

The event rate of such transits is inversely proportional to the mass of the dark matter particles, and the number density $n_{\rm DM}$ depends on $\rhoDM$:

\begin{equation}
\rhoDM = \Mdm \cdot n_\text{DM},
\end{equation}
This equation implies that the characteristic length scale $L_{\rm DM}$ associated with such objects (assuming constant density), i.e. the distance between \dmh objects, is given by:

\begin{equation}
\frac{L_{\rm DM}}{10^4~\text{km}} \simeq 1.2\times\left(\frac{M_\text{DM}}{1~\text{kg}}\right)^{1/3}.\label{eqn:Lchar}
\end{equation}
For $M_{\rm DM} = 1$ kg, this corresponds to a flux of $\Phi \sim n_{\rm DM} v_0 \sim 3 \times 10^{-10}$ km$^{-2}$ s$^{-1}$, assuming a typical DM velocity $v_0 \sim 220$ km/s. Moreover, the size of \dm, $\rdm$, would be tiny with respect to $L_{\rm DM}$: $\rdm\sim (\Mdm / (\frac{4}{3}\pi\rho_{\rm nugget}))^{1/3} \sim 3\mu\text{m}$ for $\rho_{\rm nugget}\sim 10^{17}\text{ kg/m}^3$. This implies that one transient event per year would be expected in a detector with an effective cross-sectional area\footnote{the area in the plane perpendicular to the \dmh flux within which a passing object would produce a transient} of order a few hundred km$^2$, i.e. \dm would have an impact parameter of $\sim 10$ km, which is of the same order as the length of the \ifo arms (4 km). However, gravitationally interacting \dm would not be able to produce such an event rate \cite{Hall:2016usm,Du:2023dhk}, so we will need to consider interactions of \dm with the \sm to enhance the event rate. In \cref{subsec:yukawa-int}, we will discuss the physics of Yukawa-like interactions that macroscopic \dm can have with the \sm, which will enhance the strength of the transiting \dmh object. Then, in \cref{subsec:proj-yukawa}, we show the projected constraints on Yukawa-like interactions using future data from ground-based \gwh \ifos. 

\subsection{Yukawa-like interactions}\label{subsec:yukawa-int}
If there is a Yukawa-like coupling between the \sm and \dm, allowing for self-interactions as well, \dm transiting through the \ifos could be detectable. The interaction potential between \dm and \smh particles can be described by the following form:

\begin{align}
\label{eqn:ansatz}
V_{i-j} = - M_{i}M_j \frac{G}{r}\Big(1 +(-1)^s~\delta_i\delta_j \exp[-r/\lambda] \Big)
\\ \nonumber \text{where }~ i,j =\text{SM,DM.} ~~~~~~~~~~~~
\end{align}
where $\delta_{i,j}$ represents the coupling strengths between particles $i,j$, and $i,j$ could be either \smh or \dmh particles. $r$ is a length scale, $\lambda$ is the screening length, $s$ is the spin of the interaction mediator, and $M_{i,j}$ are the two masses. By noting that the physical size of the detector and range of the interaction force are much larger than $\rdm$, but much smaller than the separation between \dmh particles $L_{\rm DM}$, the interaction can be modeled in terms of an effective geometric cross section and derived using the Born approximation \cite{Born:1926oue}:

\begin{equation}
    \frac{d\sigma}{d\Omega} = \frac{\Mdm^2}{4\pi^2}\left|\int V(\vec{r})e^{i(\vec{k}_i-\vec{k}_f)\cdot \vec{r}}d^3\vec{r}\right|^2 = \frac{\Mdm^2}{4\pi^2}\left|\tilde{V}(\vec{k}_i-\vec{k}_f)\right|^2\label{eqn:born}
\end{equation}
where $\vec{k}_i-\vec{k}_f$ is the momentum transfer between the initial and final states. The integral is simply the Fourier transform of the potential in \cref{eqn:ansatz} (considering only the interaction term):

\begin{equation}
    \tilde{V} = \frac{4\pi G \Mdm^2 \delta_{\rm DM}^2}{ (\Mdm v_0\sin({\theta/2}))^2+\lambda^{-2}} ,
\end{equation}
where $\theta$ is the scattering angle, i.e. the angle between the incoming and outgoing momenta $\vec{k}_i$ and $\vec{k}_f$, and $\vec{k}_i$ and $\vec{k}_f$ have been written in terms of $v_0$ and $\Mdm$.
Plugging in \cref{eqn:ansatz} into \cref{eqn:born} and integrating over solid angle, the cross-section can be derived:

\be
\sigma_\text{DM-DM} = 16\,\pi \times \frac{G^2\,M_\text{DM}^2\,\delta_\text{DM}^4} {v_0^4}\times \log\left[ \frac{\lambda}{r_\text{DM}} \right],\label{eqn:cross-sec}
\ee
While $|\delta_{\rm SM}|<5\times10^{-4}$ has been strongly constrained by equivalence principle experiments \cite{Adelberger:2009zz}, the \dm coupling constant could be much larger, as the only constraints come from how \dm self-interaction would have influenced structure formation \cite{Spergel:1999mh} and collisions between galaxies \cite{Harvey:2015hha}. In particular, the lack of observed deceleration of \dm in the bullet cluster collision constrained the cross-section of \dm to be less than $\sim 1$ cm$^2$/g \cite{Randall:2008ppe}, which provides limits on the coupling constant of self-interacting \dm to be:

\be
\label{eqn:deltaDM}
|\delta_\text{DM}| \lesssim 5\times 10^{9} \times \left( \frac{1~\text{kg}}{M_\text{DM}} \right)^{1/4},
\ee
in which $\log\left[ \frac{\lambda}{r_\text{DM}} \right]=5$, chosen only as a benchmark, corresponding to a force range much larger $\rdm$. Based on the current upper limits on $\delta_{\rm SM}$ and $\delta_{\rm DM}$, could allow an observation of a transit event in \gwh \ifos. Furthermore, $\delta_{\rm DM}\gg$ 1 would actually alleviate some problems with the cold \dmh scenario, including overly dense areas of dwarf galaxies found in simulations, and a prediction that the Milky Way produces more stars than are observed \cite{Boylan-Kolchin:2011qkt}.

\subsection{Projected constraints on Yukawa-like interactions}\label{subsec:proj-yukawa}

In order to search for transiting macroscopic \dm, it has been proposed in \cite{Hall:2016usm} to use \mf. To do so, firstly we must calculate the \mf \snr $\rho$:

\begin{equation}
    \rho^2 = 4\int \frac{|a(f)|^2}{S_n(f)} \, df;
\end{equation}
Here, \( a(f) \) is the Fourier transform of the differential acceleration \( a(t) \) between the test masses, and \( S_n(f) \) is the \psd of the detector. The differential acceleration \( a(t) \) arises from the \dm transiting through the \ifo test masses, and is calculated with the difference in accelerations between the input mirror and end mirror \cite{Saulson:2017jlf} in each arm. The form of $a(t)$ can be derived by taking the gradient of \cref{eqn:ansatz}:

\begin{equation}
\vec{a}(r) = - \frac{G M_{\rm DM}}{r^2} 
\left[ 1 + (-1)^s \, \delta_{\rm DM} \, \delta_{\rm SM} \, e^{-r/\lambda} \left( 1 + \frac{r}{\lambda} \right) \right] \hat{r} ,
\end{equation}
and then the individual accelerations along the $x$ and $y$ arms can be computed as:

\begin{align}
    \Delta a_{x} &= a(x=L_x,0,0)-a(0,0,0) \\
    \Delta a_{y} &= a(0,0=L_y,0)-a(0,0,0),
\end{align}
\begin{align}
\Delta a_x &= - G M_{\rm DM} 
\Bigg[
\frac{L_x - x_{\rm DM}}{r_1^{3/2}}
-
\frac{- x_{\rm DM}}{\rdm^{3/2}}
\Bigg] \nn
\\ &\times\Bigg[ 1 + (-1)^s \, \delta_{\rm DM} \, \delta_{\rm SM} \, 
\exp\Big(-\frac{r_1}{\lambda}\Big) \Big(1 + \frac{r_1}{\lambda} \Big) \Bigg], \\[1mm]
\Delta a_y &= - G M_{\rm DM} 
\Bigg[
\frac{L_y - y_{\rm DM}}{r_2^{3/2}}
-
\frac{- y_{\rm DM}}{\rdm^{3/2}}
\Bigg]  \nn
\\ &\times\ \Bigg[ 1 + (-1)^s \, \delta_{\rm DM} \, \delta_{\rm SM} \, 
\exp\Big(-\frac{r_2}{\lambda}\Big) \Big(1 + \frac{r_2}{\lambda} \Big) \Bigg],
\end{align}
where

\begin{align}
r_1 &= \sqrt{(L_x - x_{\rm DM})^2 + y_{\rm DM}^2 + z_{\rm DM}^2}, \\
r_2 &= \sqrt{x_{\rm DM}^2 + (L_y - y_{\rm DM})^2 + z_{\rm DM}^2}, \\
\rdm &=\sqrt{x_{\rm DM}^2+y_{\rm DM}^2 + z_{\rm DM}^2},
\end{align}
and $x_{\rm DM}$, $y_{\rm DM}$ and $z_{\rm DM}$ are the spatial coordinates of the \dmh object.
The acceleration depends directly on the coupling of \dm to \sm particles; thus, the \snr will be proportional to this coupling as well. In searches with \gwh data, a threshold must be set that delineates candidates that are most likely noise with those that could be of astrophysical origin. To estimate the number of detectable transiting \dm events, an \snr threshold of \( \rho > 8 \) is applied, which is standard in \gwh searches \cite{maggiore2008gravitational}, and the detection rate as a function of \dm mass, and coupling \( g \equiv \delta_{\rm SM} \delta_{\rm DM} \) is computed:

\begin{equation}
    \mathcal{R} \sim \Phi \sigma_{\rm DM-SM} A\label{eqn:ratemacdm}
\end{equation}
where $A$ is the effective area of the detector through which \dm transits. The effective area for macroscopic \dmh transits is determined by the physical area of the detector, the angle of incidence of the \dm on the detector, and the interaction cross-section between \dm and the \ifo.  For larger cross sections (larger couplings), the \snr of the signal will increase. Thus, the \snr threshold determines a minimum detectable cross section, which in turns fixes the rate of transiting \dm:

We show potential constraints on the rates of \dm transiting through advanced LIGO at design sensitivity in \cref{fig:LIGOYukawaRate}. These results demonstrate that macroscopic \dm can be detectable, depending on the parameters of the interaction.

\begin{figure*}[htbp]
    \centering
    \includegraphics[width=0.8\textwidth]{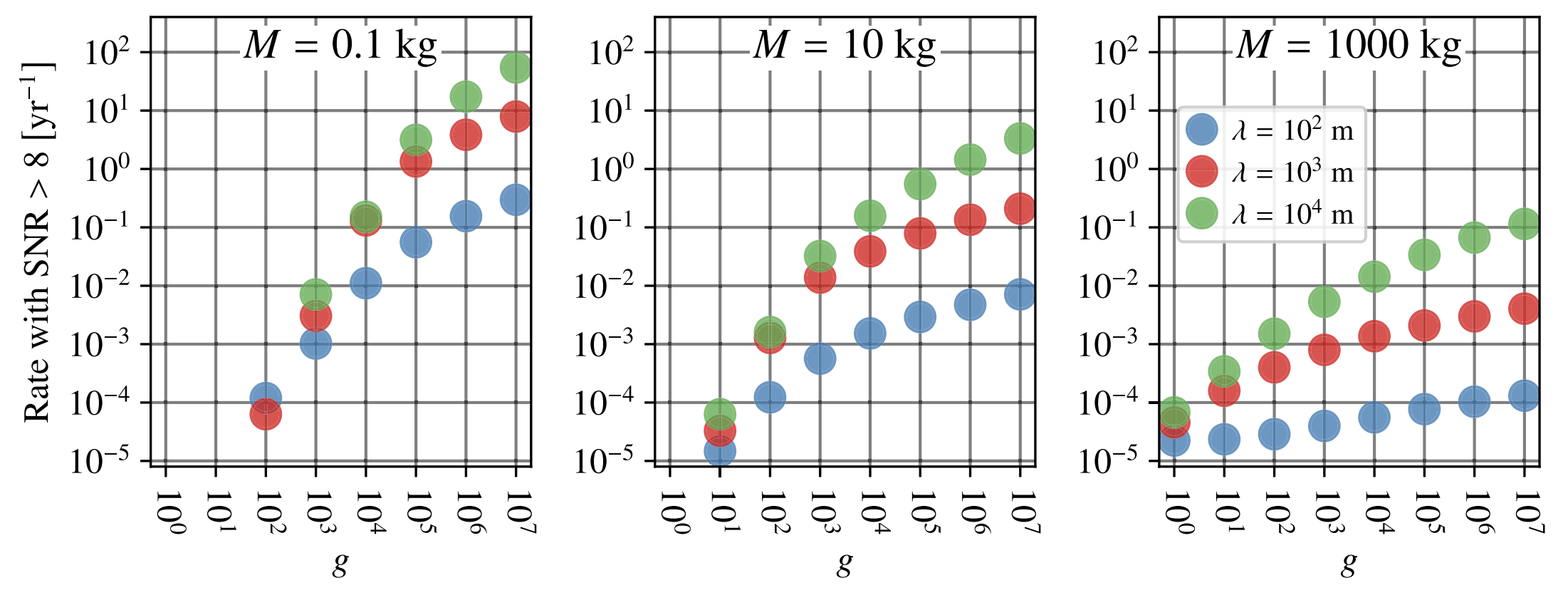}
    \caption{Taken from \cite{Hall:2016usm}. Expected number of macroscopic \dmh transit events per year detectable by advanced LIGO at design sensitivity, as a function of \dm mass, screening length \( \lambda \), and coupling \( g = \delta_{\rm SM} \delta_{\rm DM} \). The physics behind this form of \dm has been discussed in \cref{subsec:yukawa-int}. The blue, red and green colors indicate different screening lengths $\lambda$ of the Yukawa-like force. The three panels show the rate of transits for \dmh objects with three different masses. The rate is calculated using \cref{eqn:ratemacdm} and the preceding equations, assuming a detection threshold on the \snr of 8. The threshold fixes the minimum detectable coupling accessible to a \mf search. }
    \label{fig:LIGOYukawaRate}
\end{figure*}

Further work in \cite{Du:2023dhk} has expanded on this analysis, incorporating additional effects of \dm transits, such as the Doppler effect, Shapiro delay, and Einstein delay, each of which contributes to the strain observed in the interferometer. These effects arise from the interaction of the \dm object with the interferometer's test masses, leading to small deviations in the expected strain signal.

\begin{enumerate}
\item Doppler effect: As the \dm object moves relative to the detector, its velocity induces a periodic shift in the frequency of the signal received by the interferometer. This is similar to the classical Doppler shift seen in sound or light waves, but in this case, it is gravitational. The Doppler effect results in a frequency-dependent modulation of the strain signal, with the amplitude of the strain varying as the velocity of the \dm object changes. 

\item Shapiro delay: This effect arises due to the curving of spacetime near massive objects. In particular, light traveling near a massive object, such as a transiting \dm object, will experience a delay in arrival time. This delay would be observed as a time shift in the signal received by the detectors, as the light takes longer to travel through the curved spacetime near the \dm object.

\item Einstein delay: Similar to the Shapiro delay, the Einstein delay occurs due to the effects of gravitational time dilation. Time is slower near more massive objects. Thus, the passage of a massive \dm object through the interferometer’s arms would result in a slight delay in the signal, since the clocks at the test masses would tick slower as the object passes by.
\end{enumerate}
While these gravitational effects are real, they are typically very weak unless \dm interacts with \smh particles. To make the effects measurable by \gwh interferometers, Ref. \cite{Du:2023dhk} confirmed that \dm must couple with \smh particles, agreeeing with the findings of \cite{Hall:2016usm}. The strength of this coupling is given in \cite{Du:2023dhk} by the effective parameter \( \tilde{\alpha} \sim \delta_{\rm SM} \delta_{\rm DM} \), where $\tilde{\alpha}=g$ in \cite{Hall:2016usm}. Thus, detecting these weak gravitational signals requires a sufficiently large value of \( \tilde{\alpha} \), which can be constrained by observations of \dm transits.

\cref{fig:fifth_force_reach} shows the projected 90\% sensitivity on \( \tilde{\alpha} \), which can be probed by \gwh detectors for macroscopic \dm. The results, shown for two different screening lengths (\(\lambda = 1 \, \text{m}\) and \(10^6 \, \text{m}\)), are compared to existing constraints from the Bullet Cluster \cite{Spergel:1999mh}, MICROSCOPE  \cite{Berge:2017ovy}, and neutron-star kinetic heating experiments  \cite{Gresham:2022biw}. While \gwh detectors may not surpass these existing constraints, they provide complementary information, especially if part of the \dm population interacts with the \sm via a new fifth force.

\begin{figure*}[htbp]
	\includegraphics[scale=0.55]{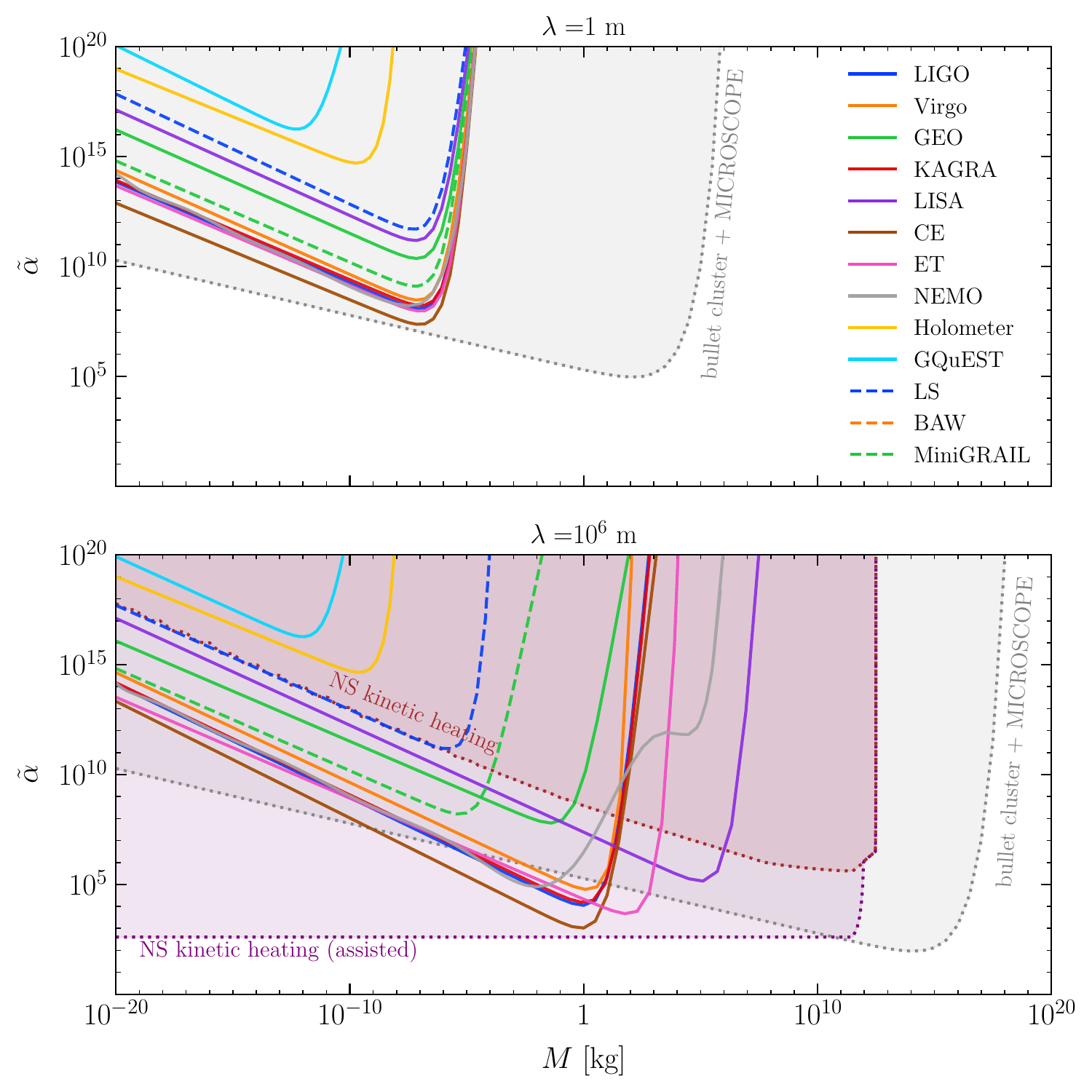} 
	\caption{Taken from \cite{Du:2023dhk}. Projected sensitivity at 90\% confidence to the \smh/\dm interaction parameter \( \tilde{\alpha} \sim \delta_{\rm SM} \delta_{\rm DM} \) from macroscopic \dm transits through several \gwh detectors. The physics behind this form of \dm has been discussed in \cref{subsec:yukawa-int}. The figure compares the projected sensitivity to existing constraints, including those from the Bullet Cluster, MICROSCOPE, and neutron-star kinetic heating, for two choices of screening length. Here, $\Tobs=1$, and $\lambda=1$ m and $\lambda=10^6$ m, and an \snr threshold of 2 are assumed. The gray line represents combined limits from Bullet Cluster observations~\cite{Spergel:1999mh, Kahlhoefer:2013dca} and the MICROSCOPE experiment~\cite{Berge:2017ovy, Fayet:2018cjy}. Purple and red lines indicate bounds from neutron star kinetic heating~\cite{Gresham:2022biw}, with and without additional short-range DM--baryon interactions, respectively.}
	\label{fig:fifth_force_reach}
\end{figure*}

While deterministic signals might be challenging to detect due to the weak gravitational interactions, there is also potential for detecting a stochastic background arising from numerous transiting \dm events. This stochastic background could be visible if the Yukawa interaction is sufficiently strong, as explored in \cite{Hall:2016usm,Du:2023dhk}. Future work will focus on whether a stochastic signal could be distinguished from the background noise in upcoming \gwh observations.

\section{\gws from ultralight boson clouds around rotating black holes}\label{sec:gwbc}

If new ultralight bosons\footnote{In the context of boson clouds around rotating black holes, this new particle could, but need not, be \dm} exist, quantum fluctuations could allow them to pop into existence near a rotating \bh. Some bosons will fall in; others will scatter off, and the extent to which either of these scenarios happens depends on the mass and spin of both the black hole and the new boson. If the following so-called ``superradiance condition'', which relates the \bh angular frequency $\Omega$ to the boson angular frequency $\omega$ via the magnetic quantum number $m$, is met \cite{Brito:2015oca}:

\begin{equation}
    \omega <m\Omega,
    \label{eqn:superradcond}
\end{equation}
the boson will extract energy from the \bh, and thus the ``bosonic wave'' will be amplified. This amplification will be maximized if the Compton wavelength of this new particle is comparable to the radius of a \bh. The massive bosons then become bound to the \bh because they sit in the potential well of the \bh, allowing for successive scatterings, thus permitting a huge number of bosons to appear around the \bh as a cloud in a given energy state (no limit on the occupation number of these particles exists). This process is called ``superradiance''.

More quantitatively, solving the Klein-Gordon equation for a Kerr \bh hole in the presence of a massive scalar field results in a Schrondinger-like equation that exhibits a $1/r$ potential, resulting from the gravitational interaction between the bosons and the \bh. The solution of this equation shows that the energy states of the boson cloud are quantized analogously to those of the hydrogen atom, allowing us to describe the boson cloud as a ``gravitational atom in the sky'' \cite{Baumann:2018vus}.

Qualitatively, superradiance will occur regardless of the spin of the boson; however, the timescales to build the cloud, as well as to deplete it, will differ. In all cases, however, as we will see, the \gwh emission timescale will be shorter than the time to build up the cloud, making these systems excellent sources of \gws. 

The cloud will continue to build up until the \bh is spun down enough such that the condition in \cref{eqn:superradcond} is no longer satisfied. In particular, the timescales $\tinst^{\rm (s)}$ and $\tinst^{\rm (v)}$ for which this occurs scalar and vector clouds, respectively, are \cite{Isi:2018pzk}\footnote{Note that we do not consider tensor \bcs in this work}:

\be \label{eq:tinst_scalar}
\tinst^{\rm (s)} \approx 27\, {\rm days} \left(\frac{\mbh}{10\, \msun}\right) \left(\frac{0.1}{\fs}\right)^9 \frac{1}{\chi_i}\, ,
\ee
 and
\be \label{eq:tinst_vector}
\tinst^{\rm (v)} \approx 2\, {\rm minutes} \left(\frac{\mbh}{10\, \msun}\right) \left(\frac{0.1}{\fs}\right)^7 \frac{1}{\chi_i}\, ,
\ee
where $\chi_i$ is the \bh spin at birth, and the \gw fine-structure constant $\alpha$ is
\begin{align}
\fs &= G \mbh m_b \\ \label{eq:alpha}
&= 0.11 \left(\frac{\mbh}{10\msun}\right)\left(\frac{m_b}{1.47\times 10^{-12}\ev}\right)
\end{align}
where $m_b$ is the mass of the boson and $\mbh$ is the mass of the \bh.

This new boson could have a spin of 0, 1 or 2, where spin-0 could correspond to the pseudo-scalar QCD axion (or axion-like) particle \cite{Brito:2015oca,Brito:2020lup}, spin-1 could indicate dark photons arising through kinetic mixing with the ordinary photon \cite{Siemonsen:2022ivj}\footnote{Note that this dark photon is \emph{not} the dark photon described in \cref{subsec:vecdm}.}, or spin-2 tensor bosons that could arise from modified theories of gravity \cite{Brito:2020lup}. After the cloud builds up in an isolated \bh system, \gws can be emitted in three different ways: (1) through boson-boson annihilation, (2) through boson energy-level transitions, and (3) after a ``bosenova'', all of which will be discussed in \cref{subsec:annil,subsec:transition,subsec:bosenova}. Furthermore, the \gwh frequency is fixed by the boson mass, which could in principle take on any ultralight mass; thus, it is important to note that \gwh \ifos, from \ptas to \hfgw instruments, will allow an extensive mass range to be probed, from $[10^{-22},10^{-9}]\ev$ \cite{Aggarwal:2020olq,Brzeminski:2024drp,Arvanitaki:2014wva,Brito:2017zvb,Ng:2020jqd}.

It has been shown, under a wide range of assumptions of \bh populations, spins, redshift distributions, etc., that such systems could be detectable by \cwh searches in advanced \lvk data and/or in future detectors, such as \et and \ce \cite{Arvanitaki:2014wva,Zhu:2020tht}; and if such sources are not seen, particular boson masses can be disfavored -- this will be the subject of \cref{sec:gwconsbc}. Furthermore, both galactic \bhs \cite{Palomba:2019vxe} and remnants of mergers of \bhs outside the galaxy could also be detected, the latter of which, in the scalar case, only being visible in \et and \ce \cite{Arvanitaki:2009fg,Arvanitaki:2014wva,Isi:2018pzk}, which will be discussed more in \cref{subsec:targets}. Annihilating vector \bc systems, however, emit stronger, though shorter, \gwh signals \cite{Brito:2015oca,Baryakhtar:2017ngi,Siemonsen:2022yyf,Jones:2023fzz,Jones:2024fpg}, and potentially may contain electromagnetic counterparts if the vector boson kinetically mixes with the ordinary photon \cite{Siemonsen:2022ivj}.

In this section, we will discuss the three main ways in which \gws can be emitted from isolated \bh / \bc systems: annihilation, transitions and bosenovae (\cref{subsec:annil,subsec:transition,subsec:bosenova}). We will also detail how self-interactions between the bosons could impact the superradiance process and the resulting \gws in \cref{subsec:self-int}. Finally, we will explain some of the rich physics that can occur when \bcs form around \bbh systems in \cref{subsec:bcbin}.

\subsection{Annihilation signal}\label{subsec:annil}


Ultralight bosons are their own anti-particles, which means that they continuously annihilate (even as the cloud is growing), producing \gwh radiation at a frequency fixed by the boson mass and slightly redshifted by the \bh mass (in the non-relativistic limit $\alpha\ll 1$, which predominately applies to the equations shown in this review). In the scalar boson case \cite{LIGOScientific:2021jlr}:

\begin{align}
f_\text{gw} &\simeq 483\,{\rm Hz} \left(\frac{m_\text{b}}{10^{-12}~\text{eV}}\right)  \nn \\ &\times \left[1-7\times 10^{-4}\left(\frac{M_\text{BH}}{10M_{\odot}}\frac{m_\text{b}}{10^{-12}~\text{eV}}\right)^2\right].
\label{eq:fgw}
\end{align}
Qualitatively, the \gws emitted are monochromatic because the mass of the boson is fixed, and all the bosons are in the same energy state.
As alluded to before, the timescales for \gwh emission are much longer than the time it takes to build the cloud \cite{Isi:2018pzk}:

\be \label{eq:tgw_scalar}
\tgw^{\rm (s)} \approx 2\times 6.5 \times 10^{4}\, {\rm yr} \left(\frac{\mbh}{10\, \msun}\right) \left(\frac{0.1}{\fs}\right)^{15}\hspace{-2pt} \frac{1}{\chi_i} .
\ee
\be \label{eq:tgw_vector}
\tgw^{\rm (v)} \approx 2\times 1\, {\rm day} \left(\frac{\mbh}{10\, \msun}\right) \left(\frac{0.1}{\fs}\right)^{11}\hspace{-2pt} \frac{1}{\chi_i} .
\ee
where we have explicitly noted a factor of 2 that is missing from the original reference \cite{Isi:2018pzk}, as pointed out in \cite{Siemonsen:2022yyf}.

As the number of bosons starts to decrease, there will also be a classical depletion of the cloud over time. This reduction in bosons causes the cloud to contract, resulting in the following positive drifts in frequency over time \cite{Isi:2018pzk}\footnote{We note that relativistic computations of the frequency drift have also been performed \cite{Siemonsen:2022yyf,May:2024npn}, which are useful for real searches of \gws arising from \bcs but not less important for our discussions here.}:

\be \label{eq:fdot_scalar}
\dot{f}_{\rm gw}^{\rm (s)} \approx 3\times 10^{-14}\, {\rm Hz/ s} \left(\frac{10M_{\odot}}{\mbh}\right)^2 \left(\frac{\alpha}{0.1}\right)^{19} \chi_i^2\, ,
\ee
\be \label{eq:fdot_vector}
\dot{f}_{\rm gw}^{\rm (v)} \approx 1\times 10^{-6}\, {\rm Hz/ s} \left(\frac{10M_{\odot}}{\mbh}\right)^2 \left(\frac{\alpha}{0.1}\right)^{15} \chi_i^2\, .
\ee
In the non-relativistic limit $\alpha\ll 1$, the signals will have the corresponding amplitudes of \cite{Isi:2018pzk}:
\be \label{eq:h0_scalar_approx}
h_0^{\rm (s)} \approx 8 \times 10^{-28} \left(\frac{\mbh}{10 \msun}\right)
\left(\frac{\alpha}{0.1}\right)^7 \left(\frac{\rm Mpc}{r}\right) 
\left(\frac{\chi - \chi_f}{0.1}\right)
\ee
\be \label{eq:h0_vector_approx}
h_0^{\rm (v)} \approx 4 \times 10^{-24} \left(\frac{\mbh}{10 \msun}\right)
\left(\frac{\alpha}{0.1}\right)^5 \left(\frac{\rm Mpc}{r}\right) 
\left(\frac{\chi - \chi_f}{0.1}\right),
\ee
where the final spin $\chi_f$ is
\be \label{eq:final_spin}
\chi_f = \frac{4 \fs_f \mb}{4\fs^2_f + \mb^2}\,.
\ee
These equations indicate interesting properties for scalar and vector boson cloud annihilation signals. For bosons whose Compton wavelengths are optimally matched to the size of the \bh, scalar bosons will annihilate over timescales that greatly exceed the observation time of ground- or space-based \gwh \ifos (\cref{eq:tgw_scalar}), and their frequencies will hardly drift over time. Thus, this signal can be thought of as a quasi-sinusoidal and persistent. On the other hand (and with the same caveat of optimal matching), the \gwh signal from annihilating vector \bcs lasts significantly shorter than the scalar one (\cref{eq:tgw_vector}), and thus can be thought of more as a ``transient'' \cw, lasting for durations for $\mathcal{O}($hours-days), i.e. much longer than mergers of black holes routinely detected now by \lvk, but much shorter than scalar boson cloud annihilation signals. Furthermore, vector signals emit much more \gwh power per unit time than scalar ones, but for scalar signals, analysis methods are able to integrate over the entire data collection period of \lvk to improve the \snr, as will be discussed in \cref{subsec:meth-bc}. However, in the vector case, it is possible to have \cwh emission for particular combinations of boson mass and \bh mass, such as large \bhs and small boson masses, where the spin-up of the signal, as described by \cref{eq:fdot_vector}, becomes comparable to that in the scalar case (\cref{eq:fdot_scalar}), thus making the signal almost monochromatic.

\subsection{Transition signal}\label{subsec:transition}

In the scalar boson case, just as electrons jump between discrete energy levels in atoms, bosons can also transition between discrete energy states, provided these states are populated. The energy states of the boson cloud are quantized, with each state corresponding to specific quantum numbers. Significant transition signals are expected when the cloud reaches a population that allows transitions between particular energy levels, such as from the 5g to the 6g states, where ``g'' refers to the orbital angular momentum quantum number of the state. These transitions between specific energy levels are expected to generate detectable gravitational wave signals that could last for durations much longer than the typical observation times of ground-based gravitational wave interferometers. The frequency of these signals is given by \cite{Arvanitaki:2014wva}:

\begin{equation}
    f_{\rm gw} \sim 15 \, \text{Hz} \left(\frac{m_b}{10^{-11} \, \text{eV}}\right)
\end{equation}

The frequency of the signal would drift over time according to the duration of the signal and the axion decay constant \( f_a \) \cite{Arvanitaki:2014wva}:

\begin{equation}
  \frac{df}{dt} \simeq 10^{-11} \, \frac{\text{Hz}}{\text{s}} \left(\frac{f_{\rm gw}}{90 \, \text{Hz}}\right) \left(\frac{M_{\rm BH}}{10 \, M_{\odot}}\right) \left(\frac{10^{17} \, \text{GeV}}{f_a}\right)^{2} \left(\frac{5 \, \text{yr}}{T_{\rm obs}}\right)^{2}
\end{equation}
We note here that the axion decay constant, arising from self-interactions, affects the spin-up of the signal. In the previous discussions surrounding \cref{eq:fdot_scalar,eq:fdot_vector}, self-interactions were assumed to be zero.

While the expected event rate of these transitions in ground-based detectors is unlikely to exceed one event in around 3.5 years of observation with advanced LIGO \cite{Arvanitaki:2014wva}, the presence of self-interactions could potentially enhance the transition rates, making these signals detectable \cite{Collaviti:2024mvh}.

In the vector boson case, transitions can also occur, and interestingly, overtone modes may grow faster than the fundamental modes, particularly for small angular numbers, fine-structure constants, and black hole spin values \cite{Siemonsen:2019ebd}. These higher overtones could saturate more quickly than the fundamental modes, allowing transitions to lower states and increasing the overall energy of the cloud, potentially leading to detectable gravitational wave emission.

\subsection{Impact of boson self-interactions on superradiance}\label{subsec:self-int}

While the formation of ultralight boson clouds around rotating \bhs is already expected to result in observable \gws, the strength of the self-interactions within the boson cloud can play a crucial role in determining the nature and detectability of these signals. Specifically, the self-interactions between bosons in the cloud can influence key dynamical processes, such as:

\begin{enumerate}
    \item Saturation of Superradiance: Self-interactions modify the growth timescale, potentially leading to a faster saturation of the superradiance process. This could result in stronger \gwh emission once the cloud reaches its maximum energy state. The timescale for superradiance saturation is sensitive to the coupling strength between bosons, and self-interactions may allow the cloud to reach a critical size faster, enhancing the \cwh signal emitted during this phase \cite{Yoshino:2012kn,Yoshino:2014wwa,Yoshino:2015nsa}.
    \item Merger Dynamics: The presence of self-interactions can affect how quickly the boson cloud dissipates or decays, influencing the coalescence of the black hole system. This decay modifies the timescale of the inspiral phase and the subsequent merger, altering the frequency evolution of the emitted \gws. A faster decay due to self-interactions could lead to a sharper rise in the frequency of the \gwh signal \cite{Boudon:2023vzl,Aurrekoetxea:2024cqd,Takahashi:2024fyq}.
    \item Enhanced Annihilation of Boson Clouds in Isolated Black Holes: In a non-interacting system, the annihilation rate of bosons is limited by the density and distribution of particles in the cloud. Self-interactions, however, can increase the density of the cloud by accelerating the process of boson accumulation and scattering. As a result, the rate of boson annihilation can be significantly enhanced. This increased annihilation rate leads to stronger and potentially more detectable \gwh signals from the cloud \cite{Baryakhtar:2020gao}.
    \item Enhanced Energy-Level Transitions: The energy levels of bosons in the cloud are quantized, similar to the energy levels of electrons in an atom. Self-interactions can facilitate the transition of bosons between higher energy states more rapidly, increasing the amplitude of the emitted gravitational waves. This could make energy-level transitions a more significant source of detectable GW signals, especially if the cloud is in a high-energy state \cite{Baryakhtar:2020gao}.
\end{enumerate}

These effects become particularly relevant when considering the potential detection of boson clouds in future \gwh observatories, such as \ce and \et \cite{Collaviti:2024mvh}. Recent searches for these signals have not accounted for these interactions, and their inclusion may shift current detection prospects by enhancing the signal strength or changing the emission characteristics. In this section, we will explore how self-interactions influence the dynamics of boson clouds and their implications for gravitational wave astronomy."

Note that while spin-0 and spin-1 \bcs are relatively well understood theoretically, spin-2 clouds are much more complicated. The superradiant instability of spin-2 fields occurs much faster than the others \cite{East:2023nsk}, and since the backreaction of the instability has not been computed due to difficulties arising from a nonlinear coupling between the spin-2 particle and gravity, it is not yet known whether the observational signatures are similar to the vector or scalar cases, and whether there is a dependency on the specific nonlinear theory chosen. Such ultralight bosons could also be used to probe the quantum nature of \bhs \cite{Mitra:2023sny}.

\subsection{Bosenova signal}\label{subsec:bosenova}

If the self-interactions between bosons become stronger than the gravitational binding energy of the cloud, the cloud begins to destabilize and cannot be described analogously to the hydrogen atom \cite{Fukuda:2019ewf,Collaviti:2024mvh,Omiya:2024xlz,Jones:2024fpg}. The cloud may undergo a significant reconfiguration, leading to a dramatic release of energy, a process referred to as a ``bosenova.'' This occurs when the potential energy of the boson cloud becomes comparable to or exceeds the self-interaction energy, leading to the rapid rearrangement of the bosonic matter.

The phenomenon of a bosenova has been observed in condensed matter systems, in which a Bose-Einstein condensate undergoes a similar process due to a change in interactions induced by an external magnetic field \cite{donley2001dynamics}. However, the occurrence of a bosenova in the context of ultralight bosons around black holes remains highly uncertain, with studies suggesting that self-interactions may limit bosenovas to a restricted region of the allowed parameter space, particularly for certain combinations of black hole and boson masses \cite{Baryakhtar:2020gao}.

In particular, in the axion case, originally proposed in \cite{Arvanitaki:2010sy}, bosenovas would likely \emph{not} occur. While early papers assumed that the \bc reached large enough amplitudes to produce \gwh emission \cite{Yoshino:2012kn,Yoshino:2014wwa,Yoshino:2015nsa}, subsequent studies in the non-relativistic \cite{Baryakhtar:2020gao} and relativistic \cite{Omiya:2022gwu} regimes confirmed that dissipative nonlinear effects would prevent large-enough cloud amplitudes from occurring to cause a bosenova. In the vector case, bosenovas could actually occur if strings are formed the Higgs-Abelian mode \cite{East:2022ppo,East:2022rsi,Brzeminski:2024drp}.

Though \gws from \bcs described above may seem independent, they could in principle be happening at the same time. In particular, if ultralight scalar or vector bosons have some self-interactions, and If these clouds are composed of string axions (as opposed to the QCD axion), a non-linear self-interaction may cause a bosenova if, during superradiance, the value of the scalar field approaches the axion decay constant $f_a$, which could be below the grand unification theory (GUT) scale \cite{Arvanitaki:2009fg,Yoshino:2012kn,Yoshino:2014wwa,Yoshino:2015nsa}. This bosenova would halt the superradiance process, as about 5\% of the \bc mass would fall back into the \bh, but superradiance would immediately start again until the next bosenova is triggered. Such periodic bosenovas could sustain superradiance for much longer durations than those in the conventional scenario, e.g. in \cite{Arvanitaki:2010sy}, permitting annihilating clouds even around extremely old \bhs, and even affect \gws from \bbh systems \cite{Xie:2022uvp} . 


\subsection{Clouds in binary black hole systems}\label{subsec:bcbin}

If \bcs form around \bbh systems, they should imprint some signature in the \gws arising from their inspiral \cite{Baumann:2018vus}. In particular, the \gwh signal will be modified due to tidal disruptions induced by the companion, the multiple moments of the cloud \cite{Baumann:2018vus,Baumann:2019eav,Baumann:2019ztm} or through a \df drag force induced by the \bcs . These kinds of signals could be visible in future space-based detectors, since the deviations primarily occur in the early inspiral stages of the system, i.e. at frequencies much lower than currently accessible by ground-based \gwh \ifos \cite{Guo:2023lbv}. In both cases, observations of the inspiral of binary systems, and in particular \emri systems \cite{Zhang:2018kib,Brito:2023pyl}, could indicate the presence of \bcs around one or both of the objects if there is some dephasing with respect to the waveform in vacuum. Additionally, once the binary reaches a certain stage of its evolution, the impact of both the spin-induced quadrupole moment and tidal forces will lessen, since the clouds will become disrupted by their companions. If we can observe variations in both the spin-induced quadrupole moment and the tidal deformability over the course of the lifetime of the binary, it would indicate the presence of a (disrupted) \bc \cite{Baumann:2018vus}.

More recently, it has been shown that as the orbital separation in a \bbh system decreases and becomes comparable to the size of the \bc, the \bc could be ``ionized'' \cite{Takahashi:2021yhy,Baumann:2021fkf,Baumann:2022pkl}. This ionization occurs when the energy from the \bbh system becomes large enough to unbind the cloud from its host \bh. As a result, the inspiral of the binary becomes significantly influenced by the dynamics of the boson cloud, rather than merely being perturbed by it.


We will now describe four different ways in which \bcs could affect \gws from \bbh systems.

\subsubsection{Spin-induced quadrupole moment}

The spin-induced quadrupole moment is often parameterized in terms of a dimensionless quantity $\kappa=\frac{Q}{J^2/\mbh}$, where $Q$ is the quadrupole moment of the \bh, that has been normalized by the angular moment $J$ and mass of the \bh. This parameter enters into the waveform at the second post-Newtonian (2PN) order, i.e.\ as corrections of order $(v/c)^4$ relative to the leading quadrupole radiation. Thus, the spin-induced quadrupole moment, regardless of its source, has a significant impact on the \gwh signal from compact objects inspiraling towards one another.

Ref. \cite{Krishnendu:2017shb} and Ref. \cite{Rahman:2021eay} showed that $\kappa$ can be used as a distinguishing parameter for different kinds of objects, and also a test of the no-hair theorem: as an example, $\kappa=1$ for \bhs \cite{Hansen:1974zz}, while $\kappa\sim[1.4,8]$ for \nss depending on the \eos \cite{Laarakkers:1997hb,Pappas:2012ns}. In the case of \bc systems, $\kappa$ will take on vastly different values, of $\mathcal{O}(1000)$, depending on the \bc and \bh masses, and the extent to which the Kerr metric is altered in the presence of a cloud with enough mass compared to the \bh. 

\subsubsection{Tidal disruptions}

Because the \bc is much less compact than the \bh, we expect that tidal disruptions from the companion could occur when the two objects in the \bbh are close enough to each other \cite{Ng:2020jqd}. To characterize the effects of tides in a binary system, the tidal deformability parameter $\Lambda$ is used, which depends on the mass and radius of the compact objects. For \bhs in vacuum, this quantity is 0, while for \nss, it can take on values of $\mathcal{O}(100-1000)$ \cite{Chatziioannou:2020pqz}. For \bcs, $\Lambda\sim 10^{7}$, which is a remarkable departure from that expected from conventional astrophysical binaries \cite{Baumann:2019eav}. In \cite{Payne:2021ahy}, the authors considered the impact that superradiance would have on hierarchical \bh mergers, and showed that the reduction of spin in \bhs implies smaller recoil velocities of \bbh mergers. Smaller recoil velocities imply that remnants could remain inside clusters, leading to higher chances of hierarchical formation and merger of \bbhs. Such disruptive effects of bosons on not just individual \bhs but also on the \bh population, imply that \bcs would significantly impact the formation and evolution histories of \bhs.

\subsubsection{Binary-induced transitions}


During the inspiral of a \bbh system that contains \bcs, the interaction between the cloud and the \bhs can cause transitions between different energy levels in the clouds. These transitions can potentially be detected by space-based \gwh \ifos if any of the following mechanisms is present:

\begin{enumerate}
    \item Hyperfine Resonance: The orbital frequency of the binary system matches the energy spacing between two adjacent energy levels in the boson cloud. This condition is analogous to a resonance in atomic physics, where the energy difference between two states is exactly matched by an external driving frequency. In this case, the system stays in equilibrium, and energy is transferred between energy levels that are in resonance \cite{Baumann:2018vus}.
    \item Bohr Resonance: In addition to hyperfine resonance, the growing mode (the mode that is being excited by the inspiral) mixes with decaying modes of the cloud. This interaction allows energy to be transferred between different states, influencing the \gwh signal \cite{Baumann:2018vus}.
    \item Landau-Zener Transitions: When the orbital frequency of the inspiral matches the energy spacing between adjacent energy levels, the system undergoes a non-adiabatic shift in energy states. This transition process, characterized by the Landau-Zener mechanism, significantly alters the frequency and amplitude evolution of the \gwh signal. These non-adiabatic transitions can provide observable changes in the signal, offering additional insights into the properties of the boson cloud \cite{Baumann:2019ztm}.
\end{enumerate}
These effects, all of which could be present simultaneously, would provide ``smoking-gun'' signatures of the presence of boson clouds around rotating black holes in a binary system \cite{Baumann:2018vus}.

\subsubsection{Ionization}

When the binary system separation becomes comparable to the size of the cloud, an ``ionization'' of the \bc occurs as the cloud becomes unbounded from its host \bh \cite{Takahashi:2021yhy,Baumann:2021fkf,Baumann:2022pkl}. The energy from the binary drives this effect, and is significantly larger than that emitted via \gws; thus, the inspiral of the binary becomes dominated by the \bc, instead of simply perturbed by it. Ionization will also tend to circularize the binary if it formed through dynamical capture, while leaving any orbital inclinations unaffected \cite{Tomaselli:2023ysb}. It has been recently found, however, that as the resonances in the \bc evolve during the orbit, the \bc is unperturbed and ordinarily visible in LISA if the cloud and binary are counter rotating; otherwise, the cloud is destroyed because of the resonances, but a distinct mark is left on the binary. That is, the binary is forced to co-rotate with the \bc, and its eccentricity is driven to a particular value, which allows the possibility of doing statistics with detected \bbh systems in LISA to probe the \bc hypothesis \cite{Tomaselli:2024dbw}.

Taken together, the various phenomena discussed above---from level transitions and resonances to the dramatic ionization of the cloud---highlight the rich interplay between \bcs and \bbhs. Each mechanism offers a distinct pathway by which the presence of a \bc can alter binary evolution and leave characteristic imprints on the emitted \gws. These theoretical insights provide the foundation for future searches with space-based \gwh data \cite{Bertone:2019irm,Bertone:2024rxe}. In the following section, however, we turn to current ground-based \gwh \ifo data, which so far constrain only annihilation signals from \bcs, as discussed in \cref{subsec:annil}.


\section{\gw constraints on boson clouds}\label{sec:gwconsbc}

The extensive theoretical background on boson clouds around rotating \bhs, coupled with the potential for them to be \dm, have motivated \gwh probes of boson cloud/\bh systems. Different \gwh \ifos probe vastly different mass regimes of \uldm \bcs. While ground-based detectors are sensitive to annihilation signals from bosons of masses between $\sim 10^{-14}-10^{-12}$ eV \cite{Arvanitaki:2014wva}, space-based detectors can probe a few orders of magnitude lower than that, $\sim 10^{-20}-10^{-16}$ eV \cite{Baumann:2018vus}, and \hfgw detectors could see systems with masses around $\sim 10^{-9}$ eV \cite{Aggarwal:2020olq}. At the moment, only ground-based \gwh \ifos exist, and so we focus on methods and search results from the most recent \lvk observing runs. In particular, only \gws from annihilating \bcs around isolated \bhs, discussed in \cref{subsec:annil}, can be searched for and constrained with current \gwh data.

This section is broken into four parts: in \cref{subsec:targets}, we will describe potential sources of \gws from annihilating \bcs that would be visible in \lvk. Next, in \cref{subsec:meth-bc}, we will discuss the \cwh methods employed to do searches for each of the targets. We will then show recent constraints from different searches for \gws from annihilating \bcs in \cref{subsec:bcsearchresults}, and conclude in \cref{subsec:bc-prospect} with prospects for future searches for \bcs with current and future data.

\subsection{Search targets}\label{subsec:targets}

In \cref{sec:gwbc}, we have explained the different mechanisms that could generate \bcs around rotating \bhs, and thus source \gws through different physical processes -- annihilation, transitions, resonances in binary systems, etc. From a practical point of view, we must also discuss where and how we should look for \bcs. Currently, only ground-based \gwh \ifos have been built; thus, we are essentially limited to looking for only one of the aforementioned \gwh emission mechanisms: annihilation, as described in \cref{subsec:annil}. Thus, the next logical question would be: what kind of systems would exhibit annihilation of \bcs, and where in the sky can they be found? 

In principle, there are four targets to consider

\begin{enumerate}
    \item Known galactic \bhs (e.g. Cygnus X-1 \cite{Fabian:1989ej})
    \item Known remnants of \bbh mergers (e.g. GW150914 \cite{Abbott:2016blz})
    \item Known \bbh inspirals and mergers \cite{LIGOScientific:2025hdt}
    \item Unknown galactic \bhs
\end{enumerate}

Each of these sources has its own benefits and drawbacks: known \bhs have many of their intrinsic parameters measured -- mass, spin, inclination angle, etc. --; however, they are \emph{old} and thus the strength of \gwh emission from annihilating \bcs would be significantly diminished with respect to that when they were first born. On the contrary, remnants of \bh mergers are extremely \emph{young}: their clouds build up quickly and they could begin emitting days or weeks after formation, as described in \cref{sec:gwbc}. Their age is thus precisely measured, as well as the final mass and spin (within $<10\%$ in the case of GW150914 \cite{Abbott:2016blz}). These estimates can be even more precise than those that arise from some x-ray binaries. However, remnants of \bbh mergers are \emph{far}: GW150914 merged at $0.410$ Mpc, and other detected systems have been seen out to Gpcs. Thus, the scaling of \gwh amplitude with distance implies that we could not detect remnants of mergers with annihilating scalar \bcs around them. We note, however, that rapidly annihilating vector \bcs could be detected out to Gpc \cite{Jones:2023fzz,Jones:2024fpg} in current-generation \gwh \ifos, making merger remnants a viable target even now.

On the other hand, with known \bbh inspirals, \bcs could alter the \gwh signal in a variety of ways, as detailed in \cref{subsec:bcbin}. Thus, using the spins and masses inferred from \lvk measurements, the allowed boson mass can be constrained: if certain \bh masses are detected with low spins, this could be evidence that \bcs have spun down \bhs. On the contrary, if \bhs with high spins are detected, this could rule out the presence of \bcs existing in a particular mass range \cite{Ng:2020ruv}. Such an analysis allows us to obtain some evidence for or rule out the presence of \bcs around rotating \bbhs, but does not necessarily permit us to claim confidently that we have detected a \bc. 

Finally, unknown galactic \bhs could take on a wide range of masses, ages and spins depending on when and how they formed. The number of expected \bhs in the galaxy is $\mathcal{O}(10^9)$, which comes from considering the age of the galaxy ($\sim 10^{11}$ years) and the supernova rate ($\sim 1$ per century) \cite{Madau:2014bja}. Thus, the \bhs we observe are such a small portion of the total number that exist in the local universe, and may not be representative of the full population. However, searches for such systems are entirely \emph{blind}: we do not know any \bh mass or spin parameters, nor do we know the location of such systems. While this may seem to be a daunting task, computationally efficient methods from the \cwh community have been leveraged to do these kinds of searches \cite{DAntonio:2018sff,Isi:2018pzk,Jones:2023fzz,Jones:2024fpg}.

We will now list the methods that have been developed to search for these sources, and then delve into the constraints that have been set using real \lvk data.

\subsection{Search methods}\label{subsec:meth-bc}


Different methods to search for long-lived \gws from isolated \bhs (or those present in a known x-ray binary system) have been developed and used in real searches in \lvk data. Detecting \gwh emission from an isolated \bh would provide smoking-gun evidence of the existence of an ultralight bosonic field around \bhs. Each has been designed to look for a quasi-monochromatic (scalar boson) or rapidly evolving long-transient (vector boson) signal arising from specific targets, such as Cygnus X-1, or originating from anywhere in the sky. While searches for known systems are significantly less expensive computationally than all-sky ones, they are limited to the few \bhs about which we have measurements of distance, spin, mass and age. Thus, a multi-pronged approach has been taken within the community to probe both known and unknown sources of \gws.

\subsubsection{SD Excess power method}\label{subsubsec:bc-excess-pow}

One approach to searching for \bcs around rotating \bhs is the excess-power method, which models the expected signal as approximately monochromatic over the entire observation time. Because the source could lie anywhere on the sky, the signal is modulated by the Earth's motion, leading to a Doppler-shifted frequency. To recover such a signal, the data are transformed into time–frequency maps and corrected for different sky positions, so that a truly monochromatic signal would appear as a narrow line \cite{DAntonio:2018sff}. This strategy is similar to that described in \cref{subsubsec:excesspower}, but here the modulation arises from the Doppler effect rather than from the random velocity distribution of individual \dmh particles. In practice, the analysis is repeated using different coherence times $\TFFT$, which control the width of the frequency bins. Shorter $\TFFT$ values are less sensitive overall but more robust to small frequency drifts, making the search less dependent on specific signal-model assumptions.


In \cref{fig:tfmaps}, we show the core stage of this method, which involves creating the time-frequency peakmap from raw time-domain strain data (discussed previously in \cref{subsubsec:excesspower}), correcting for the Doppler modulation from a particular sky location, and then integrating over time to collect all peaks at a given frequency to highlight the presence of a signal. Note that while we show the color to indicate the ratio of signal to noise power, the method does \emph{not} actually use it: it only requires that a given point in the time-frequency plane is above a given threshold. When moving from \cref{fig:afterdopp} to \cref{fig:pmproj}, all time-frequency points above this threshold are given a weight of one, and the ones are summed over time in order to create this histogram.

\begin{figure*}[htbp]
    \centering
    \subfigure[]{%
        \label{fig:beforedopp}
        \includegraphics[width=0.33\textwidth]{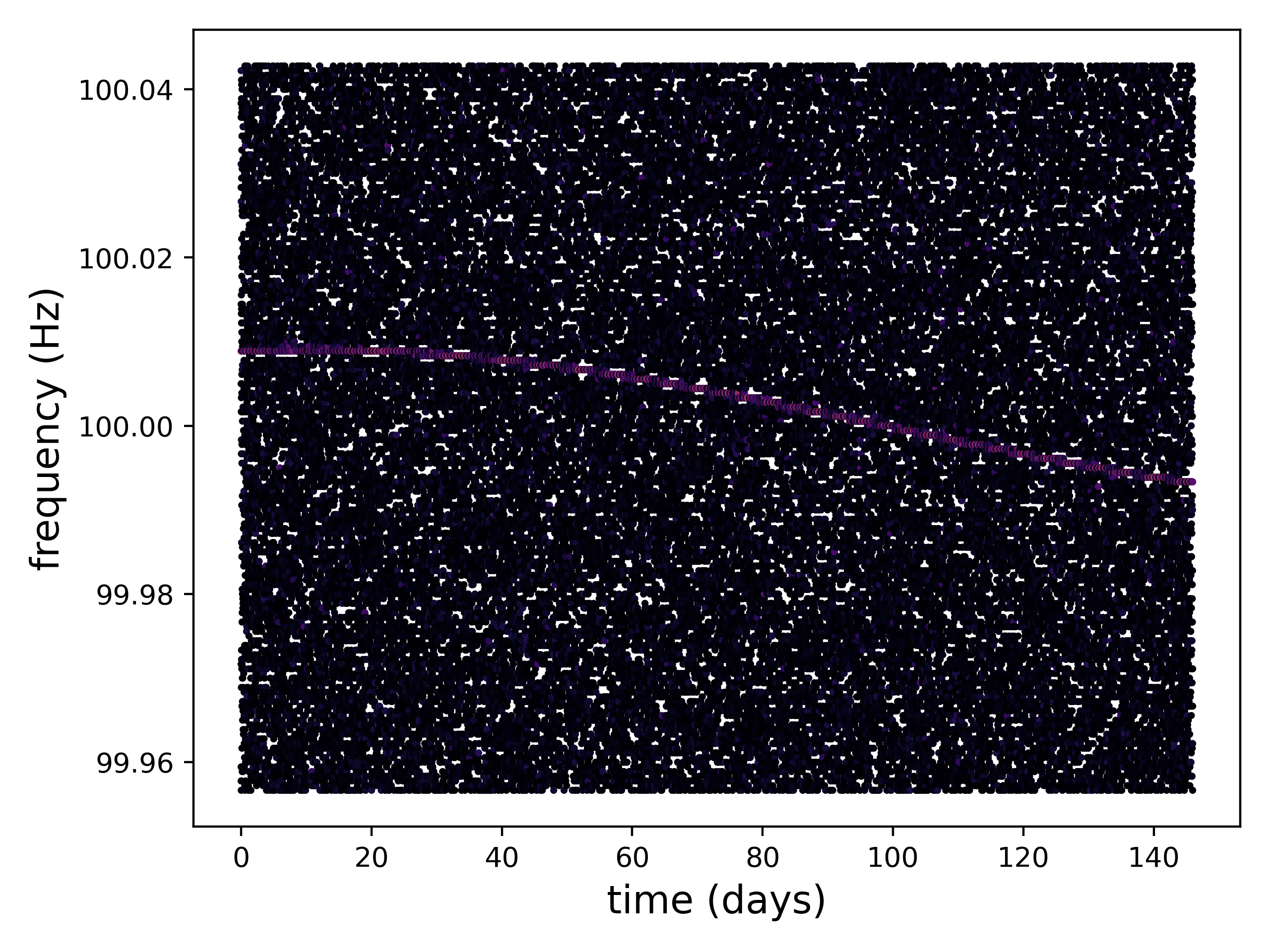}
    }%
    \subfigure[]{%
        \label{fig:afterdopp}
        \includegraphics[width=0.33\textwidth]{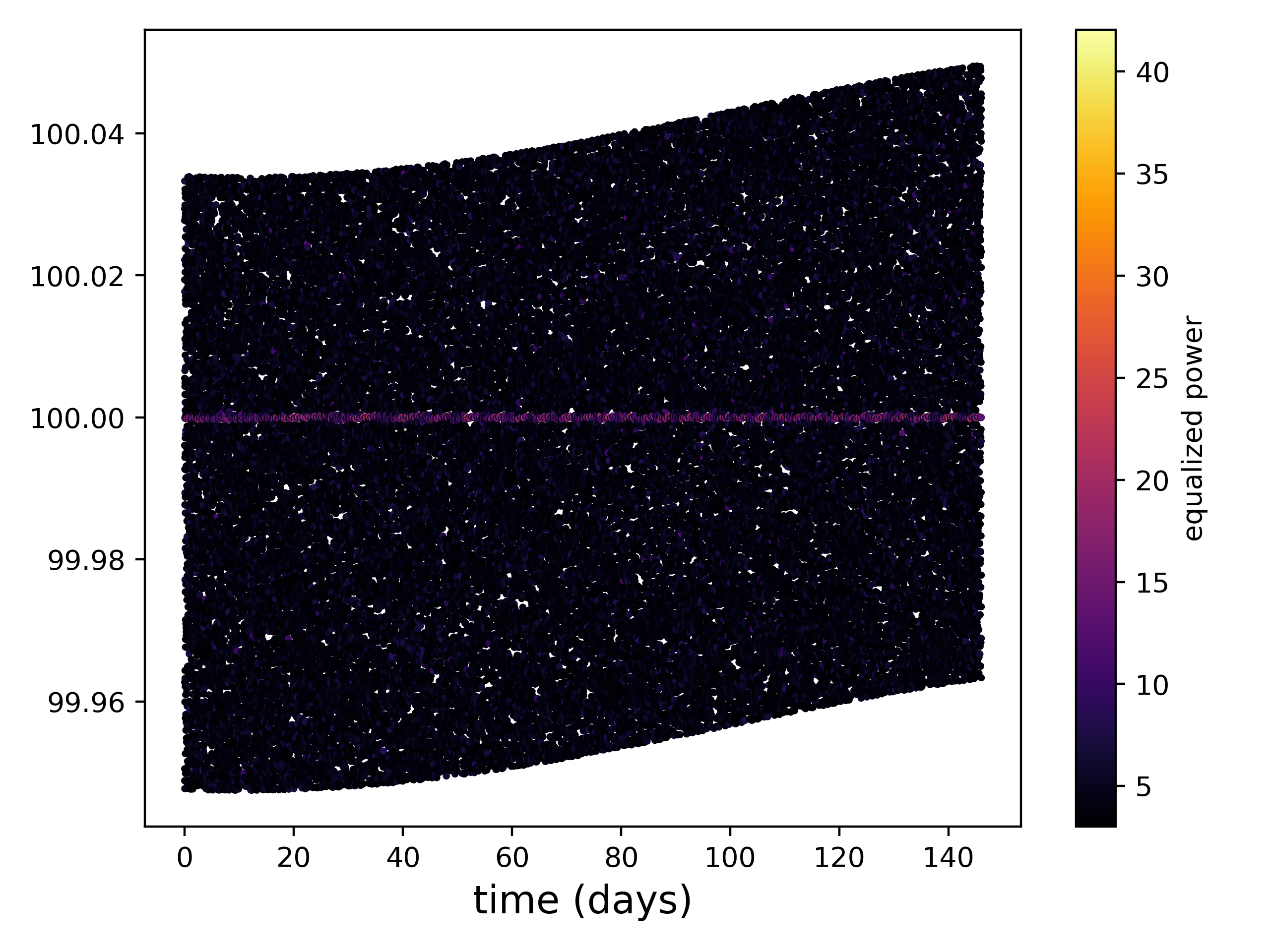}
    }%
    \subfigure[]{%
        \label{fig:pmproj}
        \includegraphics[width=0.33\textwidth]{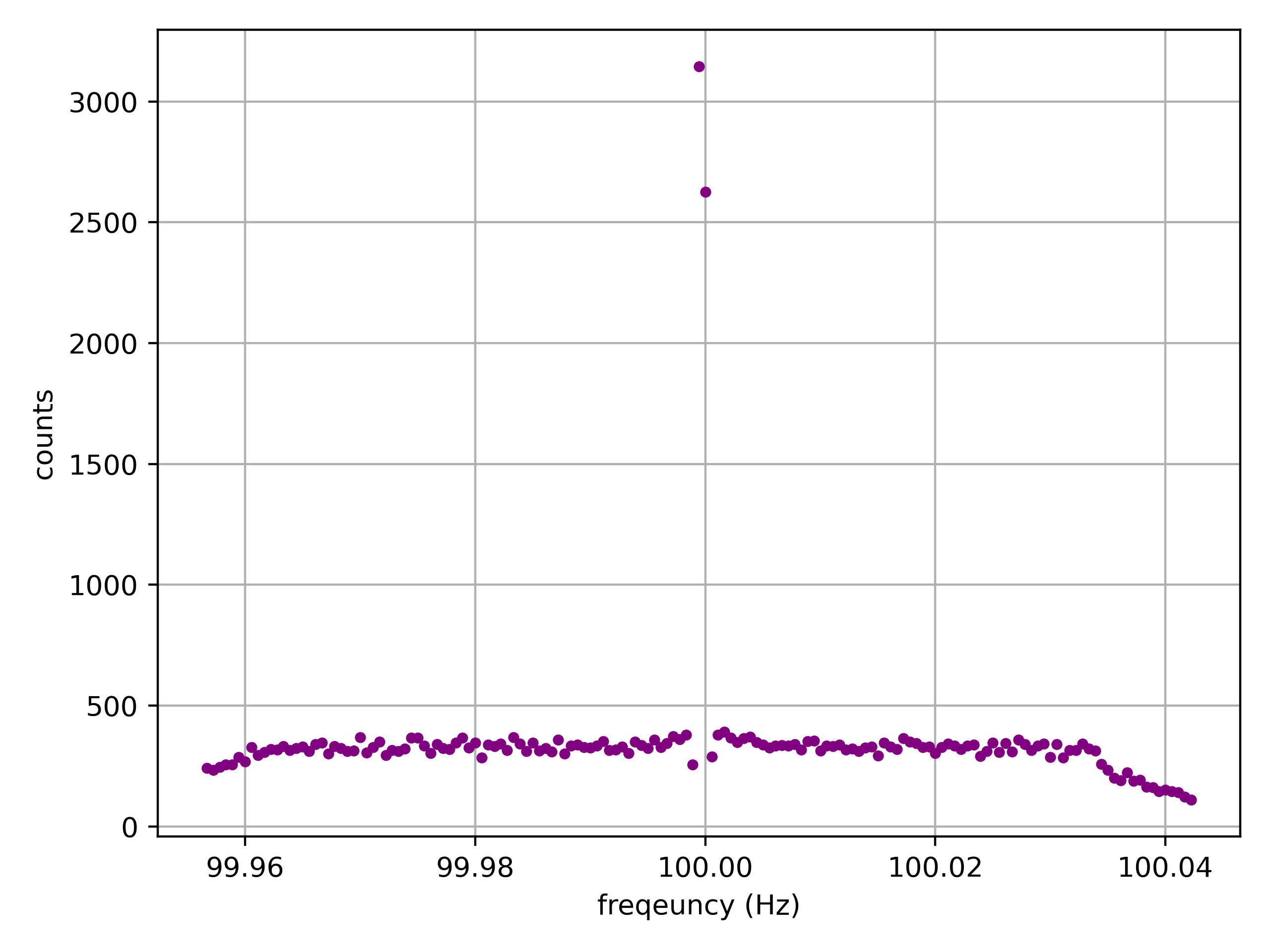}
    }
    \caption{This plot shows how the method described in \cref{subsubsec:bc-excess-pow} works for a simulated \gwh signal from an annihilating scalar \bc system. The physics behind this form of \dm has been discussed in \cref{subsec:annil}. (a) The Doppler-modulated time-frequency peakmap of the \gw signal coming from a particular position in the sky is created by Fourier transforming the \gwh strain data in chunks of length $\TFFT$, dividing the square of the FFT by the estimation of the \psd, and selecting local maxima above a threshold. (b) For a fixed sky position, the time-frequency peakmap is the Doppler corrected, which makes the signal monochromatic. (c) The projected time-frequency map is shown here, which corresponds to an integral over time in each frequency bin. Note that each time-frequency point is labeled simply as ``one'' if it is above a threshold, and is zero otherwise. This histogram thus is a count of the number of times at which the equalized power exceeded a given threshold.} 
    \label{fig:tfmaps}
\end{figure*}

\subsubsection{Viterbi}\label{subsubsec:vit}

Another method that is more robust against stochastic frequency variations relies on Hidden Markov Models (HMMs) to find tracks in the time-frequency plane. The main idea is to model a time-frequency \gwh signal probabilistically as a Markov chain of transitions between ``hidden'' (unobservable) frequency states and use a detection statistic to relate ``observed'' frequency states with the hidden ones \cite{quinn2001estimation}. In the framework for this method, the probability to jump to a new state at a given time only depends upon the previous hidden state. In total, there would be $N_Q^{N_T+1}$ possible paths through the hidden states, where $N_Q$ is the total number of hidden states and $N_T$ is the total number of times at which we have observed frequency states. The Viterbi algorithm provides a recursive, computationally efficient way of maximizing the probability that the hidden set of states is responsible for the observed sequence of states.

In \cref{fig:sample-path}, taken from \cite{Isi:2018pzk}, we show the optimal path found by the Viterbi method for injected \bc signals in Gaussian noise that has an associated random walk of the frequency, which is meant to simulate unknown theoretical features of the signal. We can see that in both cases, with less and more variability of the signal, the Viterbi algorithm can find the appropriate track.

Recently, the Viterbi algorithm has also been adapted to search for \gws from annihilating vector \bcs, which have timescales significantly shorter than in the scalar case \cite{Jones:2023fzz,Jones:2024fpg}, and for which numerical waveforms must be used to gauge sensitivity \cite{Siemonsen:2019ebd,Siemonsen:2022yyf}.

\begin{figure*}[htbp]
	\centering
	\subfigure[Less variability\label{fig:sample-path_less}]{\includegraphics[trim={0.5cm 0.5cm 1cm 2.5cm},clip,width=\columnwidth]{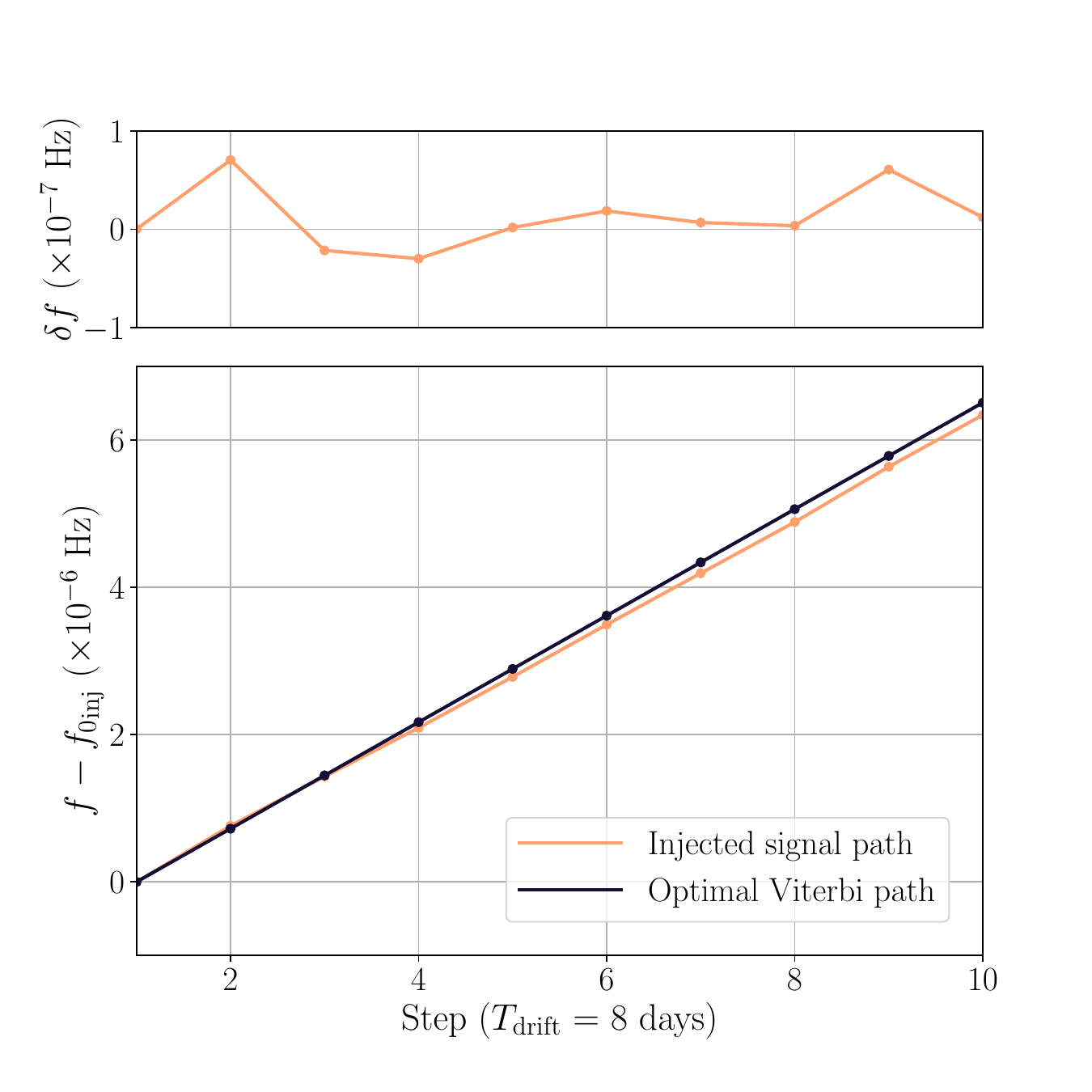}}
	\subfigure[More variability\label{fig:sample-path_more}]{\includegraphics[trim={0.5cm 0.5cm 1cm 2.5cm},clip,width=\columnwidth]{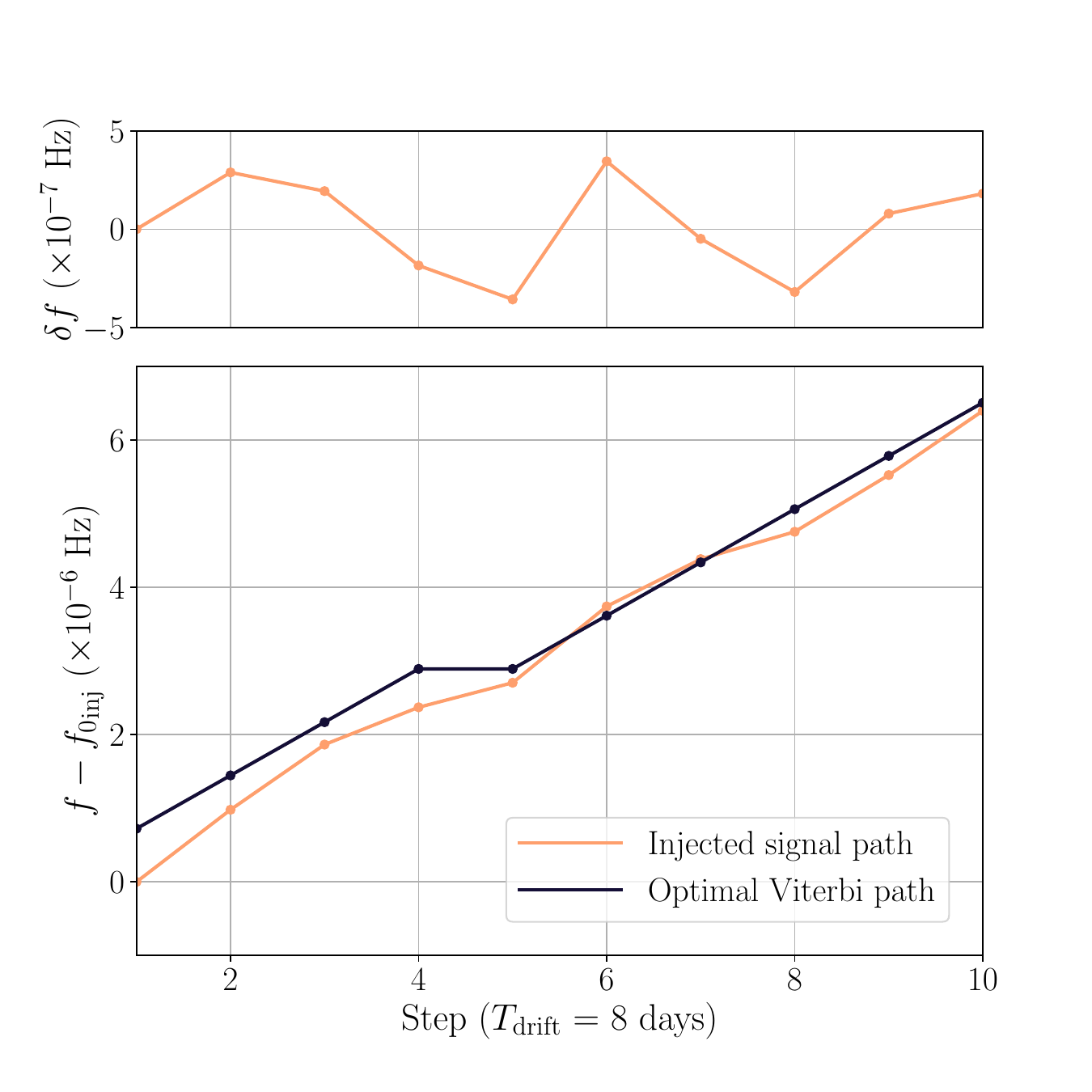}}
	\caption[]{Taken from \cite{Isi:2018pzk}. This figure illustrates how the Viterbi HMM algorithm described in \cref{subsubsec:vit} works to identify tracks in the time-frequency plane that would arise from \gwh emission from annihilating \bcs around rotating \bhs. The physics behind this form of \dm has been discussed in \cref{subsec:annil}. In particular, these plots show two examples of injected and recovered \bc signals with random walk frequency variations $|\delta f|\leq 0.1 \Delta f$ (weak variation) and (b) $|\delta f|\leq 0.5 \Delta f$ (strong variation), where $\Delta f = 1/T_{\rm drift}$ is the frequency bin size for the coherence time of the analysis $T_{\rm drift}=8$ days. The top panels show the injected random walk frequency evolutions, which cannot be seen by eye in the time-frequency tracks. The Viterbi optimal path results in a good match to the injected signal path. Each step in time on the $x$-axis is 8 days. The signal has a frequency of $201.2$ Hz, a spin-up of $10^{-12}$ Hz/s, a randomly chosen sky position, and a strain amplitude of $5\times 10^{-26}$.}
	\label{fig:sample-path}
\end{figure*}

\subsubsection{Cross-correlation and other methods}

We have already discussed in \cref{subsubsec:crosscorr} cross correlating detector data to search for \uldm that could interact directly with \gwh \ifos; however, it is important to note that such a method could also be used to search for \bcs around rotating \bhs, both deterministically and for a \sgwb composed of the superposition of cosmological or astrophysical \gwh signals. Furthermore, virtually \emph{any} method in the \cwh community \cite{Riles:2022wwz} could be tuned to look for such \bc systems because such methods already analyze almost monochromatic signals focusing on spin-downs but could also extend their parameter spaces to include spin-ups.

\subsection{Search results for scalar boson clouds}\label{subsec:bcsearchresults}

Data from the first, second and third observing runs of \lvk have been utilized to place constraints on the presence of \bcs around rotating \bhs within our galaxy. One search adopted a general approach, targeting systems across the entire sky. Another focused specifically on \gwh emission from a \bc around Cygnus X-1, and disfavored a range of boson masses. A third search investigated whether luminous dark photon clouds around stellar-mass \bhs could mimic known pulsars, setting constraints on the dark-photon mass and its kinetic mixing with the photon. A fourth analysis estimated the \sgwbs resulting from the combined effects of scalar, vector, and tensor \bcs, examining the data for each component individually. Lastly, a fifth study used the masses and spins of observed \bbh mergers to constrain the \bh spin distribution and the mass of the boson. Each of these searches is detailed in the subsections that follow.


\subsubsection{All-sky search constraints}\label{subsubsec:bc-all-sky}

The BSD excess power method described in \cref{subsubsec:bc-excess-pow} has been used to analyze LIGO O3 data to search for the presence of annihilating scalar \bcs around isolated \bhs within our galaxy \cite{LIGOScientific:2021jlr,KAGRA:2022osp}. This search was very computationally expensive, since each sky position had to be analyzed individually. 
No evidence for scalar \bc systems was found in either case, and thus upper limits were placed on the presence of \bcs with different ages, spins, and distances away from earth in two different ways \cite{Palomba:2019vxe,Dergachev:2019wqa,LIGOScientific:2021jlr}:

\begin{enumerate}
    \item Exclude the existence of \bh/\bc pairs certain distances away from us with particular spins and ages.
    \item Assume mass and spin distributions for \bhs in the universe, draw different \bhs from these distributions, and determine the maximum distance reach that could be attained for different \bc masses.
\end{enumerate}

The first way is agnostic towards whether such \bhs exist in the galaxy, while the second one contains some (uncertain) astrophysics regarding the mass and spin distributions of astrophysical \bhs, but allows for some physical intuition into the reach of such searches.

Constraints from the first way are shown in \cref{fig:exclude_chi09} for two different distances (1 kpc and 15 kpc), for \bhs with spin of 0.9 and for a few ages of the \bh/\bc system. The search on LIGO O3 data can thus exclude the existence of \bh/\bc systems that have these particular combinations of parameters at certain distances away from us.

\begin{figure*}[htbp]
    \centering
    \includegraphics[width=\columnwidth]{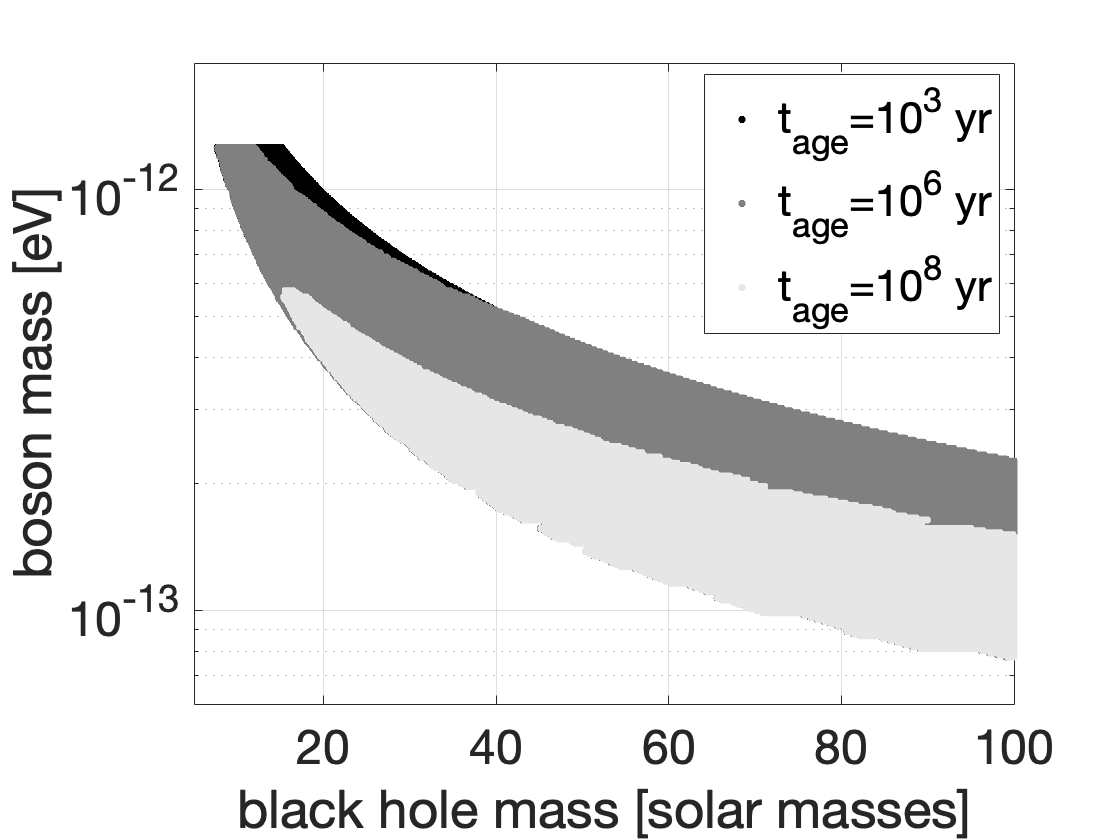}
    \includegraphics[width=\columnwidth]{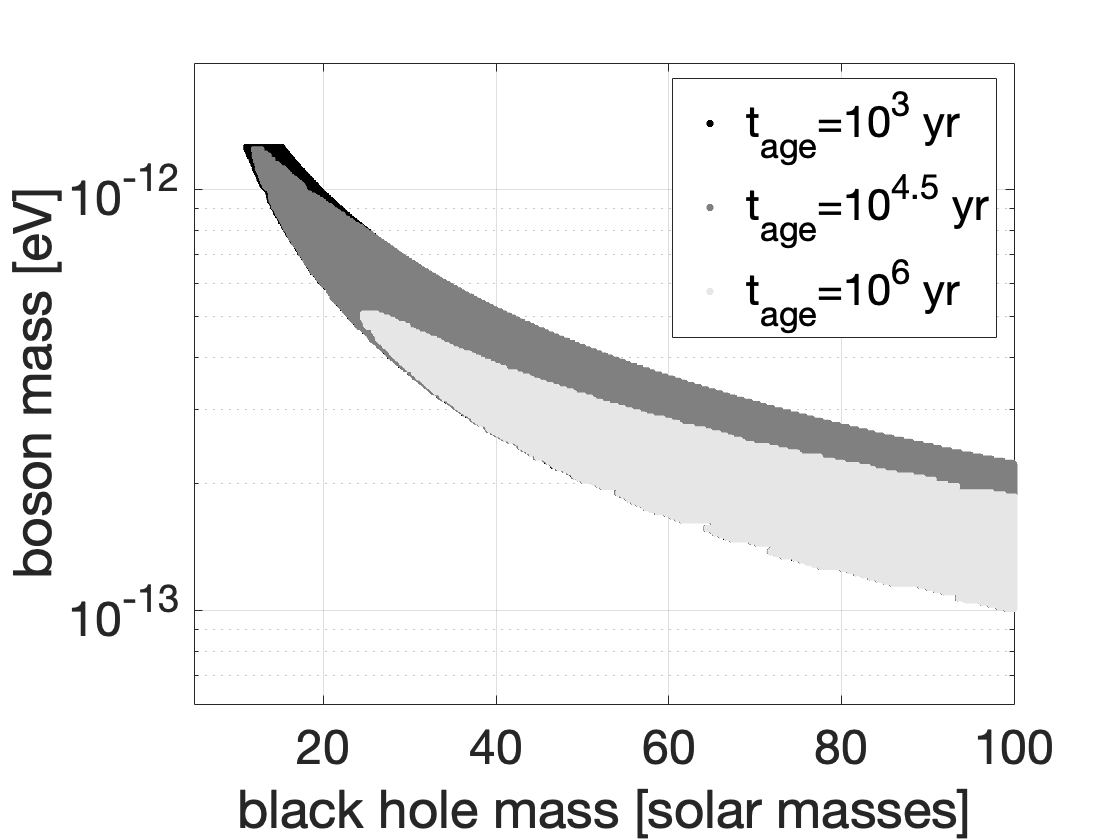}
    \caption{Taken from \cite{LIGOScientific:2021jlr}. Using upper limits from the O3 LIGO search for \gws from annihilating \bc systems, exclusion region of particular boson mass /\bh mass pairs are derived assuming different ages and distances (left: 1kpc; right: 15 kpc) for \bhs with initial spins of $\chi_i=0.9$. The physics behind this form of \dm has been discussed in \cref{subsec:annil}. This is the most model-agnostic way to probe \bcs around rotating \bhs: the limits do not assume distributions over the mass, spin, distance of \bhs in the Galaxy.
    }
    \label{fig:exclude_chi09}
\end{figure*}

Constraints from the second way are shown in \cref{fig:maxdist}, in which a Kroupa \bh mass distribution $f(m)\propto \mbh^{-2.3}$ \cite{Elbert:2017sbr} between $[5,~50]M_\odot$ and $[5,~100]M_\odot$ is assumed, as well as a uniform initial \bh spin distribution between [0.2,0.9]. For a fixed \bh age, the maximum distance reach as a function of boson mass indicates that at least 5\% of \gws from \bh/\bc systems would have been detected by the search, i.e. their strain amplitudes are larger than the minimum detectable amplitudes at a given confidence level (the upper limits) of the search (an example of this quantity is in \cref{subsubsec:excesspower}, \cref{h0min}. ).  As expected, on average when $50\msun$ is the maximum \bh mass considered, the distance that can be reached is also smaller, since the \gwh amplitudes scales strongly with the \bh mass.

\begin{figure*}[htbp]
    \centering
    \includegraphics[width=1\columnwidth]{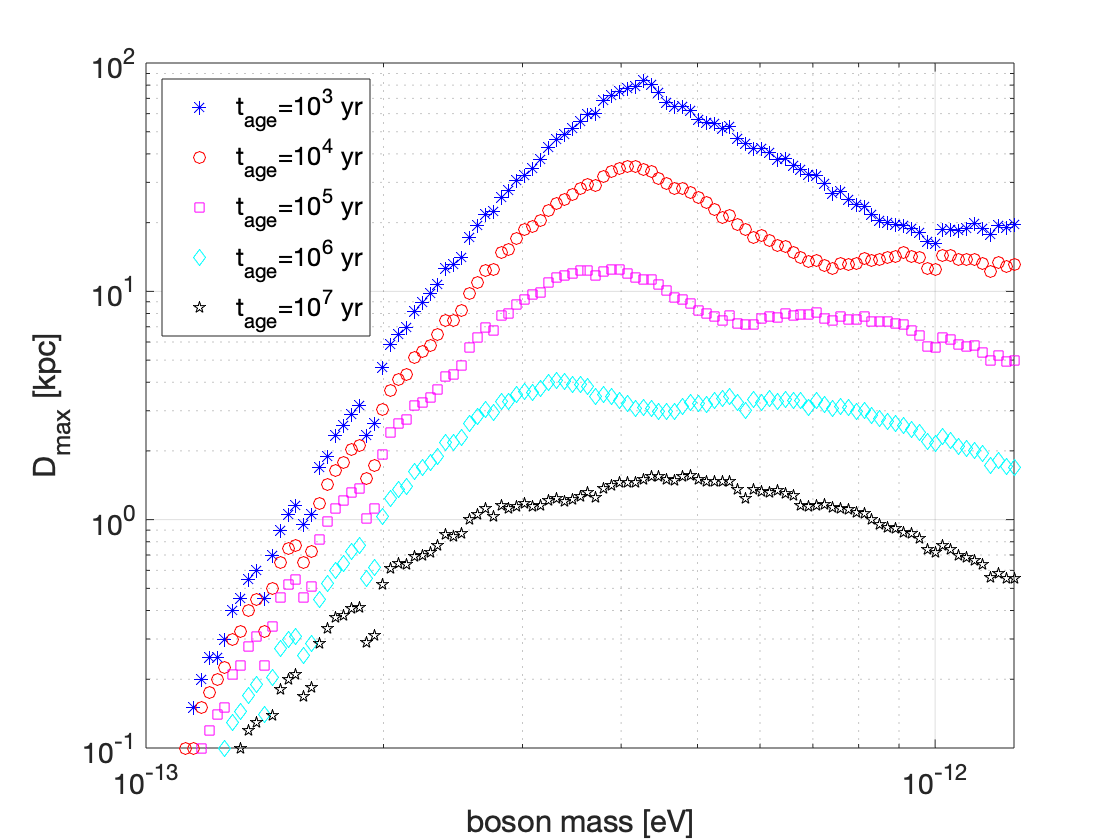}
    \includegraphics[width=1\columnwidth]{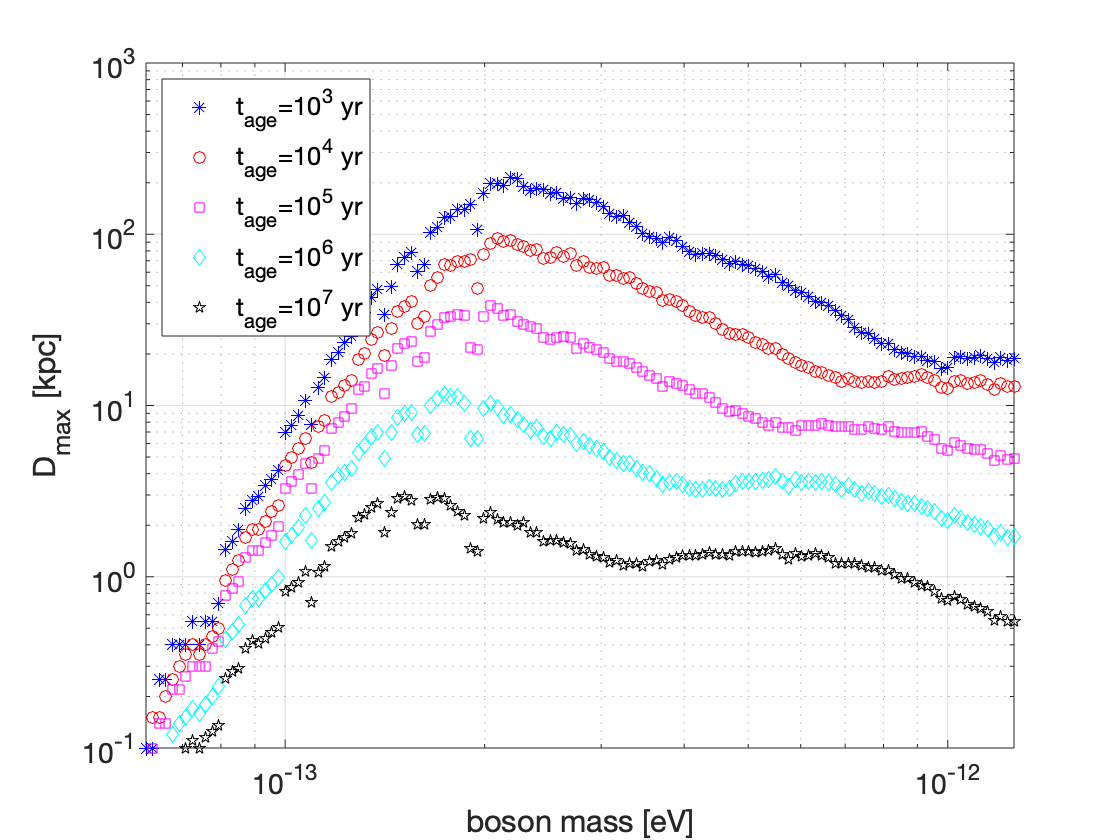}
    \caption{Taken from \cite{LIGOScientific:2021jlr}. Using upper limits from the O3 LIGO search for \gws from annihilating \bc systems, the maximum distance that the search in \cite{LIGOScientific:2021jlr} could reach at which at least 5\% of a simulated \bh population would have been detectable, that is, with a strain greater than the value of the upper limit, is shown here. The physics behind this form of \dm has been discussed in \cref{subsec:annil}. To create this plot, Ref. \cite{LIGOScientific:2021jlr} assumed a uniform spin distribution between [0.2,0.9], different ages, and a Kroupa mass distribution with maximum mass $50\msun$ (left) and $100\msun$ (right). These limits depend on the distributions assumed and are thus less agnostic than those presented in \cref{fig:exclude_chi09}.
    }
    \label{fig:maxdist}
\end{figure*}

We note that the search constraints present in, for example, \cite{Steltner:2023cfk,Ming:2024dug,Covas:2024nzs}, could also be cast in terms of all-sky or directed search constraints on scalar or vector \bcs, but with particular caveats that would have be worked out depending on the actual search parameter space. 

\subsubsection{Cygnus X-1 constraints}\label{subsubsec:cyg-x1}

Cygnus X-1 is an x-ray binary that is close by, has a large spin, $\sim0.9$, and has relatively well-measured orbital parameters, which make it a reasonable \bh to target. Since the sky position is known, searches based on the Viterbi algorithm discussed in \cref{subsubsec:vit} can use much longer coherence times than those used in \cite{LIGOScientific:2021jlr}, thus permitting exquisite sensitivity towards almost monochromatic signals, while also being robust against theoretical uncertainties in  the \gwh emission from Cygnus X-1, e.g. due to uncertainties on the measured semi-major axis, etc., due to the presence of an accretion disk, etc.

In \cref{fig:h0_estimate}, we show constraints on the strain amplitude arising from annihilating scalar \bc particles as a function of the boson mass using LIGO O2 data. The search results $h_0^{95\%}$ are given as the black line, while the two curves represent the numerically obtained strain from this system, accounting for the spin, age, mass and distance of Cygnus X-1 ($t_{\rm age} = 5 \times 10^6$\,yr and $1 \times 10^5$\,yr). The region in which the search results are below the theoretical curves denote boson masses that are disfavored as having formed around Cygnus X-1.

All constraints presented so far do not assume any boson self-interactions. If self interactions do occur, the available parameter space to probe would change. But, if the self-coupling is weak enough, standard methods discussed in \cref{subsec:meth-bc} can probe this scenario without modification. We show in \cref{fig:h0_axiverse} constraints on \gws arising from a scalar \bc around Cygnus X-1 that have self-interactions. Each panel corresponds to a particular physical mechanism: annihilation in the 211 energy level (left); annihilation in the 322 energy level (middle); and transitions between 322 and 211 energy levels (right).  The dashed curves divide the plot into different self-interacting regimes, from top to bottom: strong self-interactions, intermediate self-interactions, and the gravitational regime (negligible self-interactions). The energy level labeling correspond to the quantum numbers $n,l,m$, respectively, where $n$ is the
principal quantum number, $l$ is the total angular momentum quantum number, and $m$ is the azimuthal quantum number. The color indicates the expected signal amplitude, which, for different detectors (different colored lines), is mapped to projected constraints on the axion decay constant. The white line indicates the axion decay constant that is fixed by the axion mass. 

Constraints on self-interacting \dm using the search results on O2 data in \cite{Sun:2019mqb} are shown only in the left-hand plot, since that search was designed to analyze 211 annihilation from Cygnus X-1. Furthermore, current and future ground-based \ifos will be able to place constraints on the axion decay constant at frequencies of $\mathcal{O}(100)$ Hz (left plot), while space-based \ifos, such as DECIGO and MAGIS-Space are required to probe transition signals at frequencies below 1 Hz (right panel). The figure also highlights the possibility of multi-band \gwh astronomy: it would be possible to detect a transition from the 322 to 211 energy levels in space-based observatories while simultaneously observing the \cwh arising from annihilation of bosons in the 211 energy level in ground-based \ifos \cite{Collaviti:2024mvh}. Annihilation signals from the 322 level (middle plot) would be too weak to be detected by any of the considered current or future \gwh \ifos.


\begin{figure}[htbp]
    \centering
    \includegraphics[width=1\columnwidth]{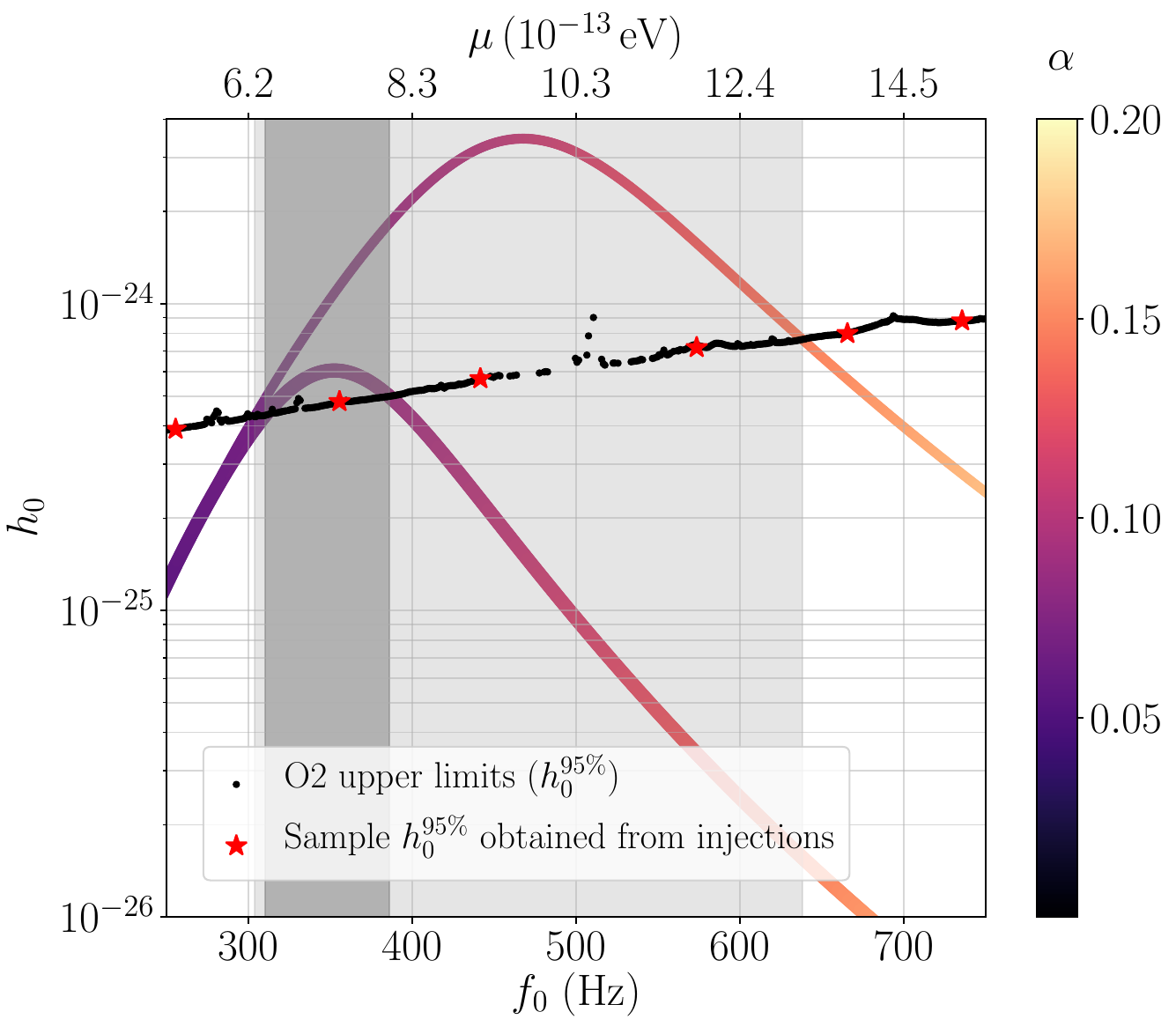}
    \caption{Taken from \cite{Sun:2019mqb}. Upper limits at 95\% confidence on the strain amplitude of annihilating scalar \bcs coming from the \bh in Cygnus X-1. The physics behind this form of \dm has been discussed in \cref{subsec:annil}. The black curve indicates the upper limits from the search on O2 data, with the red stars denoting upper limits derived in frequency bands with injections. The two different curves correspond to the signal amplitude with different choices for the (unknown) age of the system (lower curve: $10^6$ years; upper curve: $5\times 10^6$ years. The \gwh fine structure constant is colored, and the gray regions denote boson masses disfavored under the two different age assumptions. The parameters of Cygnux X-1 used to compute the amplitude curves are: $M = 14.8M_\odot$, $\chi_i=0.99$, and $d = 1.86$\,kpc.
    }
    \label{fig:h0_estimate}
\end{figure}

\begin{figure}[htbp]
    \centering
    \includegraphics[width=1\columnwidth]{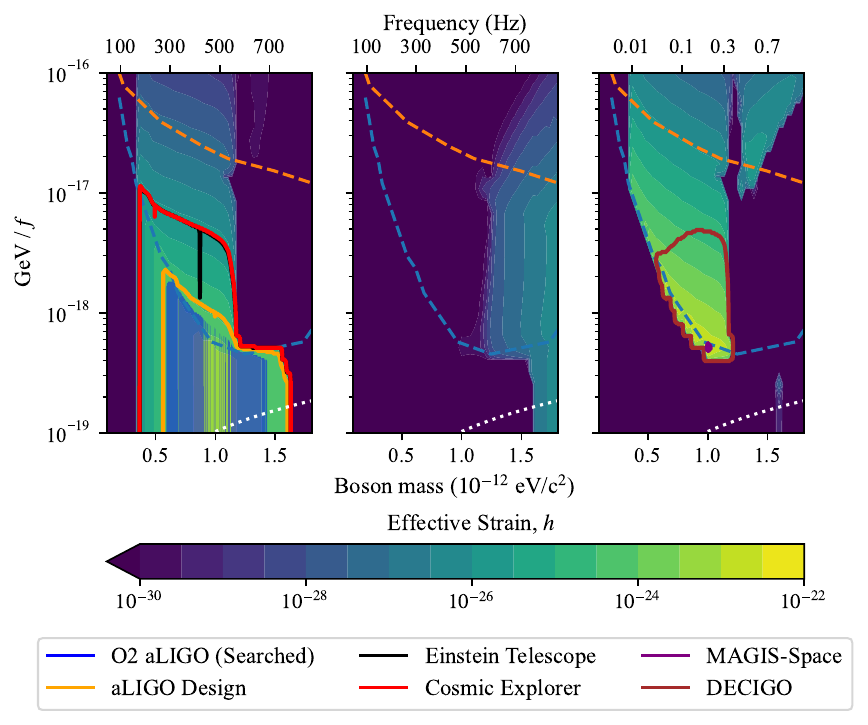}
    \caption{Taken from \cite{Collaviti:2024mvh}. 
    Parameter space reachable by current and next-generation \gwh \ifos for (left) 211 annihilation, (middle) 322 annihilation, and (right) 322 $\rightarrow$ 211 transition signals from a scalar \bc around Cygnus X-1, assuming an age of $10^5$ yr and the parameters for Cygnus X-1 given in \cref{fig:h0_estimate}. The numbers 211, 322, and the 322 $\rightarrow$ 211 transition refer to specific energy levels of the cloud. The physics behind this form of \dm has been discussed in \cref{subsec:annil,subsec:transition}. The dashed curves divide the parameter space into three regimes: (1) top, above orange-dashed line: the strong self-interaction regime, (2) middle, between dashed-orange and dashed-blue lines: the moderate self-interaction regime, and (3) bottom, below the dashed-blue line: the gravitational (negligible self-interactions) regime. The white dotted line shows the expected relation between the axion and its decay constant. Darker colors here correspond to weaker \gwh signals primarily due to smaller occupation numbers in the energy states of the \bcs. Constraints on 211 annihilation come from ground-based \ifos, while constraints from transitions will come from space-based \ifos. The 322 annihilation signal is too weak to be detected by any current our future observatory. The O2 constraints come from a real analysis of \gwh data and are thus representative of both the actual \psd of that dataset and an exclusion region on the plot.  }
    \label{fig:h0_axiverse}
\end{figure}

\subsubsection{Constraints on luminous dark photon clouds}\label{subsubsec:lum-dp-bc}

As alluded to in \cref{sec:gwbc}, if dark photon \dm kinetically mixes with the ordinary photon and forms clouds around rotating \bhs, it would produce both electromagnetic and \gwh signals, which, if detected, could enable a major multi-messenger discovery \cite{Siemonsen:2022ivj}. This additional coupling implies that soon after exponential growth of the \bc begins, electromagnetic fields act on charged particles in the vicinity of the \bh, producing electron/positron pairs essentially out of vacuum, i.e. through the photon-assisted Schwinger mechanism \cite{Sauter:1931zz,Heisenberg:1936nmg,Schwinger:1951nm}. At this point, a pair-production cascade ensues \cite{Dunne:2009gi,Monin:2010qj}, sourced by the electromagnetic fields, creating a plasma around the \bh in a state of turbulent quasi-equilibrium. This plasma emits copious amounts of electromagnetic radiation, sourced by energy injections through dissipative processes in the cloud (e.g. at magnetic reconnection sites, Landau damping, turbulence). The luminosity of these sources depends on the strength of the kinetic mixing parameter, but could be at the level of supernovae or known pulsars, and and also periodic \cite{Siemonsen:2022ivj}. If the dark photon has a mass around $\sim 10^{-12}$ eV, it will induce the formation of clouds around stellar-mass \bhs, and give off radio emission that would look remarkably similar to that from known pulsars in the ATNF catalog \cite{Manchester:2004bp}. 

The absence of observed \cws from pulsars, combined with the possibility that dark-photon clouds could act as ``pulsar mimickers,'' motivates searches for such clouds around sources in the ATNF catalog that might be misidentified as pulsars. Ref.~\cite{Mirasola:2025car} explores this scenario by analyzing a subset of spinning-up ``pulsars'' in the catalog that could instead be stellar-mass \bhs surrounded by luminous dark-photon clouds. Three analysis strategies are considered: one assuming perfect phase coherence between the electromagnetic emission and any associated \gwh signal; one allowing for small frequency deviations around the measured pulsar frequency; and one that is most robust to theoretical uncertainties but least sensitive to \gwh emission. From these methods, the authors derive upper limits on the dark-photon--photon kinetic mixing parameter as a function of boson mass for the 34 sources studied, shown in \cref{fig:dark_photon}.  

To produce these bounds, assumptions are made about both the \bh population and their electromagnetic signatures, so the precise results are not fully robust against variations in these choices. The adopted \bh distributions include a stellar-tracing spatial distribution, a Salpeter mass function, uniform spin and age distributions, and a total galactic population of $10^8$ \bhs. Even under these assumptions, the analysis requires only $\sim 10$ detectable ``events'' (i.e., \bhs with \gwh amplitudes exceeding the strain upper limit), which is conservative compared to the $\mathcal{O}(10^3)$ events expected at certain coupling--mass values (see Fig.~3 of \cite{Mirasola:2025car}). Ultimately, the study excludes dark-photon masses in the range $\sim [10^{-13},10^{-12}]$ eV with kinetic mixing larger than roughly $[10^{-9},10^{-7}]$.

\begin{figure}[htbp]
    \centering 
    \includegraphics[width=\columnwidth]{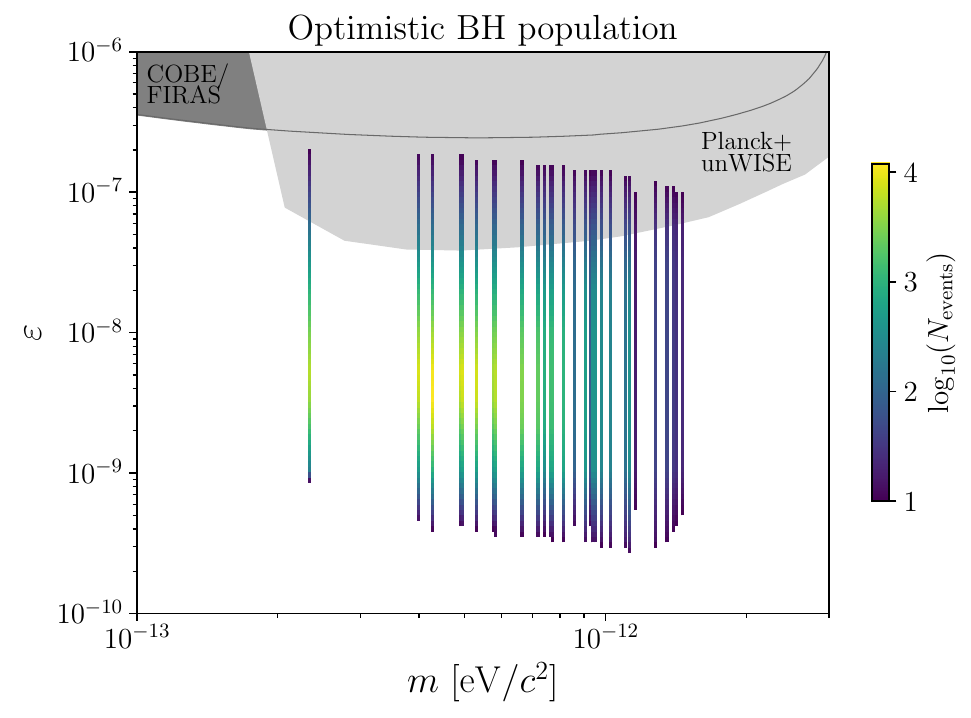}
    \caption{Taken from \cite{Mirasola:2025car}. For each of the radio pulsar mimickers targeted in \cite{Mirasola:2025car}, the upper limit at 95\% confidence on the kinetic mixing parameter $\epsilon$ is plotted as a function of the boson mass for which at least 10 systems would have been detected with signal amplitudes above the strain upper limits. Each vertical line corresponds to one radio pulsar, whose frequency fixes the mass of the dark photon. The color corresponds to the number of events that should have been detected, assuming a ``optimistic'' population model for \bhs, i.e. using a larger maximum mass and spin for the assumed \bh population with respect to the ``pessimistic'' population also constrained.  The bounds reproduced from other experiments can be found in \cite{AxionLimits}, specifically from the conversion of CMB photons into dark photons, which would cause distortions in the black body spectrum measured by COBE/FIRAS~\cite{Mirizzi:2009iz, Caputo:2020bdy} (dark gray), and from measurements by Planck and unWISE galaxies of cross correlations between temperature anisotropies in the CMB \cite{McCarthy:2024ozh} (light gray). The upper limits arising from this search surpass those from existing experiments, and indicate that between $10$ and $10^4$ events could have been detected at couplings of between $\sim [10^{-10},10^{-7}]$.
    }
    \label{fig:dark_photon}
\end{figure}

\subsubsection{Stochastic gravitational-wave background constraints}\label{subsub:sgwb}

The superposition of \gws arising from annihilating scalar \bcs could form a \sgwb detectable by current-generation \gwh \ifos. In \cite{Tsukada:2018mbp}, the authors designed two models of the source of this \sgwb: (1) from remnants of core collapse supernovae, and (2) from remnants of mergers of \bhs or \nss. The authors also performed search on LIGO O1 data using cross-correlation to look for evidence of a \sgwb from scalar \bcs \cite{Tsukada:2018mbp}. No signal was found, and no range of boson masses could be robustly excluded unfortunately, except under optimistic assumptions about \bh formation rates and spin distributions, which would disfavor the regime $2\times 10^{-13}\leq m_b\leq3.8\times 10^{-13}$ eV at 95\% confidence. Here, ``optimistic' means assuming rapidly spinning and frequently forming \bhs in the galaxy. This limitation highlighted the degeneracy between boson mass and poorly constrained astrophysical parameters, particularly \bh spins.

The search was repeated in the second observing run as well, finding no evidence for any \sgwb of \emph{vector} \bcs around rotating \bhs \cite{Tsukada:2020lgt}, in particular disfavoring vector \bc masses between $[0.8,6]\times 10^{-13}$ eV at 95\% confidence. Ref. \cite{Yuan:2022bem} has also disfavored bosons with masses between $[1.5,17]\times 10^{-13}$ eV depending upon what is assumed for the uniform spin distribution, with wider constraints present when higher-spinning \bhs dominate the priors.

Beyond scalar and vector cases, Ref.~\cite{Guo:2023gfc} investigated tensor bosons, computing the corresponding \sgwb and performing a dedicated search. Their analysis disfavored boson masses between $10^{-13.4}$ and $10^{-11.7}$ eV, again with the precise range depending on the assumed range of uniform spin distribution .  

Taken together, these studies demonstrate that the \sgwb provides a powerful probe of ultralight scalar, vector, and tensor clouds, but the strength of the resulting constraints depends strongly on assumptions about black-hole formation, mass, and spin distributions. This mirrors the caveats discussed in \cref{subsubsec:bc-all-sky}, where similar assumptions underlie the derived limits. Moving forward, it is therefore essential to develop robust and consistent methods for setting constraints that minimize reliance on specific black-hole population parameters while retaining astrophysical interpretability.

\subsubsection{Constraints from the spins of detected \bbhs}

We note that constraints on \bcs can be inferred from the spin measurements of \bbh systems \cite{Arvanitaki:2014wva,Arvanitaki:2016qwi,Brito:2017zvb,Baryakhtar:2017ngi,Cardoso:2018tly,Ghosh:2021zuf}. Recently, a Bayesian method has been developed by combining the information from each system \cite{Ng:2019jsx}, assuming a distribution for the initial \bh spins and requiring that the \bhs have had enough time to undergo superradiance before merging. Such a study has been applied to the second \gwh transient catalog (GWTC-2) \cite{LIGOScientific:2020ibl}, and has resulted in essentially a joint posterior on the scalar boson mass and the spin distribution of \bhs at their formation \cite{Ng:2020ruv}. The latter is required because \bhs born with high spins that have been spun down by superradiance is partially degenerate with \bhs simply born with low spins in a universe without superradiance. The study presented in \cite{Ng:2020ruv} tentatively rules out scalar bosons in the mass range $[1.3,2.7]\times 10^{-13}$ eV.

These constraints from measured \bh spins can disfavor particular masses or even confirm the presence of an ultralight boson; however, they rely on assumptions about the spin distribution at formation, on whether the \bcs have time to form (which could be impacted by eccentricity at formation \cite{Peters:1964zz,Wen:2005xn}), and are a function of the (strongest) \bh mergers that we see with \gws.

Thus, these kinds of analyses complement those previously that look for \cws around isolated, rotating \bhs. Both have their advantages: \cwh analyses described above allow us to probe unknown sources and known, nearby sources whose signals we expect to be stronger than those arising from remnants of mergers. Likewise, the spins and masses of \bhs formed in binary system are well measured, and the constraints obtained in \cite{Ng:2020ruv} will only improve as we detect more and more \bhs. 
 
 \subsection{Prospects for \gwh probes of \bcs}\label{subsec:bc-prospect}

At the moment, despite the vast theoretical playground of \uldm \bcs, only a few searches have actually been performed for \gws due to boson-boson annihilations. This is primarily due to the fact that the rich physics present in \bc \bbh systems, and from transitions in isolated ones, can only be probed in space-based \gwh detectors, which are expected to be operational in the 2030s. Moreover, scalar \bcs are by far easier to work with theoretically than vector or tensors, which has restricted the search parameter space primarily to what can be probed by \cwh searches. However, as the theoretical calculations of vector \bc waveforms improve \cite{Siemonsen:2019ebd,Siemonsen:2022yyf}, new methods are being developed to search for annihilating vector \bcs as well \cite{Jones:2023fzz,Jones:2024fpg}. Additionally, there are also significant astrophysical uncertainties, namely: what are the distributions of \bh spins, masses, distances and ages in the galaxy and beyond. Such known unknowns limit potential constraints to be quite model agnostic (not so astrophysically informative) or too tied to uncertain \bh population models. These conundrums are not yet resolved, and probably will not be until we start detecting significantly more \bhs than we currently do.

The chosen targets of \gwh searches should also be evaluated. While having accurate measurements of the mass, spin and age of remnants of \bbh mergers would allow us to ignore astrophysical population models \cite{Arvanitaki:2014faa}, these systems are currently too far to be detected if the clouds are composed of scalar bosons \cite{Isi:2018pzk}, though could be seen up to Gpc now if they are composed of vector bosons \cite{Jones:2023fzz}. On the flipside, \gws from vector \bcs suffer from even more theoretical uncertainties than scalars \cite{Siemonsen:2019ebd}, and so the robustness of search results by targeting remnants would have to be evaluated. 

All-sky searches for such \bc systems could also be enhanced, especially in light of boson self-interactions that could alter the \gwh signal significantly as discussed in \cref{subsec:self-int}. Potentially, new methods to track quickly-evolving waveforms, such as those in \cite{Jones:2023fzz,Jones:2024fpg}, could be useful in a variety of contexts. Furthermore, the all-sky search specifically for scalar \bc systems did not consider the possibility that the cloud could spin-up, which, in practice, fixed a maximum spin-up to which the search was sensitive (such that the spin-up did not induce a freqeuncy shift by more than one bin $\delta f$ in $\Tobs$, i.e. $\dot{f}\Tobs < \delta f \equiv 1/\TFFT$). Such modifications to searches, coupled with additional theoretical understanding of the \gwh signal itself, could increase chances of detecting a signal by exploring a wider parameter space and allowing the cloud to contract over time.

It is also difficult to say with certainty that a particular boson mass is ``ruled out'', since the constraints are either population agonistic or depend on some assumptions about \bh formation, spin or mass distributions. Ref. \cite{Hoof:2024quk} attempts to unify the way in which constraints are set in a Bayesian way by handling the subtleties associated with population assumptions. The authors in Ref. \cite{Hoof:2024quk} encourage that posteriors on mass and spin, obtained through analyses of X-ray data from \bhs, be made publicly available in order to incorporate into their future work. 

\section{\gwh probes of soliton dark matter}\label{sec:soliton}

Ultralight \dm in the mass range $\sim [10^{-24},10^{-22}]$ eV has a Compton wavelength comparable to the size of galactic halos \cite{Hui:2016ltb,Sin:1992bg}. In this regime, \dm can form self-gravitating structures, or \emph{solitons}, at the centers of galaxies, which remain stable due to quantum degeneracy pressure \cite{Schive:2014dra}.

A natural approach to probing this mass range of \dm is through \ptas, whose frequency sensitivity aligns with the nanohertz oscillation frequencies expected from \dm. In this case, solitons can exert \df on binary systems, accelerating their inspiral and suppressing the gravitational wave strain amplitude at nanohertz frequencies \cite{Aghaie:2023lan}.

Alternatively, soliton \dm can be framed as a low-frequency modulation of relatively higher-frequency \gw. The \gwh frequency is modified during its propagation through the \dm halo, undergoing a gravitational redshift or 'heterodyning', analogous to the Sachs-Wolfe effect that imprints low-frequency modulations on high-frequency photons in the cosmic microwave background \cite{Sachs:1967er}. Two \cwh sources have been considered as high-frequency carriers for this modulation \cite{Blas:2024duy}: (1) inspiraling white-dwarf binaries, observable by space-based GW interferometers, and (2) non-axisymmetric rotating neutron stars that spin down over time.

In what follows, we explore how these different probes can be used to constrain the properties of soliton. In \cref{subsec:soliton-pta}, we demonstrate how data from \ptas can be used to set constraints on soliton \dm. We then explore two other approaches in \cref{subsec:soliton-phase-bin} and \cref{subsec:soliton-dm-cw}: detecting \cws from white-dwarf binaries and isolated neutron stars, which could reveal nanohertz modulations induced by soliton \dm.

\subsection{Constraints from pulsar timing arrays} \label{subsec:soliton-pta}

The evidence for a \gwh background from NANOGrav results can be used to constrain soliton \dm \cite{Aghaie:2023lan}. Essentially, \uldm would induce dynamical friction in the inspiral of supermassive \bhs, speeding up the orbit and suppressing the \gwh strain at nanohertz frequencies. 
Using the most accurate limit at 3.92 nHz on the \gwh strain from supermassive \bhs in the right-hand panel of Fig. 1 of \cite{NANOGrav:2023hvm}, Ref. \cite{Aghaie:2023lan} derives constraints on the soliton \dm mass-to-\bh mass ratio parameter space, as shown in \cref{fig:thebound}. These constraints are obtained using Eq. 17 from \cite{Aghaie:2023lan}, which models energy loss due to \gwh emission in the presence of \uldm. This equation incorporates astrophysical inputs such as galaxy-stellar mass functions, galaxy merger rates, and galaxy pair fractions \cite{Chen:2018znx}. The constraints are obtained by integrating Eq. 17 over these distributions, taking into account variations in the assumptions about the \uldmh mass and the population of supermassive \bhs. However, these constraints are sensitive to assumptions regarding astrophysical parameters. For example, the specific values chosen for galaxy merger rates, stellar mass functions, and the fraction of galaxies with supermassive \bhs affect the final upper limits. 

\begin{figure}[htbp] 
\begin{center} 
\includegraphics[width=0.95\columnwidth]{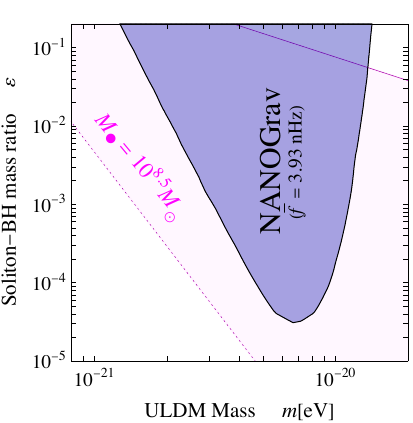}
\hfill 
\caption{Taken from \cite{Aghaie:2023lan}. Upper limits on the \dmh mass and soliton-to-supermassive black hole mass ratio parameter space, derived from NANOGrav data. These constraints are based on the notion that the \gwh background from the inspiral of supermassive \bhs, for which NANOGrav has found evidence, would be modulated by the \df exerted by soliton dark matter. Specifically, the presence of soliton \dm alters the supermassive BH inspiral dynamics, resulting in a suppression of the \gwh strain at nanohertz frequencies. The magenta lines represent two distinct sets of scaling relations derived from numerical simulations: solid lines correspond to \dmh-only simulations, where the soliton \dm is the sole interacting component, while dashed lines incorporate the influence of supermassive \bhs, suggesting a modification of the scaling relations due to the presence of these massive objects.}
\label{fig:thebound} 
\end{center} 
\end{figure}

\subsection{Dynamical friction on white dwarf binaries}\label{subsec:soliton-phase-bin}

Space-based \gwh detectors will observe the quasi-infinite inspiral of galactic white dwarf binaries and extreme-mass ratio inspirals of ordinary compact objects with much heavier \bhs \cite{Bartolo:2016ami,Babak:2017tow}. The long-durations of such signals permit the possibility of detecting a low-frequency modulation when passing through the \uldm soliton \cite{Wang:2023phr,Brax:2024yqh,Blas:2024duy}. Ref. \cite{Brax:2024yqh} has found that the impact of \df on the phase evolution of \gwh signals from white dwarf binaries exceeds that of the low-frequency \dm-induced modulation for masses above $10^{-21}$ eV, essentially providing two orders of magnitude of mass that could be probed in the future. Unfortunately, even at these masses, the density of \dm would have to be at least $10^4$ times higher than in the solar system in order to observe any changes to the phase evolution \gwh. This is broadly consistent with previous work by \cite{Wang:2023phr}, which assumed an enhancement of the \dmh density by eight orders of magnitude relative to that on earth, and showed that \dm with a mass of $\sim 10^{-23}$ eV could be detected with LISA. Furthermore, Ref. \cite{Blas:2024duy} assumes a representative population of white dwarf binaries, a fraction of which are detectable in LISA, and concludes that \dmh masses above $\sim 10^{-23}$ eV will remain unconstrained by LISA. However, the mass regime $\sim [2\times10^{-22},3\times 10^{-21}]$ eV could be constrained if \dm has stronger, quadratic couplings to ordinary matter. Again, the possibility that \dm couples to the \sm in some way is able to enhance the signal strength, as found in \cref{sec:pdm,sec:macdm} for different types of \dm.

\subsection{Nanohertz modulations of continuous gravitational waves from \nss}\label{subsec:soliton-dm-cw}

In addition to modulating the millihertz frequency of \gws from binaries in space-based \gwh detectors, soliton \uldm could induce modulations in the rotational frequency of \nss in a way distinct from ordinary spinning down, deformed \nss, thus affecting \cwh signals and allowing the possibility of detecting this effect in future ground-based detectors \cite{Blas:2024duy}. These \gws are sourced from the rotational power of the \nss, thus spinning down the \nss as \gws are emitted \cite{Riles:2017evm}. In \et and \ce, we expect to detect $\mathcal{O}(100)$ or more canonical \nss spinning down due to the emission of \cws \cite{Pagliaro:2023bvi}, which could be modulated by \uldm solitons. 

Using \cws from deformed \nss, the soliton mass regime $\sim [2\times10^{-22},3\times 10^{-21}]$ eV becomes accessible in \et for both linear and quadratic couplings, and improves with the number of \nss that are detected. The constraints will even outperform existing \pta ones, since the \snr scales linearly with the carrier \gwh frequency. This is shown in \cref{fig:lambda12res} for LISA, \et and \ce, and demonstrates the powerful probe that \cws from \nss will be of this particular \dmh model.

\begin{figure*}[htbp]
    \centering
    \subfigure[]{
    \includegraphics[width = \columnwidth]{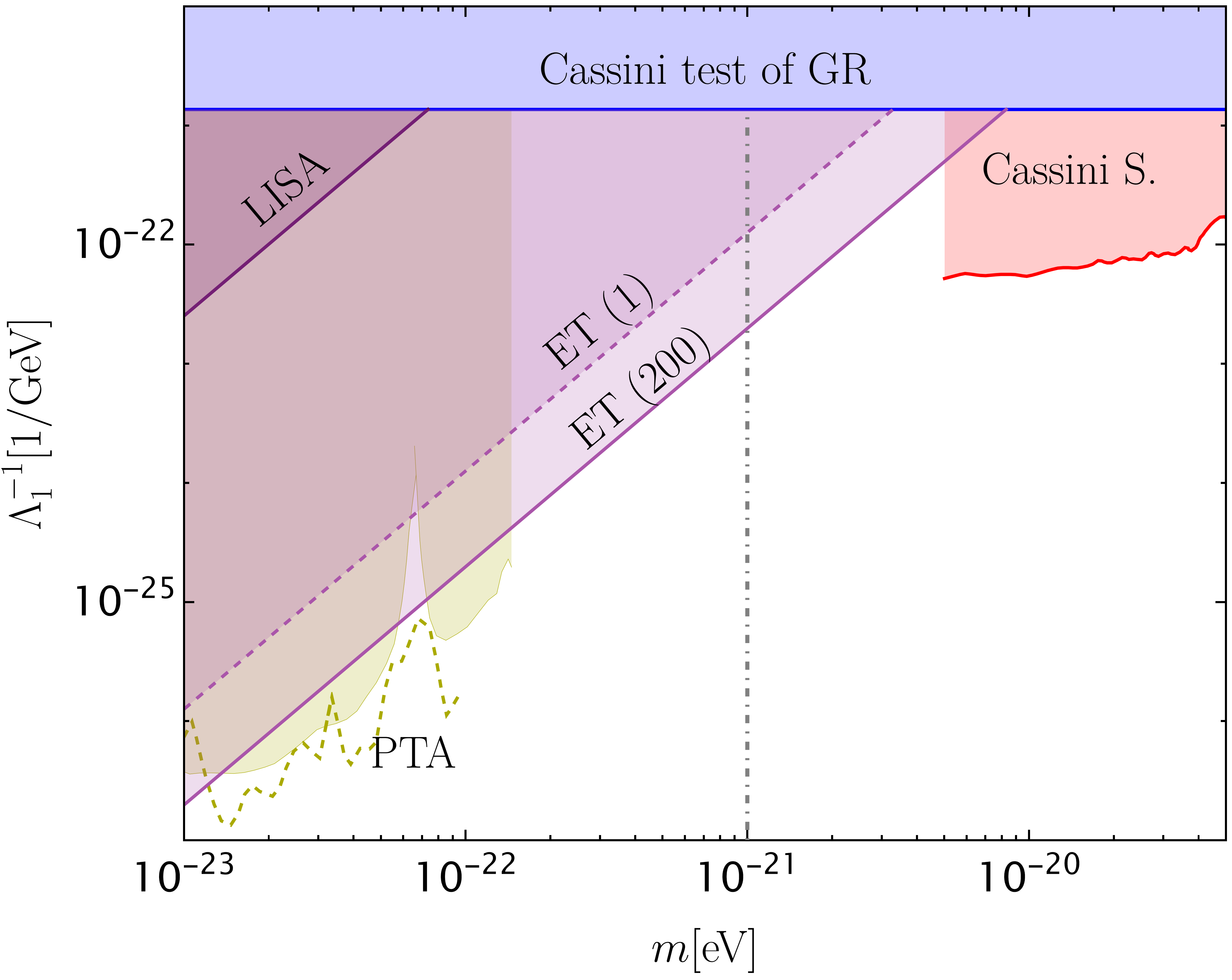}
    \label{fig:Lambda1res}}
    \subfigure[]{\includegraphics[width = \columnwidth]{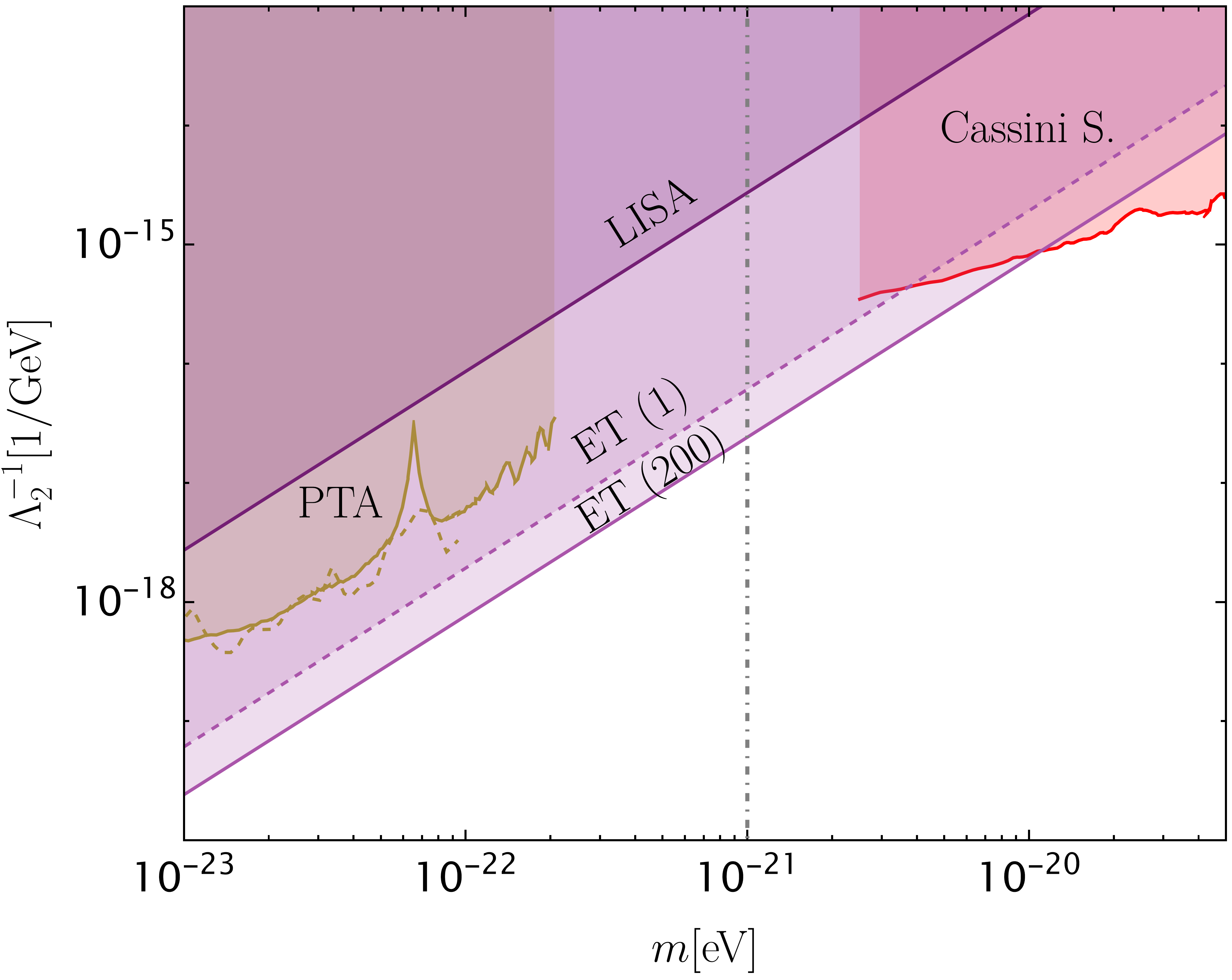}
    \label{fig:Lambda2res}} 
    \caption{Taken from \cite{Blas:2024duy}. Projected constraints on the linear (left) and quartic coupling parameters (right) for both LISA and \et from low-frequency soliton \dm modulating (``heterodyning'') high-frequency carrier \gwh signals via changes to the \moi of \nss, which alters \cws from deformed \nss (\et)  and the \gws from inspiraling \bbhs (LISA). The curves for \et assume one or 200 detections of \cws from isolated \nss. The LISA curve is determined by the impact of soliton \dm on \gwh frequencies from binary systems. The olive regions are excluded by the \pta analysis in \cite{Smarra:2024kvv}, while blue and red are constrained away via Cassini tests of general relativity (violet) and on a \sgwb (red).
    }
    \label{fig:lambda12res}
\end{figure*}

We note that in this case, \dm leaves a particular imprint on the \gwh signal itself, arising from deformed, isolated \nss. This differs from the discussion in \cref{sec:pdm}, where \dm interacts directly with the components of the \ifo, causing differential strains that are distinct from the expected signal from \gws. In contrast, here we are looking for a deviation in a canonical \gwh source due to \dm, rather than an interaction with the \ifo itself.

Furthermore, though the effect described in \cref{subsubsec:change-moi} of dilatons interacting with neutron stars is similar to that discussed here, it arises from a different physical mechanism. While dilatons alter the fundamental constants via a new coupling that leads to a oscillatory \moi of \nss, solitons interact directly with the internal structure of the \ns. This difference leads to solitons modulating the \gwh frequency, while dilatons would alter the pulse time of arrival in \pta experiments.

In summary, solitons could significantly impact the \gwh signals arising from supermassive black hole binaries, white-dwarf binaries, and asymmetrically rotating neutron stars. These signals, which are observed using \ptas, space-based \ifos, and ground-based \ifos, respectively, can be modulated by the presence of soliton dark matter. For supermassive \bh binaries, solitons can induce dynamical friction on the orbit, leading to detectable changes in the nanohertz frequency range. White-dwarf binaries, could show low-frequency modulations in their \gwh emission, potentially revealing the imprint of soliton \dm on the millihertz carrier signals. Similarly, asymmetrically rotating \nss, whose \cws are monitored by ground-based detectors, could exhibit phase shifts by passing through soliton \dm. These potential signatures across multiple observational platforms make soliton \dm a promising candidate for future studies in \gwh astronomy, and provide interesting probes of the \dm landscape.

\section{\gwh probes of WIMP dark matter}\label{sec:wimp}

While much of this review has focused on probing \uldm with \gwh \ifos, with the notable exception of macroscopic \dm discussed in \cref{sec:macdm}, it is important to also highlight the potential for using \gws to probe more conventional forms of dark matter, such as WIMPs. Although WIMPs are typically considered on different scales and their interactions are more strongly coupled than those of \uldm, they can still leave distinctive imprints on \gwh signals.

In this section, we outline two possible indirect ways to search for WIMP dark matter using \gws. First, in \cref{subsec:tbh}, we explore the possibility that WIMPs could induce the collapse of stellar-mass objects, both compact and not, into \bhs leading to unique signatures in the \gwh emission from these newly formed \bhs. Second, in \cref{subsec:gev}, we discuss how the non-detection of \cws from the \gc could help constrain the \msp hypothesis for the observed GeV excess, thereby reinforcing the alternative scenario in which annihilating WIMP \dm is the underlying cause of this astrophysical signal.

By examining these two avenues, we can see how \gw observations can provide crucial insights into the nature of WIMP \dm and contribute to the broader effort to identify and characterize a diverse set of \dmh candidates.

\subsection{Transmuted black holes in binaries}\label{subsec:tbh}

Non-annihilating WIMPs that interact with nucleons could lose energy through single or multiple scatterings off of celestial objects \cite{Bramante:2017xlb,Bell:2020lmm}, eventually becoming gravitationally bound to them over time. After WIMPs repeatedly interact with ordinary nucleons, they become thermalized, forming a compact core with a higher density than that of the celestial object. If the density of the particles becomes sufficient inside the celestial object, gravitational collapse (in the case of bosons) or Chandrasekhar collapse (in the case of fermions) of the WIMP core will occur, thus resulting in a newborn \bh that is extremely tiny with respect to the celestial object. After formation, the tiny \bh could accrete surrounding \dm, and if enough \dm becomes bound to the object quickly enough compared to the Hawking evaporation timescale, the entire object be ``swallowed'' by the growing \bh and become ``transmuted'' to what is known as a transmuted black hole (\tbh) \cite{Dasgupta:2020mqg,Ray:2023auh,Bhattacharya:2023stq,Bhattacharya:2024pmp,Bhattacharya:2024cpm}. Quantitatively, if the sum of the time to accrete new particles and the time for the WIMP core to become gravitationally unstable is smaller than the age of the universe, \tbhs should form.

\tbhs forming from inspiraling \nss were first considered in \cite{Dasgupta:2020mqg}. Since \dm would induce \nss to transmute into \bhs, from the observational perspective, there is no clear difference between electromagnetically silent \nss and low-mass \bhs. Therefore, the non-detection of low-mass \bbhs by \lvk can be used to constrain weak interaction cross-sections, since the progenitor \nss provide an extremely strong gravitational field to accrete and retain \dmh particles \cite{Bhattacharya:2023stq,Bhattacharya:2024cpm}. In \cref{fig:wimp-cbc-cons}, we show the excluded bosonic \dmh mass/cross-section parameter space based on null search results from the \lvk collaborations for \ssm compact objects \cite{LIGOScientific:2022hai}. Additionally, a projection for this constraint is shown for a 50x increase in the probable spacetime volume, which is reasonable for the future \gwh network using \et or \ce. While analyses of \lvk  data cannot yet constrain new parameter space for this model of \dm, future searches should be able to surpass the pulsar constraint, which comes from the fact that the nearby Gyr-old pulsar PSR 0437$-$4715 has not actually collapsed into a \bh in its lifetime \cite{McDermott:2011jp,Garani:2018kkd}.  

In contrast, stars, being less compact than \nss, could sustain \dmh cores with stronger interaction cross-sections, because of the weaker gravitational field around the star \cite{Bhattacharya:2024pmp}. In \cref{fig:lisa-wimp-cons}, we show the potential excluded parameter space for null detection of \tbhs arising from slowly inspiraling non-compact binaries that could be visible in space-based \gwh detectors. These binaries would be inspiraling for durations of order of the age of the universe, and thus exhibit \cwh signals in the millihertz regime. The results are parameterized in terms of $\alpha$, which denotes the fraction of stellar-mass binaries that form close enough to exist within the millihertz band. Essentially, $\alpha$ parametrizes our ignorance about how closely the population of sun-like stars form close to each other.

Comparing \cref{fig:wimp-cbc-cons} and \cref{fig:lisa-wimp-cons}, we note the vastly different interaction cross-section ranges from these two complementary \gwh probes of WIMP \dm. For both low-mass \bhs and sun-like stars, fermionic \dm has also been constrained in similar ways as bosonic \dm has, see \cite{Bhattacharya:2023stq,Bhattacharya:2024pmp} for more details.

\begin{figure*}[htbp]
	\centering
    \subfigure[]{
	\includegraphics[width=0.43\textwidth]{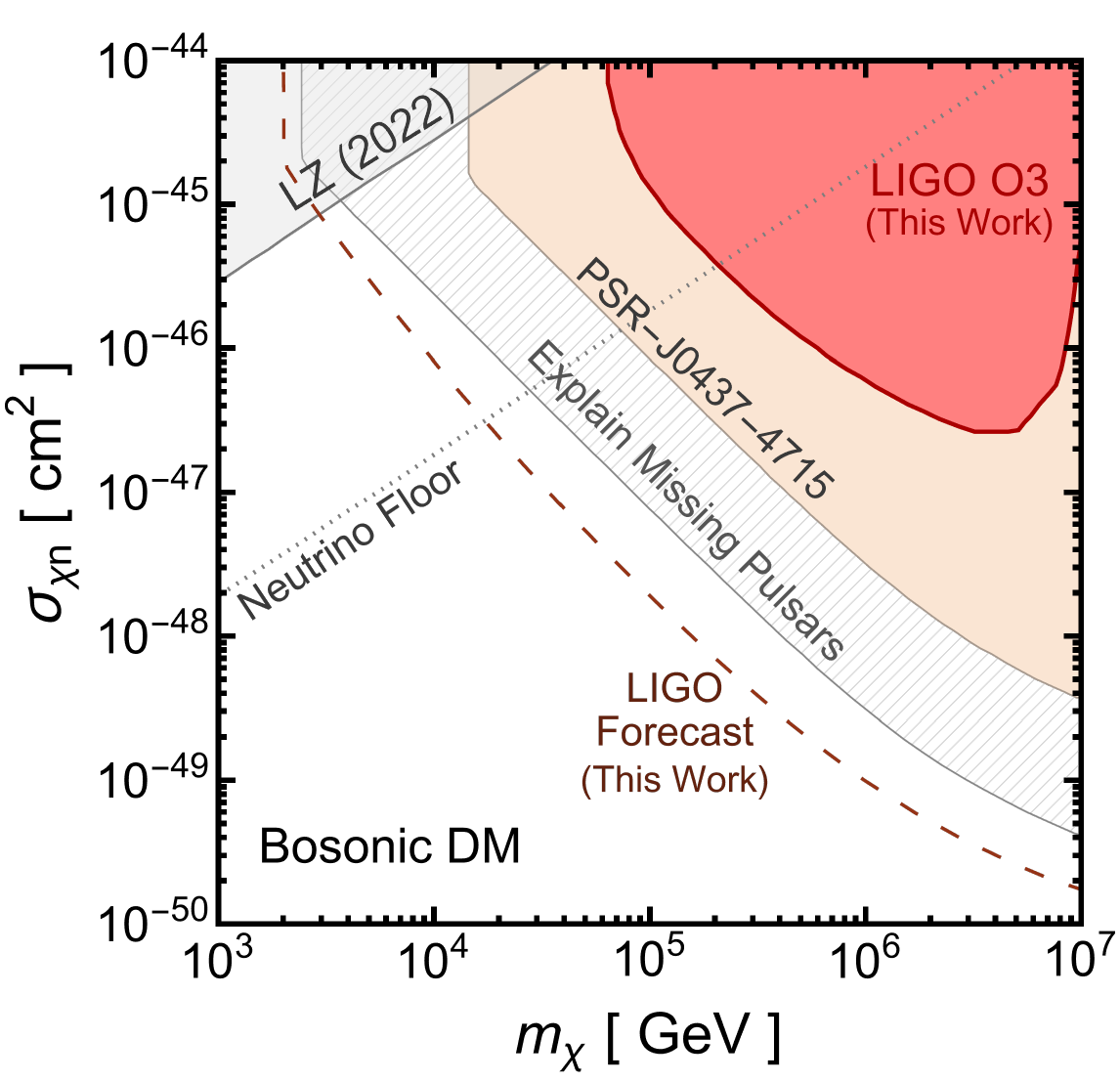}
	\hspace{1.2 cm}
    \label{fig:wimp-cbc-cons} }
    \subfigure[]{
	\includegraphics[width=0.43\textwidth]
    {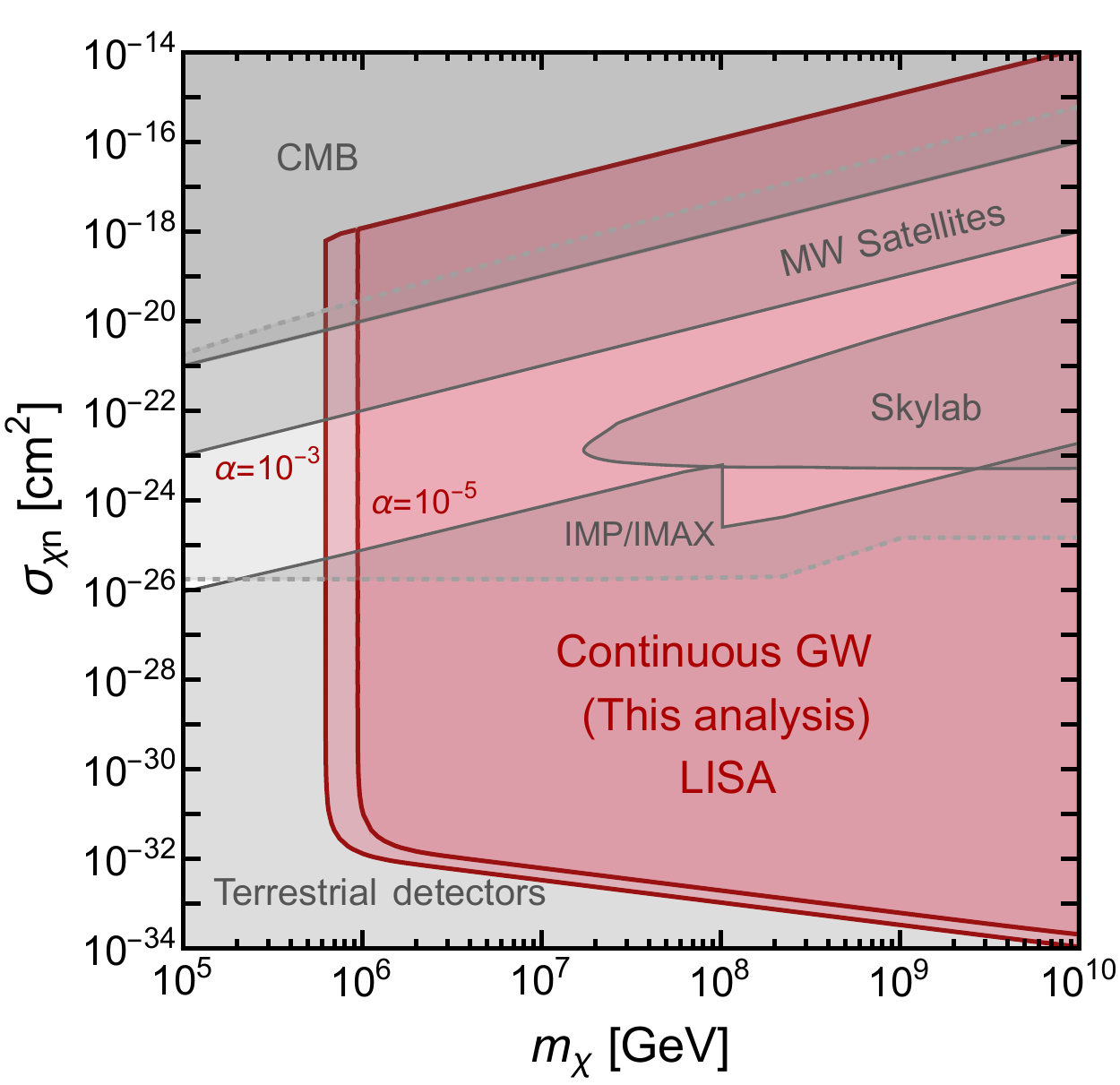}
      \label{fig:lisa-wimp-cons}}
	\caption{Taken from \cite{Bhattacharya:2023stq} and \cite{Bhattacharya:2024pmp}. (a) Constraints and (b) projected constraints in the WIMP \dmh mass ($m_{\chi}$) and interaction cross-section ($\sigma_{\chi n}$) plane for bosonic \dmh particles that could induce the collapse of celestial objects to \bhs. (a): The orange-shaded region shows the current limit from O3 data \cite{LIGOScientific:2022hai}, while the brown dashed line represents the forecast with a 50$\times$ sensitivity increase. (See Fig.1 of Ref. \cite{Bhattacharya:2023stq}). (b): The red shaded region (labeled ``Continuous GW (This analysis)'' indicates the exclusion based on the proposed sensitivity of observing quasi-monochromatic \cws from inspiraling \tbhs in LISA, with $\alpha$ being the fraction of sun-like star binaries that form close enough to be in the LISA band. There are two exclusion regions, depending on what value we assume for $\alpha$. The constraining region gets smaller as $\alpha$ decreases because less stellar-mass binaries are being assumed to form at frequencies detectable by LISA. For $\alpha < 10^{-5}$, this method excludes no parameter space. Further details on other existing constraints and other scenarios (Fermionic DM, Bose Einstein Condensate formation) are in the original references \cite{Bhattacharya:2023stq,Bhattacharya:2024pmp}.}
	\label{fig:wimps-gw-probe}
\end{figure*}



\subsection{\gwh probes of the galactic-center GeV excess}\label{subsec:gev}

Annihilating WIMP \dm and a collection of unresolved \msps are two leading explanations for the unexpected excess of GeV gamma rays observed by the Fermi-LAT experiment coming from the \gc \cite{Goodenough:2009gk,Hooper:2010mq}. At first, the GeV excess appeared spherically symmetric and well-fit by annihilating \dmh models \cite{Hooper:2011ti,Gordon:2013vta,Daylan:2014rsa,Calore:2014nla,Abazajian:2014fta}. But still, an astrophysical explanation of \msps cannot yet be ruled out \cite{Abazajian:2010zy,Calore:2014xka,Yuan:2014rca,Petrovic:2014xra,Ye:2022yxt}. 

\gws can be used to weigh in on this debate. If the \msps are deformed, they will emit \cws as they rotate. The amplitude of these \cws is determined by both the \gwh frequency and the degree of deformation, called the ellipticity, which is defined as $\epsilon\equiv\frac{I_{\rm xx}-I_{\rm yy}}{I_{\rm zz}}$ \cite{Haskell:2015psa}, where $I_{\rm xx},I_{\rm yy}$ and $,I_{\rm zz}$ are the principle \mois along each axis.

One approach to constraining the \msp hypothesis is to consider the bulk emission of a thousands or more sources of \cws that are superimposed, creating a \sgwb. Ref. \cite{Calore:2018sbp} modelled this \sgwb and projected constraints on the average ellipticity of this population as a function of the total number of \msps in the \gc. The authors showed that $\mathcal{O}(10^4)$ \msps with ellipticities of $\sim 10^{-7}$ could be detectable by current detectors and be consistent with the diffuse electromagnetic radiation coming from the \gc, as shown in \cref{fig:Ntoteps}. However, they did not go so far as to link \gwh observations with Fermi-LAT ones.

Ref. \cite{Miller:2023qph} took a different approach by analyzing the continuous gravitational wave (CGW) signals from individual neutron stars (NSs), specifically focusing on millisecond pulsars (MSPs). They explored two main scenarios for the deformation of these pulsars:

\begin{enumerate}
    \item Magnetic Field-Induced Deformations: In the first scenario, they assumed that the deformations of the \msps are sustained by the internal magnetic fields of the neutron stars, as proposed in previous work \cite{Lander:2011yr, Mastrano:2011aa, Lander:2013oea}.
    \item Model-Agnostic Deformation: In the second, more general scenario, they did not assume a specific cause for the deformation. Instead, they modeled the ellipticity as 1\% of the total electromagnetic radiation emitted by the star, making it independent of the underlying physical mechanism.
\end{enumerate}
Using these two models, Ref. \cite{Miller:2023qph} then assumed that the ellipticity and frequency distributions of unresolved population of \msps can be calcualted with knonw pulsars' parameters \cite{Manchester:2004bp}. By integrating over these distributions with upper limits on ellipticities from an O3 LIGO search for isolated neutron stars \cite{KAGRA:2022dwb}, they calculated the probability of detecting a \cw from this population for both models of ellipticity noted above

In parallel, the number of \msps in the \gc can be inferred by modeling the electromagnetic emission associated with the GeV excess. This is done by fitting luminosity functions, which describe the expected number of \msps needed to explain the GeV excess as a function of various model parameters \cite{Hooper:2016rap, Ploeg:2020jeh, Dinsmore:2021nip}. By combining the predicted number of \msps, derived from the luminosity functions, with the probability of detecting them via \cws, the authors were able to rule out certain luminosity function parameters. Specifically, they excluded models in which the number of \msps would have been large enough to produce  at least one detectable \cw, thereby exploring the possible origins of the GeV excess.


\cref{fig:gev-cons} shows the parameter space of a luminosity function that models the observed luminosity of the \gc GeV excess as following a log-normal distribution. The color show the number of \msps predicted by each choice of the two parameters of the luminosity function ($L_0,\sigma_L)$. Plotted as a light blue region is an exclusion area, which signifies that \cwh searches would have seen at least one \msp, assuming the aforementioned ellipticity and frequency distributions for the \msps in the galaxy. We can see that \gwh searches probe a complementary portion of the parameter space to what Fermi-LAT does: if there are more \msps that are dimmer electromagnetically, this means more rotational power may go into \gws, and more of them are necessary to explain the GeV excess. Thus, \gwh searches would have more \msps to find, and not seeing at least one of them rules out particular luminosity function parameters. Likewise, Fermi performs better when less \msps are more luminous.

Most recently, Ref. \cite{Bartel:2024jjj} simulated a population of $\sim 40000$ \msps in the \gc that could explain the \gc GeV excess, and found that the \gwh signal could lie between [200,1400] Hz, and would have strain amplitudes of $\mathcal{O}(10^{-30}-10^{-28})$ based on the population parameters chosen (e.g. \moi around $\sim 10^{38}$ kg$\cdot$ m$^2$, ellipticities mostly smaller than $10^{-9}$, and Boxy vs. spherical spatial distributions of the \msps). It is argued that such signals are too weak to be detected by current and future \gwh observatories; however, this conclusion is only for a stochastic background of \gws and depends heavily on the assumed ellipticity distribution and the number of \msps chosen to be in the population. If \msps appear in larger numbers than assumed here, they would each be dimmer, since the luminosity of the GeV excess is fixed. In fact, the constraints from \cite{Miller:2023qph} are strongest for larger populations than assumed in \cite{Bartel:2024jjj}. Additionally, the ellipticity distribution employed in \cite{Bartel:2024jjj} only reaches a maximum of $\sim 2\times 10^{-9}$, which is consistent with the minimum ellipticity of \nss as predicted by \cite{Woan:2018tey}, but may be too conservative. Upper limits on ellipticities at the \gc from \cwh searches have not yet constrained ellipticities to be less than $\sim 10^{-6}$ \cite{KAGRA:2022osp,KAGRA:2022dwb,Steltner:2023cfk}, which is consistent with the maximum ellipticity employed in \cite{Miller:2023qph} up to $10^{-6}$.

\begin{figure*}[htbp]
    \centering
    \subfigure[]{
    \includegraphics[width=0.85\columnwidth]{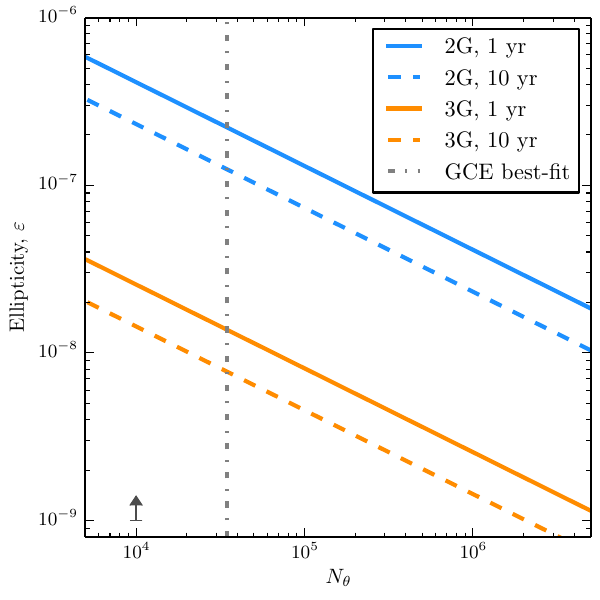} 
    \label{fig:Ntoteps}}
    \subfigure[]{
   \includegraphics[width=\columnwidth]{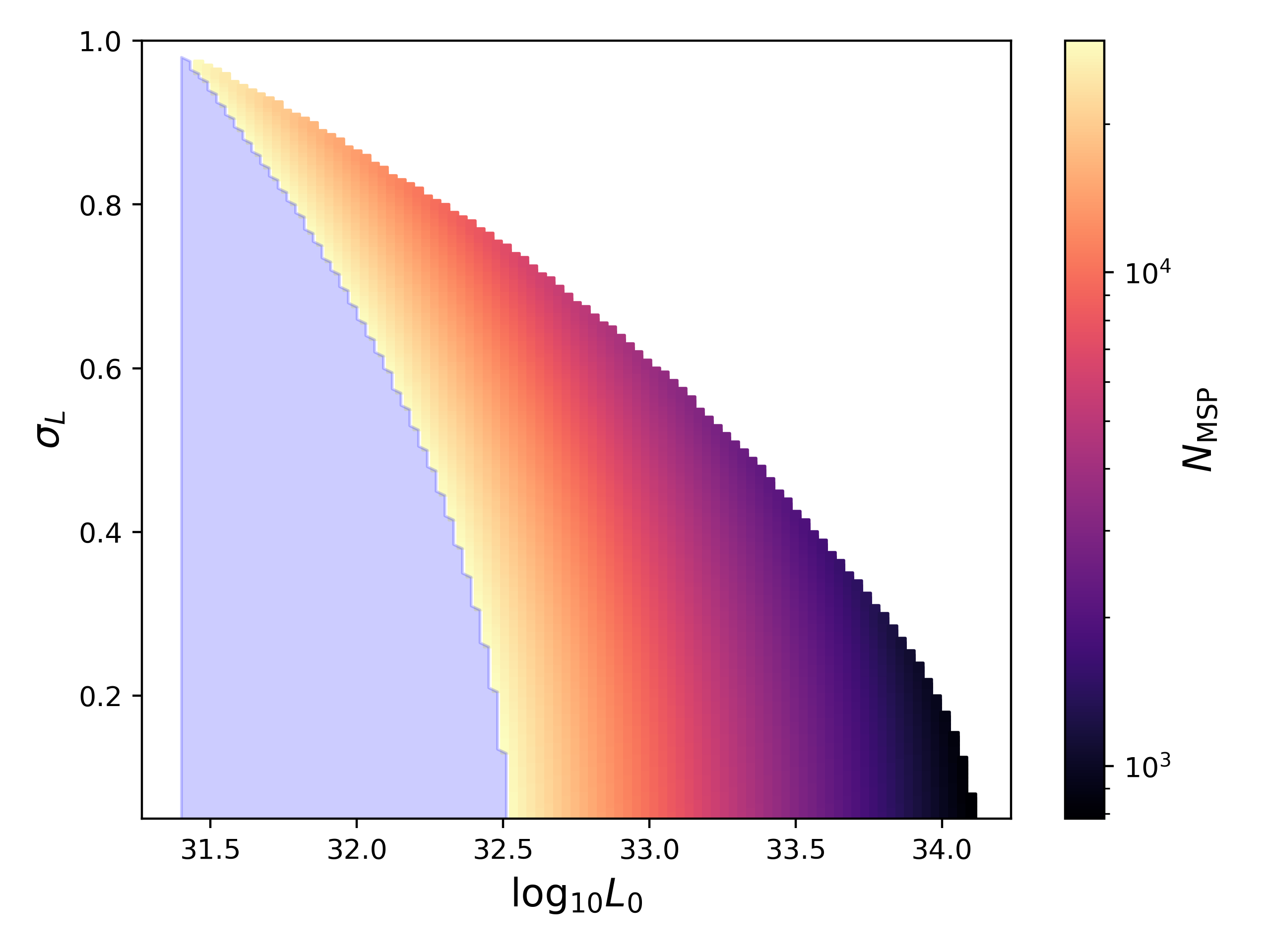}
   \label{fig:gev-cons}} 
    \caption{Taken from \cite{Calore:2018sbp} and \cite{Miller:2023qph}. (a) Projected constraints from current-generation (2G) and next-generation (3G) ground-based \gwh \ifos on the average ellipticity of a population of \msps. These are derived from a null result in a search for the \sgwb from deformed, rotating \nss across the sky, for different observation times \cite{Calore:2018sbp}. The vertical line marks the total number of \msps in the bulge required by fits to the \gc excess \cite{Bartels:2017vsx}. (b) Constraints from an all-sky search using O3 LIGO/Virgo data for isolated \nss \cite{KAGRA:2022dwb}, shown in light blue, on the portion of the luminosity function parameter space that would have yielded a population large enough for at least one \msp to be detected \cite{Miller:2023qph}. Here, $L_0$ and $\sigma_L$ denote the parameters of a log-normal luminosity function. To construct this plot, we assume: the \gc is 8 kpc away; the unknown \msps have a moment of inertia $I_{\rm zz}=10^{38}$ kg$\cdot$m$^2$; the frequency distribution follows pulsars in the ATNF catalog; and the ellipticity distribution is derived from the ATNF catalog assuming 1\% of rotational energy is converted into \gws. }
\end{figure*}

In summary, this section highlights the potential of using \gws to probe WIMP \dm, focusing on two distinct approaches. First, the formation of \tbhs from WIMP interactions with celestial objects can lead to unique \gwh signatures, such as those from inspiraling \nss and \bhs. These signals, especially in the millihertz band, offer constraints on the DM interaction cross-section and can be used to probe the properties of both bosonic and fermionic WIMPs. Second, \gwh non-observations of \cws from deformed \msps in the \gc provide a complementary method to weaken the \msp hypothesis for the GeV excess observed by the Fermi-LAT experiment, which lends strength to the annihilating WIMP hypothesis for this excess. By analyzing the number and ellipticity of unresolved \msps, \gwh searches can rule out certain luminosity function parameters, helping to distinguish between WIMP and \msps explanations for the observed excess. These two complementary methods underscore the crucial role \gws can play in advancing our understanding of WIMP \dm.





\section{\gwh probes of \dmh spikes around inspiraling black holes }\label{sec:dmspike}


It was first recognized in \cite{Gondolo:1999ef} that \bhs embedded in the \dmh halo of a galaxy could accrete \dm adiabatically, potentially forming a high-density region around the BH known as a ``spike.''  Subsequent studies demonstrated that such \dmh spikes around supermassive \bhs at centers of galaxies would have been disrupted by galactic mergers \cite{Merritt:2002vj}, if the \bh formed off-center \cite{Ullio:2001fb}, or through interactions with stars in the surrounding stellar population \cite{Merritt:2003qk,Bertone:2005hw}.  However, realistic profiles incorporating relativistic corrections have shown that these spikes can cause more \dm to accumulate closer to the \bhs in both Schwarzchild and Kerr metrics \cite{Sadeghian:2013laa,Ferrer:2017xwm,Bertone:2024rxe}, which begs the question of whether these objects could survive galactic dynamics. Though it has been shown that heavy, particle-like \dm would be dispersed in equal-mass merger events through N-body simulations \cite{Bertone:2004pz}, it is much less clear-cut for \uldm: multiple works \cite{Boudon:2023qbu,Zhong:2023xll,Cardoso:2022nzc,Cardoso:2022vpj,Vicente:2022ivh,Annulli:2020lyc,Annulli:2020ilw,Brax:2019npi,Ficarra:2021qeh,Zhang:2022rex,Choudhary:2020pxy,Yang:2017lpm} suggest that such \dmh spikes could survive, even up to ten orbits before merger \cite{Aurrekoetxea:2024cqd}. The extent to which \dmh spikes survive directly affects the amount of \df the binary will experience as it inspirals, thus altering the \gws from the inspiral.


In light of these recent developments regarding the plausibility of forming \dmh spikes around supermassive \bhs, Ref. \cite{Daniel:2025mna} has forecast upper bounds on the \dmh densities around supermassive \bhs using future LISA observations of these systems. These bounds provide new constraints on the \dmh spike formation and the nature of \dm in the central regions of galaxies, offering a unique opportunity to test the high-density \dmh environment near supermassive \bh binaries. Furthermore, Ref. \cite{Chan:2024yht} determines that the observed decay rate of the orbit can be completely explained by the existence of a \dmh spike around the supermassive \bh in  OJ 287, providing plausible evidence of the existence of \dmh spikes.

Additionally, \imbhs can sustain \dmh spikes because their smaller masses and relatively longer timescales for mergers (with respect to those of supermassive \bhs) prevent the spikes from being disrupted \cite{Zhao:2005zr,Kim:2022mdj}. Similarly, recent studies suggest that both astrophysical \bhs and \pbhs could also host \dm spikes if there is an absence of significant stellar interactions that could scatter \dm particles, and if the gravitational influence of the \imbh persists over time, especially in environments where galactic mergers or other disruptive events have not destroyed the spike \cite{Kohri:2014lza,Eroshenko:2016yve,Boucenna:2017ghj,Hertzberg:2019exb,Bertone:2024wbn,Bertone:2024rxe,Dosopoulou:2025jth}. These findings expand the range of environments where \dm spikes might form, suggesting that \imbhs could significantly influence \dm distributions.

For simplicity, despite the more realistic \dmh spike profiles recently discussed, the density of the \dmh spike is often approximated by the following equation \cite{Eda:2013gg,Cole:2022ucw,Cole:2022yzw}:

\begin{align}
 \rho(r)&= \rho_{\rm sp}\left(\frac{r_{\rm
 sp}}{r}\right)^{\gamma_s}
\,\,\,(r_{\rm min} \le r \le r_{\rm sp}),
\label{Eq:rhoDM}
\end{align}
where $\rho_{\rm sp}$ is a normalization constant for the \dmh density of the spike, $r_{\rm min}$ is the minimum separation between the two objects, often taken to the radius at the innermost stable circular orbit, and $r_{\rm sp}$ is the spike radius. Here, $\rho_{\rm sp} = 379 \ M_{\odot}/\text{pc}^{3}$ and $r_{\rm sp} =
0.33 \text{pc}$ if the mass of the \imbh is $M_{\rm BH}=10^3 M_{\odot}$ and the \dmh spike has a mass of $M_{\rm halo}=10^6 M_{\odot}$. The power-law index $\gamma_s$ denotes how the concentration of \dmh particles falls off as one moves away from the host \bh. Astrophysically speaking, measurements of $\gamma_s$ would allow us to glean important information regarding whether the host \bh evolved from the \dmh halo and/or experienced mergers in the past \cite{Eda:2013gg}.

We consider two scenarios where \df can influence the evolution of \bbhs: in \cref{subsec:imridf}, we examine how \df leads to gravitational-wave dephasing in \imri systems; in \cref{subsec:dephase-pbh}, we explore how \dmh spikes can form around \pbhs and cause dephasing in binaries with primary masses of $\mathcal{O}(1\text{--}100)\msun$ and mass ratios $\leq 10^{-3}$. Lastly, in \cref{subsec:ul-df}, we describe how observations of \bbhs can be used to place limits on \dm spikes, under the assumption that their orbits remain largely unaffected by \df.

\subsection{Dephasing due to dynamical friction of intermediate mass ratio inspirals}\label{subsec:imridf}

Ref. \cite{Eda:2013gg} first showed that the phase evolution of an intermediate-mass ratio inspiral (a binary composed to a $\mathcal{O}(1000)\msun$ and an $\mathcal{O}(1)\msun$ object) would be significantly modified at milli-hertz frequencies due to the presence of a \dmh spike around the intermediate-mass \bh. This occurs because the less massive object sweeps through the \dm spike as it orbits around the more massive one, losing energy via \df (in addition to via \gws) and speeding up the orbit \cite{Speeney:2022ryg,Mitra:2025tag}. Such effects will primarily be relevant in space-based \gwh detectors \cite{Duque:2023seg,Rahman:2023sof,AbhishekChowdhuri:2023rfv,Ghodla:2024fit}, since \imbhs inspiral and merge at much lower frequencies than stellar-mass \bhs \cite{Miller:2008fi}.

This conclusion was significantly expanded in \cite{Eda:2014kra}, in which the authors demonstrated how accurately the power-law index can be measured from a detection of an \imri signal in LISA. This accuracy is quantified in terms of the power-law index and the masses of the two objects that are inspiraling. If the \imri system consists of smaller primary and secondary masses, and a larger power-law index, the power-law index can be measured more accurately. This is because higher values of induce a larger phase difference, and smaller masses in the \imbh system ensures that it spends more cycles in the LISA band before merging. 


Recently, it has also been shown that these dephasing effects are unique enough to allow distinguishing between three \bh environments: accretion disks, \dmh spikes, and ultralight \bcs \cite{Cole:2022yzw}, as well as distinguishing between environmental effects and modified theories of gravity \cite{Yuan:2024duo}. As an illustration, we show in \cref{fig:snrlossdf} the impact of an environmental effect, the \dmh spike, on the \snr obtained when match-filtering with a vacuum template to analyze data in which a \bbh system has a \dmh spike around it. Here, $m_1=10^5\msun$, which was chosen to ensure that a plausible formation mechanism exists to form this kind of \bh \cite{Gondolo:1999ef}. Ref. \cite{Cole:2022yzw} argues that a percentage \snr loss of more than 30\% would compromise the prospects of detecting a (weak) signal, and shows that even smaller \snr losses would bias parameter estimation. However, it is important to note that in LISA, we will be in a regime in which signals of this kind have enormous \snr, and that parameter estimation and distinguishing among different \gwh signals will be a key challenge. Thus, it is more likely that parameter estimates will be biased rather than signals being completely missed. Moreover, the dephasing caused by \dmh spikes underscores the need to adapt \cwh methods, as discussed in \cref{subsec:meth-pdm} and \cref{subsec:meth-bc}, to effectively search for these systems \cite{Sethi:2025ixx}. In principle, \cwh methods could provide quick point estimates for key \bbh parameters, which would enable the development of a reduced set of templates that incorporate \dmh spikes. These templates could then be used for follow-up matched filtering searches.

\begin{figure}[htbp]
\begin{center}
\includegraphics[width=\columnwidth]{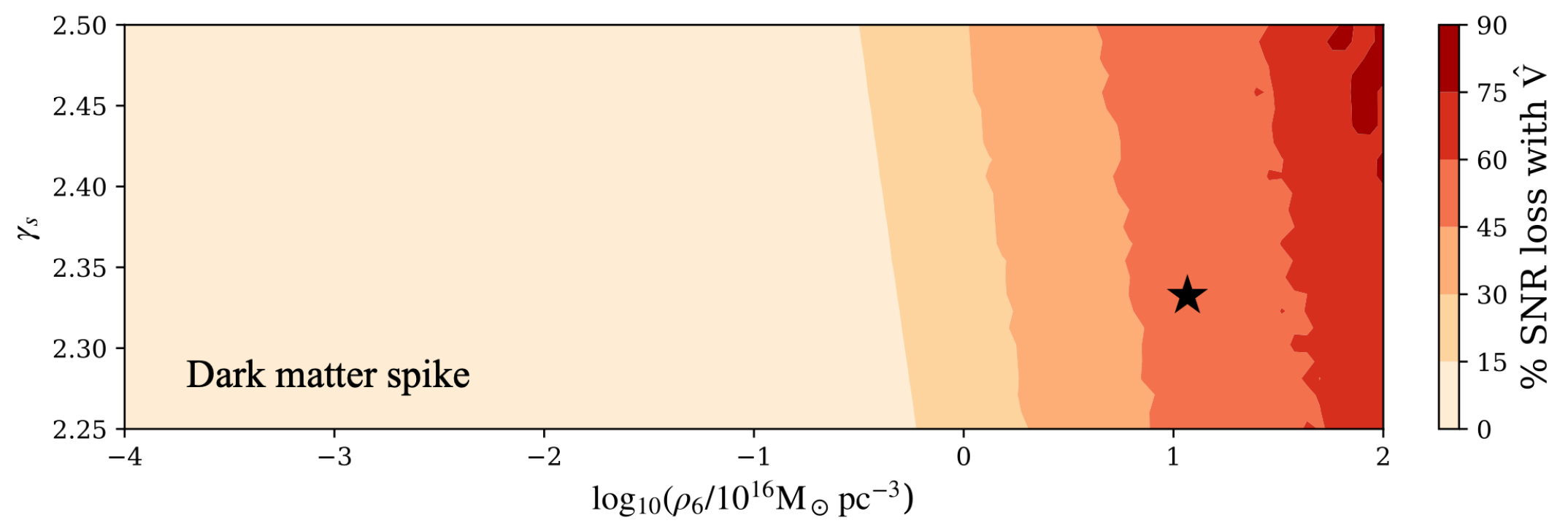}
\caption{Taken from \cite{Cole:2022yzw}.  Percentage \snr loss from matched filtering with a vacuum template when the injected \gwh signal originates from a \bbh system surrounded by \dmh spikes (the injected waveform includes the effect of dynamical friction). The $x$ and $y$ axes correspond to the parameters of the \dmh spike profile defined in \cref{Eq:rhoDM}. The assumed observation time is one year, with component masses $m_1 = 10^5 \msun$ and $m_2 = 10 \msun$. }
\label{fig:snrlossdf}
\end{center}
\end{figure}

While Ref. \cite{Cole:2022yzw} demonstrated that distinguishing between different \bh environments, including the presence of \dmh spikes, is feasible, Ref. \cite{Zwick:2024yzh} presents a different perspective, arguing that the dephasing effects caused by these environmental factors could result from a variety of unmodeled sources. Moreover, the authors argue about the practicality of measuring such dephasing effects, noting that the early inspiral is where the effects would be most pronounced, but this is also where the \gwh emission is weakest; that a constant phase offset could mimic dephasing effects ; and that \mf over such long timescales would have practical limitations. For thee reasons.

However, Ref. \cite{Wilcox:2024sqs} provides a counterpoint, suggesting that such dephasing could still be tracked within a parameterized post-Einsteinian framework, offering a potential solution to the challenges raised in \cite{Zwick:2024yzh}. Additionally, they propose more model-agnostic approaches, such as periodic forces, that could leave characteristic signatures on the \gwh signal and be applied to study the evolution of the binary system.

These differing viewpoints are crucial when considering the overall approach to detecting and interpreting \gws from systems with \dmh spikes. While \cite{Zwick:2024yzh} concerns highlight the potential difficulties in applying matched filtering to these cases, the model-agnostic approach in \cite{Wilcox:2024sqs} offer a promising alternative for tackling the problem. In light of these challenges, it is clear that both traditional and alternative methods, such as the \cwh approaches discussed in \cref{subsec:meth-pdm} and \cref{subsec:meth-bc}, should be explored in tandem to fully address the complexities involved in detecting signals from \bbhs influenced by \dmh spikes \cite{Sethi:2025ixx}. As mentioned in \cref{subsec:soliton-phase-bin}, \cwh methods could provide ways of addressing the three concerns laid out in Ref. \cite{Zwick:2024yzh} by providing point estimates of binary parameters that can be followed up by a sensitive and restricted \mf search. This approach can also be applied in lieu of traditional \mf for \pbhs in \et and \ce.

\subsection{Dephasing due to dynamical friction of inspiraling primordial black holes}\label{subsec:dephase-pbh}

\pbhs could also host \dmh spikes, since at most masses, the fraction of \dm that they could compose is less than one \cite{Green:2020jor,Carr:2023tpt}. Thus, Ref. \cite{Cole:2022ucw} showed that \et and \ce could probe the existence of \dmh spikes around binary \pbh systems, where the primary has a mass of $[1,10]\msun$ and the system has a mass ratio of $10^{-3}$. However, similarly to \imri systems discussed in the previous subsection, these exotic mergers could be missed if the \dmh spike is not taken into account in matched-filtering searches in \et and \ce. As mentioned in \cref{subsec:imridf,subsec:soliton-phase-bin}, one possible approach is to adapt \cwh methods to search for \dmh spikes around \bbhs. This could provide a quick, though less sensitive, way to obtain point estimates for important \bbh parameters, such as the chirp mass. These estimates could then be used to reduce the number of templates needed to account for \dmh spikes, enabling a follow-up matched filtering search.

\subsection{Upper limits on black-hole environments}\label{subsec:ul-df}

Because vacuum phase-evolution waveforms confirmed the detection of many \bbh mergers by \lvk, any deviations due to environmental effects, such as a \dmh spike, must be small enough to be inconsequential to the detected \gwh signals. Knowing this, Ref. \cite{CanevaSantoro:2023aol} modeled \df and other possible environmental effects as additional (negative) \pn corrections to the waveform, and placed upper limits on the density of environment around \bhs by requiring the induced dephasing to be small enough to not alter the phase evolution of the \bbh systems in vacuum. While the results were used to eliminate dynamical fragmentation \cite{Loeb:2016fzn} as a possible binary formation channel, they are generic enough to apply to \emph{any} \bh environment. Unfortunately, the upper limits on environments affected by \df, which would be relevant for the \dmh spike scenario, are, at best, about seven orders of magnitude larger than the maximum density of the spike ($10^{-6}$ g/cm$^3$) \cite{Gondolo:1999ef,Sadeghian:2013laa}. The upper limits are shown to drastically improve in DECIGO because the \df (and other environments) affect the early inspiral of the \bbh system much more than at the time of merger. These findings are consistent with those found in N-Body simulations about how the dynamics of the \dmh spike affect the inspiral, and the detectability of these effects by LISA \cite{Cardoso:2019rou,Kavanagh:2020cfn}.

\section{\gwh probes of atomic \dm}\label{sec:atomicdm}

It is possible that the dark sector is actually much more complicated than a single new particle with a single coupling to gravity or the \sm. In fact, atomic \dmh models exist \cite{Feng:2009mn,Kaplan:2009de} in which dark protons and dark electrons interact via a massless dark photon\footnote{This dark photon is different from both the dark photon that couples to baryons, discussed in \cref{subsec:vecdm}, and the one that kinetically mixes with the ordinary photon, discussed in \cref{subsubsec:lum-dp-bc}.} with a particular coupling. In such a model, \dm can form bounds states analogous to the Hydrogen atom, and dissipate energy via Bremsstrahlung, recombination, and collisions \cite{Buckley:2017ttd,Rosenberg:2017qia}. If this kind of \dm collects in small-enough regions of space, it will collapse and form ``dark \bhs'' \cite{Shandera:2018xkn}. Such dark \bhs could form in binary systems, as astrophysical \bhs do; thus, \lvk observations of \bbh mergers can constrain the existence of these objects, just as they do for \pbhs \cite{Singh:2020wiq,LIGOScientific:2021job,LIGOScientific:2022hai}. 

Here, the observable \gwh signature of dark \bhs would come from the inspiral, merger and ringdown of \bbh systems. However, this signature is not unique: there does not yet exist a way of distinguishing between astrophysical, dark \bhs and \pbhs. Thus, the constraints that we present in \cref{subsec:dark-bbh-constraints} and \cref{subsec:dark-bbh-ssm-constraints} on \bhs with $\mbh=[1,100]\msun$ and $\mbh=[0.1,1]\msun$, respectively, assume that the \bhs that are searched for in \lvk data are dark \bhs.

\subsection{Constraints from observations of \bbhs}\label{subsec:dark-bbh-constraints}

Ref. \cite{Shandera:2018xkn} first considered the prospects of detecting dark \bh mergers with advanced LIGO and \et, finding that these \ifos could set stringent constraints on the fraction of \dm that dark \bhs could compose. If the \bh merger of GW190425 \cite{Abbott:2020uma} is interpreted as the merger of two dark \bhs, the Chandrasekhar mass \cite{Chandrasekhar:1931ih} can be constrained to be below $1.4\msun$ at $>99.9\%$, bounding the lower-most value mass of the dark proton as 0.95 GeV and the spacing of energy levels in a dark atom to be near $10^{-3}$ eV \cite{Singh:2022wvw}. In order to obtain these constraints, we note that \lvk observations of \bbhs provide estimates of the chirp mass and the component masses of the system. Ref. \cite{Singh:2020wiq,Singh:2022wvw} jointly constrain, in a Bayesian way, the fraction of \dm that dark \bhs could compose and the minimum mass $M_{\rm min}$ of the distribution of dark \bhs assumed, while marginalizing over other hyperparmeters, such as the slope and range of the mass distribution. These posteriors are conditioned on the output of a \gwh search, which is essentially predicted merger rate density that depends both on what \bbhs have been observed and \psd of the \ifos.

Quantitatively, these constraints on the mass of the dark proton and the Chandrasekhar limit of the mass of dark \bhs are derived using \cite{Chandrasekhar:1931ih,Shandera:2018xkn,Singh:2020wiq}:

\begin{equation}
    M_{\rm DC} \simeq 1.4\msun \l \frac{m_p}{m_\chi}\r^2.
\end{equation}
where $m_p$ is the mass of the proton. The range of the dark proton mass $m_\chi$ can be obtained by setting $M_{\rm DC}=M_{\rm min}$ for the values of $M_{\rm min}$ for which the fraction of \dm that dark \bhs could compose is less than 1:

\begin{equation}
    \sqrt{\frac{1.4\msun}{M_{\rm min}^{\rm max}}} m_p < m_\chi < \sqrt{\frac{1.4\msun}{M_{\rm min}^{\rm min}}} m_p
\end{equation}
In \cite{Singh:2020wiq}, $M_{\rm min}=[0.054,1.50]\msun$, which implies $0.91< m_\chi/\gev < 4.76$ when considering both GW190425 and GW190814 \cite{Abbott:2020khf} as dark \bbh mergers.

Beyond a range of implied $M_{\rm min}$ and $m_\chi$, constraints can be placed on the fraction of \dm taht \pbhs compose as a function of the average mass of the \bbh system. These constraints are shown in \cref{fig:CODM}, assuming that (1) none of the detected \bhs by \lvk are dark, (2) half are dark, and (3) GW190425 and/or GW190814 are dark. Here, we limit ourselves to case (1) and case (3). These constraints, along with those on the dark proton mass, rely on an assumed mass distribution of dark \bhs based on Population III star formation works \cite{Bromm:2003vv}, which is chosen to be $P(m)\propto m^{-b}$ along with a mass range of $[M_{\rm min},rM_{\rm min}]$. The parameters $b=[-1,2],r=[2,100]$ determine the mass distribution of dark \bhs. Assuming all \bhs are not dark, the constraint beats existing ones by several orders of magnitude; assuming that GW190425 and GW190814 are dark \bhs, a region in this parameter space is carved out of possible average mass/ \dmh fractions. More observations of dark \bhs would be needed to further narrow this region. However, it must be emphasized that while GW190425 and GW190814 being dark \bhs are not yet excluded by other experiments, this work does not provide a way of distinguishing astrophysical \bhs from dark \bhs.

\begin{figure}[htbp]
\includegraphics[width=\columnwidth]{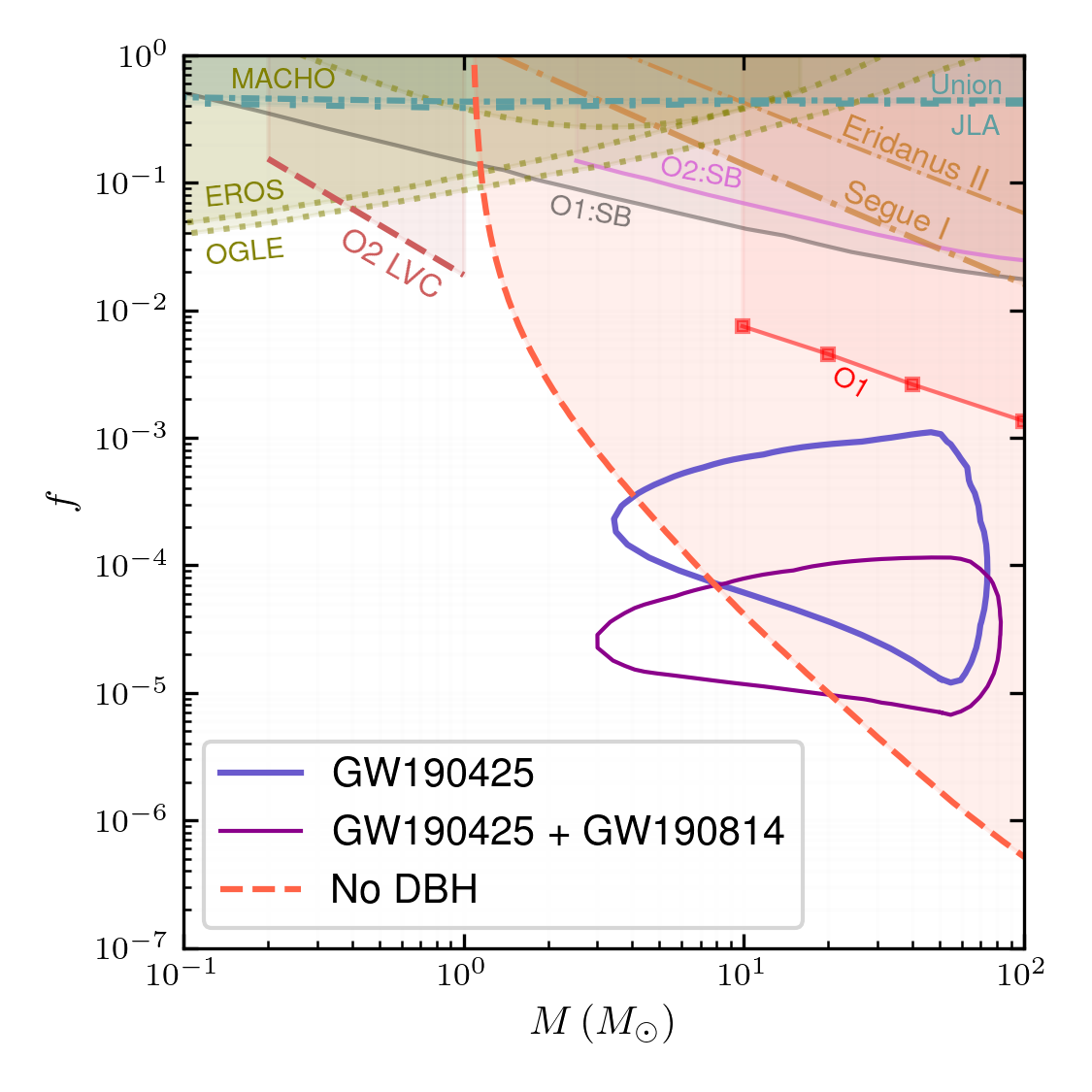}
\caption{\label{fig:CODM} Taken from \cite{Singh:2020wiq}. Constraints on the fraction of \dm that dark \bhs could compose relative to other experimental upper limits using all \bbhs detected in the first three observing runs of \lvk. These plots assume an atomic model of \dm that forms dark \bhs. The constraints are obtained in three cases: (1) assuming none of the detected \bhs are dark; (2) assuming only GW190425 is a dark \bh; and (3) assuming both GW190425 and GW190814 are dark \bhs. The latter two cases essentially arise from a joint posterior distribution conditioned on the rate densities inferred from the search results, which also constrain the minimum allowed mass in the assumed population distribution (see text for details). Even in the absence of any dark \bhs, the constraints obtained in \cite{Singh:2020wiq} significantly outperform existing microlensing experiments \cite{Macho:2000nvd,EROS-2:2006ryy,Niikura:2019kqi} and those from dynamics of dwarf galaxies \cite{Brandt:2016aco,Koushiappas:2017chw}. Additional constraints arise from \pbh searches \cite{LIGOScientific:2019kan,Gaggero:2016dpq,Raidal:2017mfl,Raidal:2018bbj,Ali-Haimoud:2017rtz}. SB stands for stochastic background and refers to analyses of the first and second observing runs of \lvk data for stochastic \gwh backgrounds from generic systems.
}
\end{figure}

\subsection{Constraints from \ssm \bbh searches}\label{subsec:dark-bbh-ssm-constraints}

The lack of smoking-gun evidence for dark \bhs in current \lvk observations motivates the need to consider \ssm \bhs, whose origins, if detected, cannot be explained by astrophysical means. Searches for \gws from \ssm compact objects in binary systems have grown in popularity in recent years, covering mass ranges of roughly $[10^{-11},1]\msun$ \cite{Nitz:2021mzz,Phukon:2021cus,LIGOScientific:2021job,Miller:2021knj,Nitz:2022ltl,LIGOScientific:2022hai,KAGRA:2022dwb,Miller:2024fpo}, and, traditionally, have put constraints on the fraction of \dm that \pbhs could compose. However, as discussed in \cite{Singh:2020wiq}, these searches do not assume a particular formation mechanism for these compact objects, and the binaries can thus be interpreted as arising from not only \pbhs, but also dark \bhs.

In the most recent analysis of \lvk data from the second half of the third observing run, constraints on dark \bhs with masses below $1\msun$ were placed, as shown in \cref{fig:dissipativeDM}, using different matched filtering methods, again assuming a power-law dependence for the mass distribution of dark \bhs and as a function of the minimum mass of this distribution. These limits do not go below $\sim 10^{-2}\msun$ because the maximum mass probed is between $[2,1000]\msun$, and so the distribution tends to disfavor smaller masses. Moreover, the limits do not go above a solar mass because the search is only for \ssm objects. 

From this work \cite{LIGOScientific:2022hai}, the mass of the dark proton has been constrained to be [0.66,8.8] GeV, which is actually broader than the constraint in \cite{Singh:2020wiq} of [0.91,4.76] GeV since there are no assumed detections of dark \bhs. 

Additionally, constraints can be placed on the energy spacing between dark atomic levels. In atomic \dmh models, the collapse of dark gas into compact stars (and hence into dark \bhs) requires efficient radiative cooling, which is governed by the energy level spacing  
\begin{equation}
    \Delta E \sim \tfrac{1}{2}\,\alpha_\chi^2 m_{e_\chi},
\end{equation}
where $\alpha_\chi$ is the dark fine-structure constant and $m_{e_\chi}$ the mass of the dark electron. By combining the Chandrasekhar mass relation with assumptions about the dark fine-structure constant, the dark fermion masses, and the Jeans mass for collapse, one can use Eq.~(1) of \cite{Singh:2020wiq} together with Eq.~(4) of \cite{Shandera:2018xkn} to infer the allowed range of $\Delta E$. For the GW190425 interpretation, this yields a characteristic value of order $\sim 10^{-3}\,\mathrm{eV}$.


\begin{figure}[htbp]
\includegraphics[width=\columnwidth]{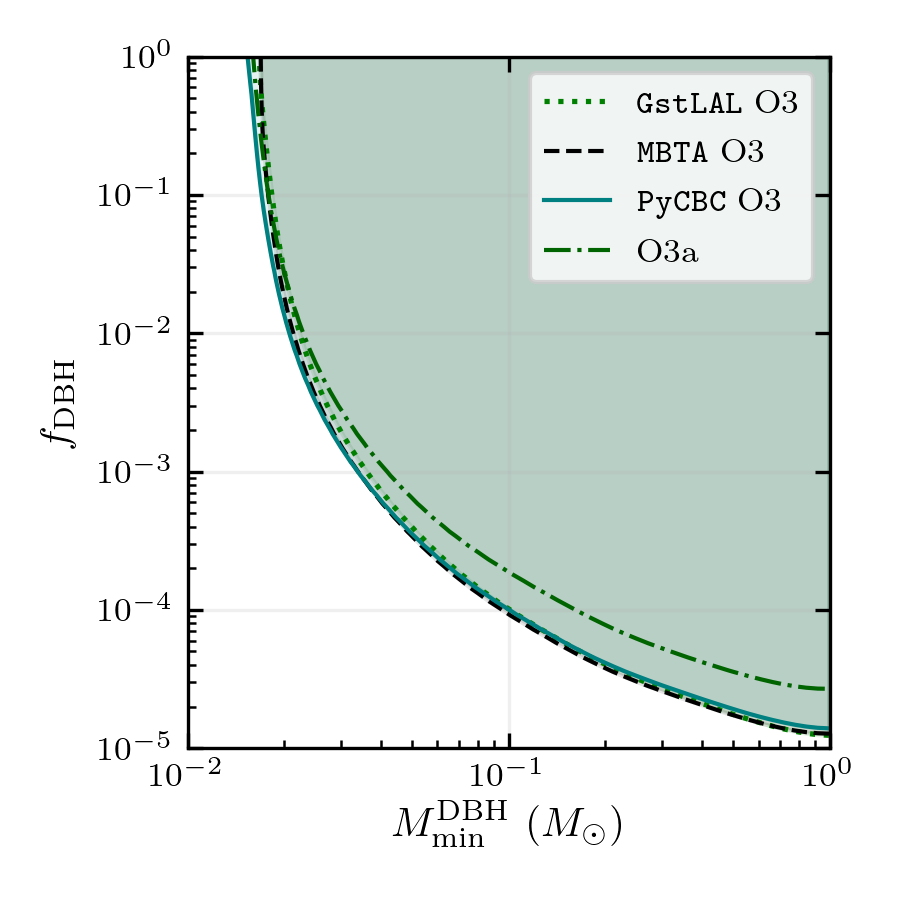}
\caption{\label{fig:dissipativeDM} Taken from \cite{LIGOScientific:2022hai}. Constraints on the fraction of \dm that dark \bhs could compose in the atomic \dm scenario. A mass distribution for dark \bhs is assumed with the shown minimum mass on the $x$-axis, and the constraints are shown for three different matched filtering algorithms: \texttt{GstLAL} (\textit{dotted}), \texttt{MBTA} (\textit{dashed}) and \texttt{PyCBC} (\textit{solid}). Constraints from a previous search for \ssm compact objects in the first half of the third observing run of \lvk are also shown ~\cite{LIGOScientific:2021job}. This plot differs from \cref{fig:CODM} primarily in the mass range that is considered in the \ssm search versus the ones in, e.g. \cite{LIGOScientific:2021djp}. At lower masses, the constraints on $f_{\rm DBH}$ weaken with respect to those at higher masses. }
\end{figure}

Although \ssm searches for dark \bhs have placed promising constraints, they are typically limited to masses above $\sim 10^{-2}\msun$, largely due to the prohibitive computational cost of conducting \mf searches for lighter-mass binaries. In contrast, \cwh methods—which search for time-frequency tracks rather than relying on \mf—can access lower-mass regimes \cite{Miller:2020kmv,KAGRA:2022dwb,Miller:2024khl,Alestas:2024ubs}. While originally developed for \pbh searches, these techniques are equally applicable to testing the dark \bh hypothesis. As such, \cwh approaches offer a valuable path toward exploring a broader parameter space than is feasible with traditional \mf methods.

\section{Conclusions}\label{sec:concl}

There is no shortage of observable signatures that \dm could leave on \gwh sources when the particle, mass and couplings of \dm the \sm are varied. We do not assume to have covered every single \dmh model in this work, but we hope to have provided an overview for the interested reader that wishes to pursue more details about different ways that \gws can be used to probe particle \dm. 

In this work, we have explored (1) how \uldm could couple to \gwh \ifos, (2) how macroscopic \dm can transit through the \ifos, (3) how \bc systems formed through a superradiance process around rotating \bhs can emit \gws by annihilation, (4) how soliton \dm can induce changes in \gws arising from binary systems and isolated \nss, (5) how WIMP \dm can collect around celestial objects and induce them to collapse into \bhs detectable via \gwh emission, (6) how environmental effects of \dmh spikes around \bhs can affect \gwh signals in future detectors, and (7) how \gwh observations and non-observations of \bbhs can be used to probe atomic models of \dm. Some of these types for \dmh have been extensively constrained through analysis of terrestrial \gwh detector data; others will require the advent of space-based and next-generation ground-based \gwh \ifos to obtain insight into these types of \dm.

Despite the ``WIMP miracle'' announced decades ago, WIMPs have not yet been detected, and are only one of numerous possible ways that \dm could exist in the Universe. The lack of detection of WIMPs over the years has opened up both the experimental and theoretical playgrounds for new probes of \dm, one of which is through its impact on \gwh signals, or as a signal in \ifo data itself. Moreover, it has been assumed in this work, and almost all those cited, that \dm is \emph{the} way to solve the puzzling cosmological observations on a variety of scales. However, modified theories of gravity are real contenders to solve this problem as well. These two communities -- \dm and modified gravity -- act entirely independently, choosing to break, respectively, one of the two assumptions of ``classical'' physics: (1) the matter in the universe is mostly baryonic and luminous, and (2) general relativity and, in the limit of slow-moving objects, Newtonian theory, are correct \cite{Martens:2020lto}. It may be that these assumptions need to be broken simultaneously, or that these two distinct fields have more in common than it appears, as argued in \cite{Martens:2020lto}.

We only briefly mentioned \pbhs in this review, which have been reviewed extensively elsewhere \cite{Green:2020jor,Miller:2024rca}. \pbhs could also comprise a fraction of or the totality of \dm, and recent observations of \bbhs  by the \lvk with low spins whose merging rates are consistent with \pbhs have reignited interested in this hypothesis \cite{Bird:2016dcv, Clesse:2016vqa,Sasaki:2016jop,Croon:2022tmr}. While analyses of \pbhs typically receive criticism for using monochromatic mass functions to place constraints, all of the results presented here implicitly assume that \dm is \emph{only} composed of particles that are being constrained\footnote{The exception to this statement is superradiance probes of \bhs -- these bosons need not be \dm.}. It could be that multiple forms of \dm co-exist, e.g. in the atomic \dmh models, or that there is a whole dark sector. 

Nevertheless, \gws provide a new window to probe the origins of \dm in a variety of ways. Most of the actual analyses performed on \gwh data do not assume particular models for \dm, and only constrain them later when one interprets results in terms of a \dmh model. \gwh searches are therefore powerful, relatively model-agnostic probes of \dm that permit the detection of a variety of \dmh signatures.


\section*{Acknowledgments}

We thank William East, Alexandre Gottel, Dana Jones, Jun'ya Kume, Sachiko Kuroyanagi, Ling Sun, and Yue Zhao for reviewing this manuscript. We also thank the three anonymous referees that provided valuable feedback on this review article.

This material is based upon work supported by NSF's LIGO Laboratory which is a major facility fully funded by the National Science Foundation

This research has made use of data, software and/or web tools obtained from the Gravitational Wave Open Science Center (https://www.gw-openscience.org/ ), a service of LIGO Laboratory, the LIGO Scientific Collaboration and the Virgo Collaboration. LIGO Laboratory and Advanced LIGO are funded by the United States National Science Foundation (NSF) as well as the Science and Technology Facilities Council (STFC) of the United Kingdom, the Max-Planck-Society (MPS), and the State of Niedersachsen/Germany for support of the construction of Advanced LIGO and construction and operation of the GEO600 detector. Additional support for Advanced LIGO was provided by the Australian Research Council. Virgo is funded, through the European Gravitational Observatory (EGO), by the French Centre National de Recherche Scientifique (CNRS), the Italian Istituto Nazionale della Fisica Nucleare (INFN) and the Dutch Nikhef, with contributions by institutions from Belgium, Germany, Greece, Hungary, Ireland, Japan, Monaco, Poland, Portugal, Spain.

We would like to thank all of the essential workers who put their health at risk during the COVID-19 pandemic, without whom we would not have been able to complete this work.

\bibliographystyle{apsrev4-1}
\bibliography{biblio}

\end{document}